\documentclass[11pt,reqno]{amsart}
\usepackage{graphics,graphicx,amsmath,amssymb,mathtools,epsfig,stmaryrd,color,flafter,xspace,lscape,threeparttable,pdfpages,csquotes,amsfonts,amsthm,geometry,array,booktabs,latexsym,caption,subfig,multirow,tabularx,rotating,cmap}
\usepackage{makecell}
\usepackage{booktabs}
\usepackage{tabularx}
\usepackage{array}
\newcolumntype{C}[1]{>{\centering\arraybackslash}m{#1}}
\usepackage[font=tiny,labelfont=bf]{caption}
\usepackage{changepage,enumitem}
\usepackage{natbib}
\DeclareCaptionFont{tiny}{\tiny}
\everymath{\displaystyle}
\RequirePackage{setspace,indentfirst,epsfig,psfrag,ifthen,ifpdf}

\newtheorem{theorem}{Theorem}
\theoremstyle{plain}

\newtheorem{assumption}{Assumption}
\newtheorem{subassumption}{Assumption\normalfont}[assumption]

\newenvironment{assumption*}
{\ifnum\value{subassumption}=0 \stepcounter{assumption}\fi\subassumption}
{\endsubassumption}
\newenvironment{assumption+}[1]
{\renewcommand{\thesubassumption}{#1}\subassumption}
{\endsubassumption}
\theoremstyle{definition}
\newtheorem{assump}{Assumption}

\newtheorem{definition}{Definition}
\newtheorem*{definition*}{Definition}
\newtheorem{example}{Example}

\newtheorem{lemma}{Lemma}

\newtheorem{proposition}{Proposition}
\newtheorem{remark}{Remark}

\newenvironment{continued}[1][continued]{\begin{trivlist}
\item[\hskip \labelsep {\bfseries #1}]}{\end{trivlist}}

\newcommand\independent{\protect\mathpalette{\protect\independenT}{\perp}}
\def\independenT#1#2{\mathrel{\mathbb{R}lap{$#1#2$}\mkern2mu{#1#2}}}
\numberwithin{equation}{section}
\newcommand{\norm}[1]{\left\Vert #1\right\Vert }

\newcommand{\ud}{\mathrm{d}}

\newcounter{steps}

\DeclareMathOperator*{\argmax}{arg\,max}

\def\independenT#1#2{\mathrel{\rlap{$#1#2$}\mkern2mu{#1#2}}}

\begin{document}
\title{Finite-sample inference in partially identified and incomplete models}
\author{Lixiong Li and Marc Henry}
\address{Johns Hopkins and Penn State}
\thanks{The first version is of February 14, 2022. This version is of \today. The authors thank Ju Hyun Oh for excellent research assistance. The authors also gratefully acknowledge helpful comments from St\'ephane Bonhomme, Xavier d'Haultfoeuille, Moyu Liao, Francesca Molinari, Ulrich M\"{u}ller, Charles Murry, and seminar audiences at Brown, CEME 2022, Chicago, Cornell-PennState 2022, GNYEC 2022, the John Morgan Memorial Conference, Johns Hopkins, the Kantorovich Initiative, MEG 2022, NASMES 2022, the Paris Econometrics Seminar, TSE, UCSC, UKyoto, UTokyo, and Yale. The usual disclaimer applies. Corresponding author: Lixiong Li: \texttt{lixiong.li@jhu.edu}
Department of Economics, Johns Hopkins University, Wyman Park Building 5th Floor, 3100 Wyman Park Drive, Baltimore, MD 21211.}

\begin{abstract}

We propose confidence regions for the parameters of incomplete models with exact coverage of the true parameter in finite samples. Our confidence region inverts a test, which generalizes Monte Carlo tests to incomplete models. The test statistic is a discrete analogue of a new optimal-transport formulation of the structural model. Both test statistic and critical values rely on simulation draws from the distribution of latent variables and are computed using solutions to discrete optimal transport, hence linear programming problems. 
We also propose a fast preliminary search in the parameter space with an alternative, more conservative yet consistent test, based on a parameter-free critical value. We show that this test is maximally consistent in the following sense: under infinite exchangeability, it consistently rejects every data-generating process that some test with finite-sample size control consistently
rejects.
We compare size and power of our procedure with competing proposals in simulations based on a regression with interval valued regressors. A second simulation based on latent consideration sets shows that dependent selection can severely distort inference calibrated under i.i.d. sampling, while our procedure retains finite-sample size control. Finally, we apply our methodology to the model of airline entry and price competition in \cite{CMT:2021}.

\vskip20pt

\noindent\textit{Keywords}: Incomplete models, set prediction, multiple equilibria, data/model compatibility, simulation-based testing, finite-sample inference, optimal transport.

\vskip10pt

\noindent\textit{JEL codes}: C15, C57, C61

\end{abstract}

\maketitle

\section*{Introduction}

In this paper, we study a class of incomplete econometric models that combines (i) a restriction on the support of the random variables involved in the model specification, and (ii) a restriction on the distribution of those variables in the model, that the analyst cannot observe. The support restriction is implied by economic theory, and usually involves the implications of behavioral assumptions, equilibrium concepts and structural features of the economic environment. A game of perfect information with a pure-strategy equilibrium concept, as in \cite{Jovanovic:89} and \cite{Tamer:2003}, is a prime example. Other examples include models of choice with limited attention, as in \cite{BCMT:2021}, discrete choice with endogeneity, as in \cite{CRS:2011}, auction models, as in \cite{HT:2003}, network formation, as in \cite{dPRST:2018}, and structural vector autoregressions, as in \cite{GK:2021} and \cite{GKR:2021}. \cite{Molinari:2020} and \cite{CR:2020} provide comprehensive surveys of the literature on incomplete structural models. The procedure we propose in this paper applies to parametric incomplete structural models: Both the support restriction and the distribution of unobserved heterogeneity are known up to a finite-dimensional parameter vector. It is ill-suited to semiparametric extensions, where either the support restriction or the distribution of unobserved heterogeneity depends on an infinite-dimensional parameter.

Incomplete structural models are called {\em incomplete} because the model structure predicts a set of possible values for the outcome variables. Incompleteness arises because of multiple equilibria in games, unobserved heterogeneity in choice sets in limited attention models, interval predictions in auctions, and unknown sample selection mechanisms. Model incompleteness generally leads to partial identification, where more than one value of the model parameter could have given rise to the true data-generating process for the observed variables. However, model incompleteness and partial identification are distinct concepts.

The current state of the art in deriving confidence regions for the parameters of incomplete structural models involves the \cite{BMM:2011}-\cite{GH:2011} characterization of the empirical content of the model as a collection of conditional moment inequality restrictions, and the application of one of the existing inference methods with conditional moment inequality models, surveyed in \cite{CS:2018}, \cite{Molinari:2020} and \cite{shi2025inference}. This method, however, results in a very large, possibly infinite, number of conditional moment inequalities. Even in cases, where the endogenous variables are discrete, such as discrete games, the number of moment inequalities increases exponentially in the number of strategy profiles. 

The challenge is both computational and statistical, as the number of inequalities may be much larger than the sample size, requiring new methods, such as \cite{CCK:2019}. Basing inference on a reduced non-sharp collection of inequalities leads to low power and loss of robustness to misspecification. See \cite{li2024discordant} for a discussion. Methods to reduce the number of conditional moment inequalities without losing sharpness exist. They are based on core determining classes, as proposed in \cite{GH:2011} and further developed in \cite{CRS:2011}, \cite{CR:2017e}, \cite{LW:2017}, \cite{MM:2018} and \cite{Ponomarev:2022}.\footnote{\cite{GRS:2022} characterize the empirical content of incomplete models using minimal relevant partitions, as an alternative to core determining classes.} However, these methods are complex, model specific, and only partially alleviate the problem. 
In addition, when the conditional moment inequalities are transformed into unconditional ones, as in \cite{AS:2013}, sharpness is preserved only when the number of moment inequalities increases with sample size, which induces an extra layer of computational burden.\footnote{The dimensionality of the conditioning set in such models generally precludes the alternative approach to conditional moment inequalities, which involves estimating them, as in \cite{CLR:2009} and \cite{linton2023testing}.} Moreover, inference methods in moment inequalities rely on asymptotic arguments and some user-chosen tuning parameter to preselect inequalities that are close to binding in the sample and thereby avoid overly conservative inference.

We propose an alternative method to construct confidence regions for the parameters of incomplete structural models that circumvents the many moments and conditioning issues, allows for continuous outcome variables, and avoids tuning parameters and asymptotic arguments. As is customary with moment inequality models, we construct our confidence region by inverting a test. The test we propose is based on a new test statistic and it controls size in finite samples. Our testing procedure relies on two key ingredients. First, the test statistic is based on an optimal-transport characterization of data/model compatibility, inspired by formulations in \cite{GH:2006} and \cite{EGH:2010}. As a result, the test statistic is the solution of a discrete optimal-transport problem, which is a special kind of linear programming problem, the computation of which has a long history. Second, the test generalizes Monte Carlo tests of \cite{Dwass:57} and \cite{Barnard:63}\footnote{See also \cite{Dufour:2006} and \cite{DK:2001}.} to incomplete models to control size in finite samples. The test statistic and critical values are based on simulation draws from the known conditional distribution of latent variables.

Our test controls size, and therefore controls the coverage probability of the confidence region for any finite-sample size. To the best of our knowledge, this is the first procedure to derive confidence regions for the true value of the structural parameters of incomplete models that are valid and exact in finite samples. In that respect, the closest related paper in the literature is \cite{KZ:2019}, which proposes a small-sample minimax optimal test of simple hypotheses in incomplete models with finite outcome space~$\mathcal Y$. When the outcome and covariate spaces are finite, the model specification in \cite{KZ:2019} is the same as ours.
Non-asymptotic results in the broader partial identification literature include \cite{CS:2022}, who provide an inference method for a class of partially identified models that requires no tuning parameter, and achieves exact finite-sample size in normal models. \cite{CLR:2009} and \cite{CCK:2019} derive nonasymptotic bounds on the rejection probabilities of their confidence regions. These bounds are useful to derive asymptotic rates of convergence, not for finite-sample inference. \cite{rosen2025finite} provide finite-sample inference for the maximum score.
\cite{CCT:2018} is also related to our procedure, as their asymptotically exact set inference (for the full parameter vector or subvectors) is based on Monte Carlo simulations from quasi-posteriors.

Finite-sample validity has several advantages, beyond the obvious benefit of avoiding reliance on often questionable asymptotic approximations. First, the support constraint and the dimension of the vector of unobservables may change with sample size, as would arise in applications to games on networks and network formation games. Second, our finite-sample validity result requires no additional restriction on the dependence between endogenous observations in the sample. This property is particularly desirable with incomplete models. As discussed in \cite{EKS:2016}, it is hard to reconcile the customary stationarity and independence or mixing assumptions across units of observation with total ignorance of the mechanism that selected each realization from the model prediction set. The degree of dependence between observations does not affect size control of our procedure. 
We investigate asymptotic power of our test and show that it consistently rejects a candidate value~$\theta$ whenever the empirical distribution of the observations remains bounded away from the set of one-observation distributions compatible with the model at~$\theta$. Our consistency result allows general triangular arrays and does not require stationarity or ergodicity of the observed outcomes. We complement it with an maximal consistency result: when the observations are the first~$n$ coordinates of an infinitely exchangeable process, our test consistently rejects every data-generating process that some test with finite-sample size control consistently rejects.

Our method requires a search in the space of parameters. At each value of the parameter in the search, we must compute a test statistic and a critical value. This computational burden is shared by inference methods in partially identified models, where the objective is coverage of the true value of the parameter. 
In order to accelerate the search, we also propose a conservative superset of our confidence region. The conservative superset is based on a parameter-free critical value. Once this conservative confidence region is computed, all values of the parameter that lie outside of it can be excluded a priori from the exact confidence region in our main proposal.

We provide two extensive simulation studies of our procedure. The first, in section~\ref{sec:MC-interval}, is based on our pilot interval-valued covariate model. We investigate the effect of the choice of discrepancy in the definition of the test statistic, and compare our procedure in terms of size, power, and computing time with the current state of the art procedures proposed for the case of many moment inequalities in \cite{CCK:2019} and \cite{andrews2017inference}. The second, in section~\ref{sec:MC-consideration}, examines a latent consideration-set model in which a common selection rule induces dependence across observations. In this design, moment-inequality inference calibrated under i.i.d. sampling overrejects, whereas our procedure retains finite-sample size control while still yielding informative confidence regions.

The study of power is particularly important, since our theoretical results on power are limited to consistency of the test and maximal consistency in the sense that under infinite exchangeability, our procedure ultimately rejects any parameter value that some procedure with finite sample size control ultimately rejects. Recent progress has been made on inference in incomplete models with optimality properties in \citeauthor{KZ:2019} [\citeyear{KZ:2019,kaido2025universal}]. \cite{KZ:2019} derive minimax optimal tests for simple hypotheses in incomplete models with finite outcome space and derive their asymptotic local power. Minimax optimality in the context of simple hypothesis testing for parameters of incomplete models means that the test maximizes worst case power under the alternative (i.e., among all possible DGPs predicted under the alternative parameter value), while controlling worst case size under the null (i.e., such that the largest rejection probability among DGPs predicted under the null parameter value is smaller than or equal to nominal size). \cite{KZ:2019} also derive asymptotic local power properties of their minimax optimal tests, with extensions to composite hypotheses. \cite{CK:2022} apply the theory of \cite{KZ:2019} to test model incompleteness, thereby leveraging the completeness of the model under the null hypothesis. \cite{kaido2025universal} apply the framework of \cite{KZ:2019} to develop a finite-sample valid and asymptotically exact subvector inference procedure for simple hypotheses in incomplete models with finite outcome space~$\mathcal Y$. In related work, \cite{KM:2024} propose misspecification robust inference in incomplete models based on a relative entropy projection of the empirical distribution on the set of predicted data-generating processes. 

Finally, we illustrate the implementation of our procedure in the structural model of airline entry and price competition in \cite{CMT:2021}. We apply our procedure to the empirical model of \cite{CMT:2021} without modification, and we use exactly the same data set. \cite{CMT:2021} apply the methodology in \cite{CHT:2007}. However, they recenter their criterion function around its minimum and report a parameter region that can be loosely interpreted as a confidence region for the pseudo-true value of the parameter vector. In contrast, we derive a confidence region with finite-sample valid and exact coverage of the true value and we project it on each dimension for comparison with the region reported in \cite{CMT:2021}.

\subsubsection*{Notation and preliminaries}

All random elements considered in the paper are defined on a common complete probability space~$(\Omega,\mathcal F,\mathbb P)$. All vectors are written as \emph{row} vectors throughout. We use~$^\top$ to denote transposition and omit it when the format is clear from the context. Throughout the paper, $(Y,X,U)$ will denote a random vector on~$\mathcal Y \times \mathcal X\times \mathcal U$, and~$\theta \in\Theta$ a fixed parameter vector, where~$\mathcal Y\subseteq \mathbb R^{d_Y}$, $\mathcal X\subseteq \mathbb R^{d_X}$, $\mathcal U\subseteq \mathbb R^{d_U}$, and~$\Theta\subseteq \mathbb R^{d_\theta}$. We assume that~$\mathcal Y$, $\mathcal X$, and~$\mathcal U$, endowed with their relative Euclidean topologies, are Polish spaces. Their finite products are therefore Polish as well. 
We will denote~$\mathcal Q$ and $\mathcal P$ the collections of Borel probability measures on~$\mathcal U\times \mathcal X$ and~$\mathcal Y\times \mathcal X$ respectively. Both are endowed throughout with the topology of weak convergence and the corresponding Borel sigma-field. Since $\mathcal U\times\mathcal X$ and~$\mathcal Y\times\mathcal X$ are Polish, so are $\mathcal Q$ and~$\mathcal P$.
$\mathcal M(Q,P)$ is the set of probability measures on $(\mathcal U\times \mathcal X)\times(\mathcal Y\times \mathcal X)$ with marginals~$Q$ on~$\mathcal U \times \mathcal X$ and~$P$ on~$\mathcal Y \times \mathcal X$. We denote~$Q^{\otimes n}:=Q\otimes\cdots\otimes Q$ the~$n$-fold product measure. The notation~$X\sim P$ indicates that the random element~$X$ is distributed according to~$P$. The collection of closed subsets of a set~$\mathcal A$ is denoted~$\mathcal G(\mathcal A)$. The convex hull of a set~$A$ is denoted~co$A$. We denote~$\mathcal M_n^+$ the set of~$n\times n$ nonnegative matrices, and~$\Pi_n$ the subset of~$\mathcal M_n^+$ containing matrices~$\pi$ such that~$n\pi$ is doubly stochastic, i.e., such that $\Sigma_i\pi_{ij}  = 
\Sigma_j\pi_{ij} = 1/n$,  for all $i,j \leq n.$ Finally,~$\lfloor a \rfloor$ (resp.~$\lceil a \rceil$) denote the component-wise integer part (resp. ceiling) of a vector~$a$. 

\subsubsection*{Overview}

Section~\ref{sec:model} defines the model and the target of inference. Section~\ref{sec:inference} presents the finite-sample inference procedure, together with its finite-sample size control, its consistency, and its maximal consistency under exchangeability. Section~\ref{sec:refine} proposes refinements of the procedure. It discusses the case of parametric latent variables and discrete outcomes. Section~\ref{sec:MC} contains two simulation exercises, the first based on a regression with interval censored covariates in section~\ref{sec:MC-interval}, the second on a latent consideration-set model with dependent selection in section~\ref{sec:MC-consideration}, and section~\ref{sec:empirical} illustrates the procedure on the model and data from \cite{CMT:2021}. Proofs are collected in appendices~\ref{sec:proofs} and~\ref{sec:lemmas}. Appendix~\ref{sec:add} gives a version of the size control result conditional on~$X^{(n)}$, appendix~\ref{sec:num} describes the numerical implementation, and appendix~\ref{sec:cs-appendix} collects the size-distortion result and the computational reductions for the consideration-set simulation.

\section{Theoretical framework}
\label{sec:model}

\subsection{Theoretical structural model} 

We restrict attention to the class of parametric incomplete structural models introduced in \cite{Jovanovic:89}.
The vector of variables of interest~$(Y,X,U)\in\mathcal Y \times \mathcal X \times \mathcal U$ satisfies support constraint $(Y,X,U) \in \Gamma(\theta)\subseteq \mathcal Y \times \mathcal X \times \mathcal U$, and~$U$ has fixed and known distribution~$Q_U$.\footnote{This is without loss of generality as we explain in section~\ref{sec:parametric}.} The object of inference is the finite-dimensional parameter~$\theta\in\Theta$. Both vectors of variables~$Y$ and~$X$ are observed, in the sense that available data consists in a sample~$((Y_1,X_1),\ldots,(Y_n,X_n))$. The variables in vector~$U$ are unobserved. The variables in vector~$X$ are exogenous\footnote{We show in section~\ref{sec:parametric} that the distribution of~$U$ may depend on~$X$ as long as it is known up to a finite-dimensional parameter vector.} in the sense that~$U\independent X$. All endogenous variables are subsumed in vector~$Y$. 

The model is incomplete in that multiple values of endogenous variables may be consistent with a single value of exogenous and unobserved variables. This is reflected in the fact that
the set $\{ y\in \mathcal Y: (y,x,u)\in\Gamma(\theta) \}$ may not be a singleton for all~$(u,x)\in\mathcal U\times\mathcal X$. This corresponds to the fact that the model fails to produce a unique prediction.

\subsection{Examples} 

Incomplete models as described above encompass examples as diverse as static simultaneous move games with complete information and pure strategy equilibrium concepts, choice models with limited attention or partially observed consideration sets, and auctions with independent private values. Section~3 in \cite{Molinari:2020} gives a detailed account of such incomplete structural models with extensive references. 
In what follows, we concentrate on two recent structural examples and then introduce a simple running example used throughout the paper.

\begin{example}[Discrete choice with unobserved heterogeneity in consideration sets]
\label{ex:consideration}
We set out the structural model in \cite{BCMT:2021} in their notation, before translating it into our framework.
Consider a finite set of alternatives~$\mathcal D$ for a decision maker to choose from. A decision maker is characterized by
observed covariates~$X$, unobserved random vector~$\nu$ with distribution\footnote{Section~\ref{sec:parametric} shows how to handle the case, where the distribution of the unobserved random vector~$\nu$ depends on an unknown parameter vector.}~$P_\nu$, and a latent choice set~$G\subseteq\mathcal D$.
Let~$\delta$ be a fixed unknown parameter vector\footnote{The notation~$\delta$ is taken from \cite{BCMT:2021}. It will not be used outside this example, in order not to confuse it with our notation for the discrepancy function in the rest of the paper.}. The decision maker chooses~$d$ to maximize utility. Formally,~$d^\ast(G,X,\nu;\delta):=\mbox{arg}\max_{c\in G}W(c,X,\nu;\delta)$, where the maximization of discrete choice $c$ is over the latent choice set~$G$.
This unobserved heterogeneity in choice sets is the driver of incompleteness in this model. It is disciplined by the assumption that the realized choice set~$G\subseteq\mathcal D$ under consideration satisfies~$\mathbb P(\vert G \vert \geq \kappa)=1$, for some fixed $\kappa\geq 2$ imposed by the researcher. 
The model therefore stipulates that the observed choice~$d$ must be in 
\begin{eqnarray*}
D_\kappa^\ast := \bigcup_{G\subseteq \mathcal D: \vert G \vert \geq \kappa}\left\{ d^\ast(G,X,\nu;\delta) \right\} = \bigcup_{G\subseteq \mathcal D: \vert G \vert = \kappa}\left\{ d^\ast(G,X,\nu;\delta) \right\},
\end{eqnarray*}
where the equality follows from Sen's property~$\alpha$, as shown in \cite{BCMT:2021}.\footnote{Sen's property~$\alpha$ is the independence of irrelevant alternatives of individual choice theory.}
This example fits into the current framework, with~$Y:=d$, $U:=\nu$, $\theta=\delta$,
$\Gamma(\theta):=\{ (y,x,u): y\in D_\kappa^\ast\}$, and~$Q_U:=P_\nu$.
\end{example}

\begin{example}[Market Structure and Competition in Airline Markets]
\label{ex:airline}
Once again, we set out the structural model in the notation of \cite{CMT:2021}, before translating it into our framework.
Six firms, indexed by~$j$ decide whether to enter a market based on the profit they expect under optimal pricing. If firm~$j$ enters, it faces demand~$s_j(P,A,y,\xi;\beta)$, which is a function of the vector of endogenous prices~$P$, the vector of exogenous demand relevant firm characteristics~$A$, the binary entry decisions~$y$ of all firms, unobservable demand shocks~$\xi$ and parameter vector~$\beta$. Fixed costs of entry for firm~$j$, is $F(Z_j,\nu_j;\gamma)$, and marginal unit cost of production is~$c(W_j,\eta_j;\delta)$, where~$W$ and~$Z$ are the vectors of exogenous observed cost shifters, $\nu$, $\eta$ are unobserved cost shifters, and~$\gamma,\delta,$ are parameters.
Structural model constraints include for each firm~$j$: equality of predicted and realized demand share
\begin{eqnarray}
\label{eq:demand}
S_j = s_j(P,A,y,\xi;\beta),
\end{eqnarray} 
an entry condition, namely $y_j=1$ if and only if
\begin{eqnarray}
\label{eq:entry}
\pi_j := (P_j-c(W_j,\eta_j;\delta))M s_j(P,A,y,\xi;\beta)-F(Z_j,\nu_j;\gamma) \geq0,
\end{eqnarray}
and zero otherwise, where~$M$ is observed market size, and an equilibrium pricing condition in case of entry
\begin{eqnarray}
\label{eq:pricing}
(P_j-c(W_j,\eta_j;\delta))\frac{\partial s_j}{\partial p_j}(P,A,y,\xi;\beta) + s_j(P,A,y,\xi;\beta) & = & 0.
\end{eqnarray}
This example fits into the current framework with the following notation correspondence: $(y,yS,yP)$ is the endogenous vector~$Y$, $(M,A,W,Z)$ is the vector of covariates~$X$, and $\Sigma^{-\frac{1}{2}}(\xi,\eta,\nu)$ is the vector~$U$ of latent variables with multivariate standard normal distribution~$Q_U:= N(0,I)$. The parameter vector~$\theta$ includes~$\beta,\gamma,\delta$, and~$\Sigma$. The structural model correspondence is~$\Gamma((\beta,\gamma,\delta,\Sigma)):=\{ (Y,X,U) : Y=(y,yS,yP),X=(M,A,W,Z),U=\Sigma^{-\frac{1}{2}}(\xi,\eta,\nu) \mbox{ and } (\ref{eq:demand})-(\ref{eq:pricing}) \mbox{ hold for all }j \}$.
\end{example}

The final example in this section is a simple parametric regression with interval censored covariates, which will serve as a running example to illustrate several aspects of our inference procedure. 

\begin{example}[Regression with interval censored covariates]
\label{ex:interval}
Consider the regression model $Y=\theta W^\ast+U$. Assume~$U\sim N(0,1)$. Assume~$\theta\geq0$ to fix ideas\footnote{In the simulations, we allow negative true values for~$\theta$.}. Covariate~$W^\ast$ is unobserved, but known to belong to~$[\underline W,\overline W]$, with~$(\underline W,\overline W)$ observed. This model fits into our theoretical framework with~$X:=(\underline W,\overline W)$ and structural model correspondence~$\Gamma(\theta):=\{(Y,X,U): X=(\underline W,\overline W); Y\in[\theta\underline W+U,\theta\overline W+U]\}$. The model extends straightforwardly to the more general regression~$Y=\alpha+\beta W^\ast+\varepsilon$ with~$\varepsilon\sim N(0,\sigma^2)$, in which case the model fits in our framework with~$U:=\varepsilon/\sigma$ and~$\theta:=(\alpha,\beta,\sigma^2)$. However, since this model is used for illustrative purposes throughout the paper, we minimize notation in what follows and stick to the simplified version.
\end{example}

\subsection{Target of inference}
\label{sec:sharp}

We consider a sample of size~$n$ of observed variables,
\[
(Y^{(n)},X^{(n)}):=((Y_1,\ldots,Y_n),(X_1,\ldots,X_n)),
\]
with unknown true distribution~$P_0^{(n)}$ on~$\mathcal Y^n\times\mathcal X^n$. We refer to~$P_0^{(n)}$ as the true \emph{sample-level} data-generating process.
An incomplete structural model imposes restrictions on the true sample-level data-generating process~$P_0^{(n)}$. These restrictions are known up to an unknown vector~$\theta$ of structural parameters. For each value of~$\theta$, the structural restrictions induce a set~$\mathcal P_\theta^{(n)}$ of sample-level observed-data distributions consistent with the model, defined as follows.

\begin{definition}[Structural model]
\label{def:DGP}
For each~$\theta\in\Theta$, $\mathcal P_\theta^{(n)}$ is the set of distributions~$P^{(n)}$ on~$\mathcal Y^n\times \mathcal X^n$ for which there exists a random vector~$(Y^{(n)},X^{(n)},U^{(n)})$ satisfying:
\begin{enumerate}
\item Data compatibility: $(Y^{(n)},X^{(n)})\sim P^{(n)}$;
\item Support restriction: for each~$i\leq n$, $(Y_i,X_i,U_i)\in\Gamma(\theta)\subseteq\mathcal Y\times\mathcal X\times\mathcal U$ almost surely;
\item Latent-variable restriction: $U^{(n)}\sim Q_U^{\otimes n}$ and~$U^{(n)}\independent X^{(n)}$.
\end{enumerate}
\end{definition}

If the model is correctly specified, there exists a true value~$\theta_0\in\Theta$ such that the model restrictions are satisfied, which is to say that the true sample-level data-generating process~$P_0^{(n)}$ belongs to the model-predicted set~$\mathcal P_{\theta_0}^{(n)}$. Ultimately, the object of interest is the true structural parameter~$\theta_0$. However, there may be other parameter values~$\theta'\in\Theta$ such that~$P_0^{(n)}$ also belongs to~$\mathcal P_{\theta'}^{(n)}$. In that case,~$P_0^{(n)}\in\mathcal P_{\theta_0}^{(n)}\cap\mathcal P_{\theta'}^{(n)}$, and~$\theta'$ is observationally equivalent to~$\theta_0$. Any parameter value~$\theta$ such that~$P_0^{(n)}\in\mathcal P_\theta^{(n)}$ could be the true value; no other parameter value could. Thus, to conduct inference on~$\theta_0$, we test the validity of the model structure at each candidate value~$\theta\in\Theta$. For a fixed~$\theta$, this amounts to testing the null hypothesis of {\em data/model compatibility},
\begin{eqnarray}
\label{eq:null}
H_0(\theta):\; P_0^{(n)}\in\mathcal P_\theta^{(n)},
\end{eqnarray}
namely the hypothesis that the observed data could have been generated by one of the data-generating processes predicted by the structural model at~$\theta$. This is a joint test of the behavioral assumptions (represented by the support restriction in condition~(2) of Definition~\ref{def:DGP}), the distributional assumptions on the latent variables (represented by condition~(3)), and the specific parameter value~$\theta$ under consideration. The confidence region we construct in section~\ref{sec:inference} collects all values of~$\theta$ at which the null hypothesis~(\ref{eq:null}) is not rejected. It thereby covers each parameter value that is compatible with the true sample-level data-generating process---in particular the true value~$\theta_0$ and every value observationally equivalent to it---with prescribed probability.

When~$n=1$, we drop the superscript and write~$\mathcal P_\theta:=\mathcal P_\theta^{(1)}$ for the set of single-observation distributions on~$\mathcal Y\times\mathcal X$ that are compatible with the model at~$\theta$. Compatibility of a distribution~$P_0$ with the model at~$\theta$, i.e., the statement~$P_0\in\mathcal P_\theta$, is the notion of data/model consistency underlying the analysis of incomplete models in \cite{BMM:2011} and \cite{GH:2011} and the subsequent literature. If the true data-generating process~$P_0^{(n)}$ is i.i.d., compatibility of the sample-level distribution reduces to compatibility of the single-observation marginal.\footnote{For completeness, we prove this statement as proposition~\ref{prop:iid} in the appendix.}

In general, however, the definition of~$\mathcal P_\theta^{(n)}$ does not restrict how each endogenous outcome~$Y_i$, $i=1,\ldots,n,$ is selected from the set of predicted outcomes~$\{y\in\mathcal Y:(y,X_i,U_i)\in\Gamma(\theta)\}$. 
We refer to the mechanism through which~$Y_i$ is selected from the predicted set as the {\em selection mechanism} for the~$i$th observational unit. The joint selection mechanisms for~$(Y_1,\ldots,Y_n)$ may be arbitrarily dependent and heterogeneously distributed across observational units, conditionally on being supported on corresponding set of predicted outcomes. In Example~\ref{ex:consideration}, this allows the latent consideration sets of different individuals to depend on a common latent factor or to depend on peers' preferences on a latent network. In Example~\ref{ex:airline}, it allows equilibrium selection in one airline market to be related to equilibrium selection in other markets.\footnote{In the same incomplete model framework, \cite{EKS:2016} also recognize that the structure of definition~\ref{def:DGP} is compatible with correlated and nonstationary outcomes. However, their inference is based on asymptotic arguments.} 
In most applications of structural models, dependence across observational units arises naturally. This dependence can induce size distortions in finite samples if it is not taken into consideration when designing inference procedures, as we show in the simulation exercise of section~\ref{sec:MC-consideration}.

\section{Finite-sample inference}
\label{sec:inference}

The objective of this section is to provide a confidence region for the parameters of interest~$\theta$. We are going to provide a confidence region~$CR_n$ that covers each value of the parameter~$\theta$ compatible with the true sample-level data-generating process, i.e., each~$\theta$ such that~$P_0^{(n)}\in\mathcal P_\theta^{(n)}$, with probability at least as large as the nominal level~$1-\alpha$. Additionally, we want exact inference, by which we mean that there exists a data-generating process under which the confidence region~$CR_n$ covers the true value~$\theta_0$ with probability exactly equal to the nominal level. 

The confidence region~$CR_n$ is obtained by test inversion, as in \cite{AR:49}. For each value of~$\theta$, we test the null hypothesis of data/model compatibility~$H_0(\theta)$ defined in~(\ref{eq:null}).
The hypothesis is rejected and~$\theta$ deemed outside the confidence region if and only if the test statistic~$T_n(\theta)$, a function of the sample~$((Y_1,X_1),\ldots,(Y_n,X_n))$, is larger than a corresponding critical value. We consider two kinds of critical values: A parameter-free critical value~$c_{n,1-\alpha}^0$ and parameter dependent critical values~$c_{n,1-\alpha}(\theta)$. Hence, we consider two confidence regions
\begin{eqnarray}
\label{eq:CR0}
CR_n^0 & := & \{ \theta \in \Theta: T_n(\theta) \leq c_{n,1-\alpha}^0\}\\
\label{eq:CR}
CR_n & := & \{ \theta \in \Theta: T_n(\theta) \leq c_{n,1-\alpha}(\theta)\}.
\end{eqnarray}
We achieve finite-sample valid inference with parameter-free critical values and exact inference with parameter-dependent critical values by extending the traditional Monte Carlo tests of \cite{Dwass:57} and \cite{Barnard:63} to incomplete models. Our proposed inference on the true value of the structural parameter combines the parameter-free critical value with the parameter dependent critical value. The former allows very fast initial search in the parameter space, since the critical value is only computed once for all values of~$\theta$. Then, given that~$c_{n,1-\alpha}(\theta)\leq c_{n,1-\alpha}^0$ the parameter dependent critical value~$c_{n,1-\alpha}(\theta)$ need only be computed for~$\theta\in CR_n^0$, which greatly reduces the computational burden needed to achieve exact finite-sample inference.
The rest of this section is devoted to constructing the test statistic and the critical values.

\subsection{Test statistic}\label{sec:test}

\subsubsection*{Optimal-transport characterization of data/model compatibility}

Our test statistic is based on a characterization of data/model compatibility in terms of the value of an optimal-transport problem. As such, it is related to the characterizations in \cite{GH:2006} and \cite{EGH:2010}. However, our treatment of exogenous covariates~$X$ differs crucially from previous proposals. We encode the role of~$X$ directly through a reformulation of the model's support constraint. 
Define the correspondences~$\Gamma_u$ and~$\Gamma_y$ between~$\mathcal Y \times \mathcal X$ and~$\mathcal U \times \mathcal X$ by:
\begin{eqnarray}
\label{eq:Guy}
\begin{array}{lcl}
\Gamma_y(u,x;\theta) & = & \{ (y,x^\prime)\in\mathcal Y\times \mathcal X: x^\prime=x \mbox{ and } (y,x,u)\in\Gamma(\theta)\},\\
\Gamma_u(y,x;\theta) & = & \{ (u,x')\in\mathcal U\times \mathcal X: x'=x \mbox{ and } (y,x,u)\in\Gamma(\theta)\}.
\end{array}
\end{eqnarray}
Correspondence~$\Gamma_y$ defines the set of model predictions for the endogenous variables, whereas correspondence~$\Gamma_u$ defines the set of latent variables that can rationalize the data. We define the correspondences between~$\mathcal Y\times\mathcal X$ and $\mathcal U\times\mathcal X$ instead of simply~$\mathcal Y$ and~$\mathcal U$ in order to avoid conditioning on~$X$. 

Using the notation in~\eqref{eq:Guy}, write~$v:=(u,x)\in\mathcal U\times\mathcal X$ and~$w:=(y,x^\prime)\in\mathcal Y\times\mathcal X$. A pair~$(v,w)$ satisfies the model support restriction at~$\theta$ if and only if~$v\in\Gamma_u(w;\theta)$. This encodes the support restriction~(2) of Definition~\ref{def:DGP}. We measure violations of this restriction by a discrepancy~$\delta$ that vanishes exactly on the graph of~$\Gamma_u$.\footnote{The construction and choice of discrepancy are discussed below.}
\begin{assumption}\label{ass:discrepancy}
    For all~$\theta\in\Theta$, the function~$\delta(\cdot,\cdot;\theta):(v,w)\in(\mathcal U\times\mathcal X)\times(\mathcal Y\times\mathcal X)\mapsto \delta(v,w;\theta)$ is nonnegative, lower semi-continuous, and equal to zero if and only if~$v\in\Gamma_u(w;\theta)$.
\end{assumption}
A discrepancy function satisfying assumption~\ref{ass:discrepancy} exists if and only if~$\Gamma(\theta)$ is closed for all $\theta\in \Theta$. For completeness, we add a proof of this statement as proposition~\ref{prop:closed} in the appendix.

In addition to support restriction~(2), Definition~\ref{def:DGP} also imposes a data compatibility restriction~(1) and a latent variable restriction~(3). These are marginal restrictions on the set of joint distributions for~$(Y,X,U)$. Let~$Q$ (resp.~$P$) be a probability distribution on~$\mathcal U\times\mathcal X$ (resp.~$\mathcal Y\times\mathcal X$). Let~$\mathcal M(Q,P)$ be defined as the set of joint distributions with marginals~$Q$ and~$P$ (see notations and preliminaries). If there exists a pair~$(V,W)$ with~$V\sim Q$ and~$W\sim P$ and~$V\in\Gamma_u(W;\theta)$ a.s., then the following must hold:
\begin{eqnarray}
\label{eq:char1}
\mathcal D(Q,P;\theta) & := & \inf_{\pi\in\mathcal M(Q,P)} \int \delta\left( v,w;\theta\right)d\pi(v,w) \; = \; 0.
\end{eqnarray} 
The quantity~$\mathcal D(Q,P;\theta)$ is the value of an optimal transport problem (see \cite{Villani:2003}) with cost function~$(v,w)\mapsto \delta(v,w;\theta)$. 
The following proposition shows that when~$P_0^{(n)}$ is i.i.d. with marginals~$P_0:=P_{Y\vert X,0}\times P_{X,0}$, (\ref{eq:char1}) is necessary and sufficient for~H$_0(\theta)$.

\begin{proposition}\label{prop:char1}
Under assumption~\ref{ass:discrepancy}, when~$P_0^{(n)}$ is i.i.d. with marginals~$P_0:=P_{Y\vert X,0}\times P_{X,0}$, H$_0(\theta)$ is true if and only if
$\mathcal D(Q_U\times P_{X,0},P_0;\theta) = 0.$
\end{proposition}

There are two essential differences between our characterization and those in \cite{EGH:2010} and \cite{GH:2011}. First, we define the model correspondence as~$\Gamma_y:\mathcal U\times\mathcal X\rightrightarrows\mathcal Y\times\mathcal X$, rather than as a correspondence from~$\mathcal U\times\mathcal X$ to~$\mathcal Y$. Second, we replace the binary discrepancy used in \cite{EGH:2010} and \cite{GH:2011} with a more general discrepancy. 
Taken together, these two modifications allow us to measure violations of the model at given covariate values while also incorporating nonparametric matching on covariates. This formulation avoids conditioning on~$X$, which can increase computational burden and complicate inference.

\subsubsection*{Construction of the test statistic}

Our test statistic is based on a sample analogue
of the optimal-transport criterion~$\mathcal D(Q_U\times P_{X,0},P_0;\theta)$ of proposition~\ref{prop:char1}, in which the distribution~$P_0$ of the observables is replaced by the empirical distribution~$\hat P_n:=\hat P_{Y\vert X,n}\times \hat P_{X,n}$ of the sample~$((Y_1,X_1),\ldots,(Y_n,X_n))$. We rely on an approximation\footnote{In section~\ref{sec:discrete} we propose a variant of the procedure in case the outcome space~$\mathcal Y$ is finite, which involves no approximation.} of the product~$Q_U\times \hat P_{X,n}$, based on~$(\tilde u^{(n)},X^{(n)})$. In the latter, $X^{(n)}$ is the sample of covariate observations and~$\tilde u^{(n)}:=(\tilde u_1,\ldots,\tilde u_n)$ is either an i.i.d. sequence of simulated draws from~$Q_U$ or a low discrepancy sequence that approximates~$Q_U$ (see appendix~\ref{sec:num} for details). We denote by~$\hat P_{UX,n}$ the empirical distribution of the simulated pairs,
\begin{eqnarray}\label{eq:PUXn}
    \hat P_{UX,n}(A) & := & \frac{1}{n}\sum_{i=1}^n\mathbf 1\{(\tilde u_i,X_i)\in A\},
\end{eqnarray}
which plays the role of the approximation to~$Q_U\times\hat P_{X,n}$. Our chosen test statistic~$T_n(\theta)$ is the discrete optimal-transport solution
\begin{eqnarray*}
T_n(\theta) = \mathcal D_n(C(\theta)),\mbox{ where:}
\end{eqnarray*}
\begin{enumerate}
\item The $n \times n$ cost matrix~$C(\theta)$ has entries
\begin{eqnarray}
\label{eq:CM}
C_{ij}(\theta) & := & \delta((\tilde u_i, X_i),(Y_j,X_j);\theta), \mbox{ for each }i,j\leq n.
\end{eqnarray}
\item For any cost matrix~$C\in\mathcal M_n^+$,~the program~$\mathcal D_n$ is defined by
\begin{eqnarray}
\label{eq:OT} 
\mathcal D_n(C)   :=   \min_{\pi\in\Pi_n} \; \sum_{i,j=1}^n\pi_{ij}C_{ij};
\end{eqnarray}
where~$\Pi_n$ is the set of~$n\times n$ nonnegative matrices~$\pi$ such that $\Sigma_i\pi_{ij}  = 
\Sigma_j\pi_{ij} = 1/n$,  for all $i,j \leq n,$ as defined in the {\em notations and preliminaries} section.
\end{enumerate}
Equivalently, $T_n(\theta)=\mathcal D(\hat P_{UX,n},\hat P_n;\theta)$: the couplings
of two $n$-point empirical distributions with uniform weights are exactly the
matrices in~$\Pi_n$, so that~(\ref{eq:OT}) is the optimal-transport
problem~(\ref{eq:char1}) with~$Q$ and~$P$ replaced by~$\hat P_{UX,n}$
and~$\hat P_n$. This is the form in which the statistic is analyzed in
section~\ref{sec:consistency}.
Computation of the test statistic is discussed in appendix~\ref{sec:num}. For now, note that~(\ref{eq:OT}) solves a discrete optimal-transport problem, which is a special kind of linear programming problem.

Beyond the i.i.d. case, the null hypothesis~(\ref{eq:null}) restricts the joint distribution of the sample rather than the marginal distribution of a representative observation, and the criterion of proposition~\ref{prop:char1} no longer characterizes it. Two properties nonetheless justify the use of the same test statistic when the selection mechanisms are dependent or heterogeneous across observations. First, the finite-sample size control result of theorem~\ref{thm:size} below holds without any restriction on the joint selection mechanism. Second, section~\ref{sec:consistency} gives a consistency result for general triangular arrays whenever the empirical distribution remains bounded away from the one-observation compatibility set~$\mathcal P_\theta$ with probability approaching one.

\subsubsection*{Choice of discrepancy}
The test statistic also requires a specific choice of discrepancy~$\delta$ such that~$\delta(v,w;\theta)\ge 0$ with equality if and only if~$v\in\Gamma_u(w;\theta)$. Such a discrepancy can be constructed in multiple ways. Unless specified otherwise, we recommend constructing the discrepancy as follows:
\begin{eqnarray}\label{eq:discrepancy}
    \delta(v,w;\theta) & := & \inf_{v^\prime\in\Gamma_u(w;\theta)}
    (v-v^\prime)\hat\Sigma^{-1}(v-v^\prime)^{\top}.
\end{eqnarray}
In the expression above, the matrix~$\hat\Sigma$ is an invertible approximation of the covariance matrix of~$V=(U,X)$.
Let~$\Sigma_U$ denote the (known) covariance matrix of the random vector~$U$ with distribution~$Q_U$ and~$\hat\Sigma_X$ the empirical covariance matrix of the sample~$(X_1,\ldots,X_n)$. Since $U$ and $X$ are assumed to be independent in the model, a recommended choice for~$\hat\Sigma$ is
\begin{eqnarray*}
    \hat\Sigma & := & \left( 
    \begin{array}{cc}
        \Sigma_U & 0 \\
        0 & \hat\Sigma_X
    \end{array}\right).
\end{eqnarray*}

A useful alternative definition of the discrepancy~$\delta$ is the analogue of~(\ref{eq:discrepancy}) in the outcome space, namely
\begin{eqnarray}\label{eq:alternative discrepancy}
    \delta(v,w;\theta) & := & \inf_{w^\prime\in\Gamma_y(v;\theta)}
    (w-w^\prime)\tilde\Sigma^{-1}(w-w^\prime)^{\top}.
\end{eqnarray}
In the expression above, the matrix~$\tilde\Sigma$ is an approximation of the covariance matrix of~$W=(Y,X)$ by simulation. An advantage of this alternative discrepancy is that~$\Gamma_y$ may be easier to characterize than~$\Gamma_u$, as will be the case in the empirical illustration of Section~\ref{sec:empirical}. The major drawback is that~$\tilde\Sigma$ is often more complicated to obtain than~$\hat\Sigma$. It is also likely to be a poor approximation of the covariance of~$(Y,X)$ because of the incompleteness of the model. Indeed, simulating values of~$Y$ from the model requires assuming an equilibrium selection mechanism (a way to select within~$\Gamma_y$), which may be incorrect. In the latter case, however, our inference procedure remains valid.

\begin{continued}[Example~\ref{ex:interval} continued:]
In the regression with censored covariates, the correspondence~$\Gamma_u$ takes values~$\Gamma_u(y,(\underline w,\overline w);\theta)=\{(u,(\underline w^\prime,\overline w^\prime)): (\underline w^\prime,\overline w^\prime)=(\underline w,\overline w); u\in[y-\theta\overline w,y-\theta\underline w]\}$.

The discrepancy defined in \eqref{eq:discrepancy} is the value of a quadratic program over a closed interval:
$$
    \delta((u, x), (y, x');\theta) = \min_{u'\in [y - \theta\overline{w}', y - \theta\underline{w}']} (u-u'\hspace{1em} \underline{w} - \underline{w}' \hspace{1em} \overline{w} - \overline{w}') \hat{\Sigma}^{-1} 
    \begin{pmatrix}
        u-u'\\
        \underline{w} - \underline{w}'\\
        \overline{w} - \overline{w}'
    \end{pmatrix}
$$
It admits a closed-form solution:
\begin{eqnarray}\label{eq:closed}
\delta((u, x), (y, x');\theta) = \delta_u(u, y, x';\theta) +  (\underline{w} - \underline{w}'\hspace{1em} \overline{w} - \overline{w}') \hat{\Sigma}_X^{-1} 
    \begin{pmatrix}
        \underline{w} - \underline{w}'\\
        \overline{w} - \overline{w}'
    \end{pmatrix},    
\end{eqnarray}

where 
\[
\delta_u(u, y, x';\theta) = 
\begin{cases}
(u - y + \theta\underline{w}')^2& \text{if } u > y - \theta\underline{w}',\\
    0 &\text{if } y - \theta\overline{w}' \le u \le y - \theta\underline{w}',\\
    (u - y + \theta\overline{w}')^2 &\text{if } y - \theta\overline{w}' > u.
\end{cases}
\]
The test statistic therefore takes the form~$T_n(\theta)=\mathcal D_n(\mathcal C(\theta))$ where~$\mathcal D_n$ is defined by~(\ref{eq:OT}), and~$\mathcal C(\theta)$ is defined by~(\ref{eq:CM}) with closed form discrepancy~(\ref{eq:closed}).

The discrepancy in \eqref{eq:alternative discrepancy} is also the value of a quadratic program over a closed interval:
$$
    \delta((u, x), (y, x');\theta) = \min_{y'\in [ \theta\underline{w} + u,\, \theta\overline{w} + u]} (y-y'\hspace{1em} \underline{w} - \underline{w}' \hspace{1em} \overline{w} - \overline{w}') \tilde{\Sigma}^{-1} 
    \begin{pmatrix}
        y - y'\\
        \underline{w} - \underline{w}'\\
        \overline{w} - \overline{w}'
    \end{pmatrix}
$$
and likewise admits a closed-form representation. We omit the expression for brevity.
\end{continued}

\subsection{Critical values}

Our critical values rely on simulated samples of unobservables.

\begin{definition}[Monte Carlo latent samples]
\label{def:MC}
A Monte Carlo latent sample~$\tilde U^{\prime(n)}$ is a collection~$(\tilde U_1^\prime,\ldots,\tilde U_n^\prime)$ of mutually independent vectors with identical distribution~$Q_U$ independent from~$X^{(n)}$. 
\end{definition}

A Monte Carlo latent sample~$\tilde U^{\prime(n)}$ is designed to be generated from the same distribution as the true latent variables~$U^{(n)}$. Repeated Monte Carlo latent samples are used to approximate the distribution of the worst case test statistic under the null hypothesis. 

\subsubsection{Parameter dependent critical value}

To ensure valid coverage of the true parameter with confidence region~$CR_n$, we choose as critical value~$c_{n,1-\alpha}(\theta)$, the~$1-\alpha$ quantile of a distribution that first order stochastically dominates~$T_n(\theta)$ for each~$n$. We then show exact coverage by constructing a data-generating process in~$\mathcal P_\theta^{(n)}$ such that~$T_n(\theta)$ has~$1-\alpha$ quantile~$c_{n,1-\alpha}(\theta)$. 
Let~$\tilde U^{\prime(n)}$ be a Monte Carlo latent sample.
Let~$\tilde y^{(n)}=(\tilde y_1,\ldots,\tilde y_n)$ be the notation of a generic vector in~$\mathcal Y^n$. Define the cost matrix~$C(\tilde y^{(n)};\theta)$ with elements
\begin{eqnarray}\label{eq:cost}
    C_{ij}(\tilde y^{(n)};\theta) & := & \delta((\tilde u_i,X_i),(\tilde y_j,X_j);\theta).
\end{eqnarray}
Finally, define the set\footnote{Note that there is a slight abuse of notation here, in the sense that~$\Gamma_y$ defines a set of pairs~$(y,x)$, whereas~$\Gamma_y^{(n)}$ defines a set of vectors~$(y_1,\ldots,y_n)$ without covariates~$x$.}
\begin{eqnarray}\label{eq:Gamma(n)}
    \Gamma_y^{(n)}(\tilde U^{\prime(n)},X^{(n)};\theta) & := & \{\tilde y^{(n)}: (\tilde{y}_j,X_j)\in\Gamma_y(\tilde U^\prime_j,X_j;\theta) \mbox{ all } j\leq n\}.
\end{eqnarray}

The critical value we propose is the $1-\alpha$ quantile~$c_{n,1-\alpha}(\theta)$ of the distribution of
\begin{eqnarray}
\label{eq:critical}
\tilde T_n(\theta) & = & \sup_{\tilde y^{(n)}\in\Gamma_y^{(n)}(\tilde U^{\prime(n)},X^{(n)};\theta)}\mathcal D_n(C(\tilde y^{(n)};\theta)).
\end{eqnarray}
The statistic~$\tilde T_n(\theta)$ of equation~(\ref{eq:critical}) differs from the test statistic~$T_n(\theta)$ in two critical ways. First, the sample realization~$Y_j$ in the element~$C_{ij}$ of the cost matrix is replaced with a value~$\tilde y_j$. This value~$\tilde y_j$ is constrained by the Monte Carlo draw~$\tilde U_j^\prime$ from~$Q_U$ and the constraint~$(\tilde{y}_j,X_j)\in\Gamma_y(\tilde U^\prime_j,X_j;\theta)$ to enforce the null hypothesis~$H_0(\theta)$. Second, the supremum in expression~(\ref{eq:critical}) ensures that~$\tilde y_j$ is chosen to achieve the worst-case scenario, i.e., the largest possible value of~$\tilde T_n(\theta)$ under the null~$H_0(\theta)$.

\begin{continued}[Example~\ref{ex:interval} continued:]
In the regression with interval censored covariates, 
the critical value statistic~$\tilde T_n(\theta)$ is given by~(\ref{eq:critical}), where the optimization is over the set
\[
\Gamma_y^{(n)}(\tilde U^{\prime(n)},X^{(n)};\theta)=\{\tilde y^{(n)}: \forall j, \;\theta\underline W_j+\tilde U^{\prime}_j\leq\tilde y_j\leq\theta\bar W_j+\tilde U^{\prime}_j\},
\]
and the cost matrix~$C(\tilde y^{(n)};\theta)$ of~(\ref{eq:cost}) is computed with closed form discrepancy~(\ref{eq:closed}). See appendix~\ref{sec:num} for details on the implementation. 
\end{continued}

\subsubsection{Parameter-free critical value}\label{sec:parameter-free}

For computational convenience, we also provide a more conservative confidence region~$CR_n^0$ based on a critical value~$c_{n,1-\alpha}^0$ which is independent of the parameter value~$\theta$. 
Let~$\delta^0$ be a discrepancy on~$\mathcal U\times\mathcal X$ that satisfies the following.
\begin{assumption}\label{ass:discrepancy0}
    The function~$\delta^0(\cdot,\cdot):(v,v^\prime)\in(\mathcal U\times\mathcal X)\times(\mathcal U\times\mathcal X)\mapsto \delta^0(v,v^\prime)$ is nonnegative, continuous, bounded, and equal to zero if and only if~$v=v^\prime$. In addition, discrepancy~$\delta^0$ dominates~$\delta$ from section~\ref{sec:test} in the sense that
    \begin{eqnarray}\label{eq:compatible}
        \forall(v,v^\prime)\in(\mathcal U\times\mathcal X)^2, 
        \; \; \sup_{\theta\in\Theta} \; \sup_{w\in \Gamma_y(v';\theta)}\delta(v, w;\theta) \; \le \; \delta^0(v, v').
    \end{eqnarray}
\end{assumption}
The recommended choice of discrepancy is the following:
\begin{eqnarray*}
\delta^0(v,v^\prime) & := & (v-v^\prime)\hat\Sigma^{-1}(v-v^\prime)^{\top},
\end{eqnarray*}
where~$\hat\Sigma$ is defined in~(\ref{eq:discrepancy}). With a suitable truncation in case~$\mathcal X$ is not bounded, this choice satisfies assumption~\ref{ass:discrepancy0} when~$\delta$ is chosen according to~(\ref{eq:discrepancy}).
Denote~$C^0$ the cost matrix with elements
\begin{eqnarray*}
C_{ij}^0 = \delta^0((\tilde u_i,X_i),(U_j^\prime,X_j)),
\end{eqnarray*}
where~$(\tilde u_i)_{i\leq n}$ is the same sequence as in the definition of~$T_n(\theta)$, and~$(U_j^\prime)_{j\leq n}$ is a Monte Carlo latent sample of definition~\ref{def:MC}.
The critical value~$c_{n,1-\alpha}^0$ is chosen to be the $1-\alpha$ quantile of the distribution of
\begin{eqnarray*}
\tilde T_n^0 & = & \mathcal D_n(C^0).
\end{eqnarray*}
Under assumption~\ref{ass:discrepancy0}, (\ref{eq:compatible}) holds and hence, for any~$\tilde y$ such that~$(\tilde y,X_j)\in\Gamma_y(U_j^\prime,X_j;\theta)$, we have~$\delta((\tilde u_i,X_i),(\tilde y,X_j);\theta)\leq C_{ij}^0$. It follows that for all~$\theta\in\Theta$, $\tilde T_n(\theta)\leq\tilde T_n^0$, and, therefore:
\begin{eqnarray}
\label{eq:conserv}
\sup_{\theta\in\Theta}c_{n,1-\alpha}(\theta) \leq c_{n,1-\alpha}^0 & \mbox{and} & CR_n \subseteq CR_n^0 \; .
\end{eqnarray}

From Statement~(\ref{eq:conserv}), we deduce two advantages of the outer confidence region~$CR_n^0$. First, the critical value is independent of the parameter value. Hence, it needs to be computed only once, and only the test statistic~$T_n(\theta)$ needs to be computed for each value of the parameter~$\theta$. 
Second, given that~$CR_n\subseteq CR_n^0$, the computation of the confidence region~$CR_n$ can be performed with a search limited to~$CR_n^0$ as opposed to the whole parameter space~$\Theta$.

Constructing~$\delta^0$ directly as
\begin{eqnarray*}
\delta^0(v, v^\prime)
=
\sup_{\theta\in\Theta} \; \sup_{w\in \Gamma_y(v^\prime;\theta)}
\delta(v, w;\theta)
\end{eqnarray*}
would yield the smallest parameter-free critical value within this class.
However, computing this discrepancy may entail substantial computational cost, since it requires solving the supremum over both parameter values and model-predicted outcomes. This cost may offset the computational advantage of using a parameter-free critical value, unless the structural model at hand admits a closed-form expression for such $\delta^0$.

\subsection{Valid and exact finite-sample inference results}
\label{sec:exact}

The next theorem shows finite-sample validity and exactness of our procedure. Validity is achieved because both~$\tilde T_n(\theta)$ and~$\tilde T_n^0$ first order stochastically dominate the test statistic~$T_n(\theta)$. Hence, both confidence regions~$CR_n^0$ and~$CR_n$ have valid coverage in finite samples. Exact finite-sample inference is achieved because we can construct a data-generating process under which both~$T_n(\theta)$ and~$\tilde T_n(\theta)$ have the same distribution. Hence, our proposed confidence region~$CR_n$ has the correct coverage probability in finite samples.

\begin{theorem}
\label{thm:size}
Under assumption~\ref{ass:discrepancy}, for all~$\theta\in\Theta$ such that~$\mathcal P_\theta^{(n)}\ne\varnothing$, all~$\alpha\in(0,1)$, 
\begin{eqnarray}
\label{eq:size}
\inf_{P^{(n)}\in\mathcal P_\theta^{(n)}} P^{(n)} \left( \; T_n(\theta)\leq c_{n,1-\alpha}(\theta) \; \right) & \geq & 1-\alpha.
\end{eqnarray}
If the cumulative distribution function of~$\tilde T_n(\theta)$ is continuous and increasing in a neighborhood of~$c_{n,1-\alpha}(\theta)$, then~(\ref{eq:size}) holds as an equality.
\end{theorem}

The formal proof of Theorem~\ref{thm:size} is given in the appendix, where we also give a version of the theorem conditional on~$X^{(n)}$ (see appendix~\ref{sec:add}). Proof heuristics are as follows.
By construction, under the null hypothesis, the Monte Carlo latent sample~$\tilde U^{\prime(n)}$ has the same distribution as the true latent sample~$U^{(n)}:=(U_1,\ldots,U_n)$. Now, if the process~$P^{(n)}$ generating~$(Y^{(n)},X^{(n)})$ is in~$\mathcal P_\theta^{(n)}$, then each realization~$(Y_j,X_j)$, $j\leq n$, falls in~$\Gamma_y(U_j,X_j;\theta)$ almost surely (according to the support restriction in the model). Hence, the test statistic~$T_n(\theta)$ is no larger than~$\sup\{ \mathcal D_n(C(\tilde y^{(n)};\theta)): \tilde y^{(n)}\in\Gamma_y^{(n)}(U^{(n)},X^{(n)};\theta)\}$. Since the latter is identically distributed to~$\sup\{ \mathcal D_n(C(\tilde y^{(n)};\theta)): \tilde y^{(n)}\in\Gamma_y^{(n)}(\tilde U^{\prime(n)},X^{(n)};\theta)\}$, size control follows.
To see that the inequality in~(\ref{eq:size}) is an equality, we find~$(Y^{(n)},X^{(n)})$ that achieves a value arbitrarily close to the maximum of~$T_n(\theta)$ under the constraint~$(Y_i,X_i)\in\Gamma_y(\tilde U^\prime_i,X_i;\theta)$.

\begin{remark}
    If we make the additional assumption that the true data generating process is i.i.d., then theorem~\ref{thm:size} guarantees valid but not exact size control. However, we can achieve exact size control for exchangeable data generating processes, providing we use an exchangeable low discrepancy sequence. See section~2 of \cite{li2022finite}.
\end{remark}

\subsection{Consistency}
\label{sec:consistency}

Theorem~\ref{thm:size} controls size over the whole~$\mathcal P_\theta^{(n)}$ under every selection mechanism the model permits, and it does so at every sample size. What remains to be established is informativeness, namely that parameter values incompatible with the model and the data are eventually excluded. Since size control imposes no sampling scheme on the observations, we propose a power analysis that does not impose one either. We therefore work with triangular arrays, in which the sample-level distribution~$P^{(n)}$ of the observed row may change with~$n$, and we ask along which sequences~$(P^{(n)})_{n\geq1}$ our test rejects with probability approaching one. We prove the result for the conservative outer region~$CR_n^0$; since~$CR_n\subseteq CR_n^0$ by~(\ref{eq:conserv}), the conclusion applies a fortiori to our proposed confidence region~$CR_n$.

The test statistic~$T_n(\theta)$ is a sample analogue of the optimal-transport criterion of proposition~\ref{prop:char1}, evaluated at the empirical distribution~$\hat P_n$ of the observed pairs, whereas the parameter-free critical value~$c_{n,1-\alpha}^0$ collapses to zero as the sample grows (lemma~\ref{lem:critical-value-collapse} in the appendix). Rejection therefore occurs eventually as soon as the criterion remains bounded away from zero along the array. Since the criterion vanishes exactly on~$\mathcal P_\theta$, this is a statement about how far the empirical distribution remains from the set of one-observation distributions compatible with~$\theta$. The next assumption states that requirement directly.

Recall that~$\mathcal Q$ and~$\mathcal P$ denote the collections of Borel probability measures on~$\mathcal U\times\mathcal X$ and~$\mathcal Y\times\mathcal X$, respectively. Let~$\rho_{\mathcal P}$ and~$\rho_{\mathcal Q}$ be any bounded metrics that metrize weak convergence on~$\mathcal P$ and~$\mathcal Q$. For any nonempty subset~$\mathcal A\subseteq\mathcal P$, define
\begin{eqnarray*}
\operatorname{dist}_{\mathcal P}(P,\mathcal A)
\coloneqq
\inf_{P'\in\mathcal A}\rho_{\mathcal P}(P,P').
\end{eqnarray*}

\begin{assumption}[Separation and tightness]
\label{ass:triangular_array}
Fix the candidate value~$\theta$, and suppose $\mathcal P_\theta\ne\varnothing$. Assume $(P^{(n)})_{n\ge 1}$ satisfies the following conditions:
\begin{enumerate}[label=\textnormal{(\roman*)}]
\item \emph{Separation:}
\begin{eqnarray}
\label{eq:law-separation}
\lim_{\varepsilon\downarrow0}\limsup_{n\to\infty}
P^{(n)}
\left\{
\operatorname{dist}_{\mathcal P}(\hat P_n,\mathcal P_\theta)
\leq\varepsilon
\right\}
& = & 0.
\end{eqnarray}
\item \emph{Tightness:} for every~$\eta>0$, there exists a weakly compact set~$\mathcal K_\eta\subseteq\mathcal P$ such that
\begin{eqnarray}
\label{eq:law-tightness}
\liminf_{n\to\infty}
P^{(n)}\{\hat P_n\in\mathcal K_\eta\}
& \geq & 1-\eta.
\end{eqnarray}
\end{enumerate}
\end{assumption}

Condition~(\ref{eq:law-separation}) requires the empirical distribution to remain asymptotically bounded away from~$\mathcal P_\theta$, in the sense that for every~$\eta>0$ there is a neighborhood of~$\mathcal P_\theta$ that~$\hat P_n$ enters with probability at most~$\eta$ in the limit. We emphasize three features. First, it is a restriction on the empirical distribution, not on the sample-level law: our test sees the sample only through~$\hat P_n$, and~$\mathcal P_\theta=\mathcal P_\theta^{(1)}$ is the set of distributions that rationalize~$\theta$ for a single observation, which coincides with the model-predicted set in the i.i.d.\ case by proposition~\ref{prop:iid}. Second, the condition does not require~$\hat P_n$ to converge. When it does converge weakly to some~$P_\infty$, as under i.i.d.\ sampling or, more generally, stationarity and ergodicity, condition~(\ref{eq:law-separation}) reduces to the familiar requirement that the limit be incompatible with the model, $P_\infty\notin\mathcal P_\theta$, because~$\mathcal P_\theta$ is weakly closed by lemma~\ref{lem:mixture}(i). The value of the formulation in~(\ref{eq:law-separation}) is that no such limit needs to exist: the array may be nonstationary and the selection mechanisms arbitrarily dependent across observational units, so long as the empirical distribution does not drift into the compatibility set. Third, some such condition is unavoidable for a procedure built on~$\hat P_n$. There are sample-level distributions outside~$\mathcal P_\theta^{(n)}$ whose empirical distributions nonetheless merge into~$\mathcal P_\theta$. Since $\hat{P}_n$ is the only 
feature of the sample that our statistic records, such samples become indistinguishable from compatible ones. Section~\ref{sec:optimality} shows that, under infinite exchangeability, no test with finite-sample size control can reject them either.

Condition~(\ref{eq:law-tightness}) prevents the empirical distributions from escaping to infinity. It is automatic when~$\mathcal Y\times\mathcal X$ is compact, and it holds more generally under a moment bound that is uniform across the array.

We also need the following regularity conditions on the discrepancy~$\delta$ and on~$(\tilde{u}^{(n)})_{n\ge 1}$.

\begin{assumption}[Regularity]
\label{ass:reg}
The discrepancy~$\delta$ in~(\ref{eq:char1}) is fixed across sample sizes, continuous and bounded\footnote{The recommended quadratic discrepancy can be truncated to satisfy boundedness. If its weighting matrix is estimated, a sufficient extension is uniform convergence of the resulting discrepancies to a fixed bounded continuous discrepancy, together with uniform nondegeneracy of the weighting matrices.} on~$(\mathcal U\times \mathcal X) \times (\mathcal Y\times \mathcal X)$ for all~$\theta\in\Theta$.
\end{assumption}

Continuity and boundedness of the discrepancy make the optimal-transport value continuous for weak convergence, which is what converts separation in the metric~$\rho_{\mathcal P}$ into separation of the OT criterion from zero. Boundedness is innocuous here: assumption~\ref{ass:discrepancy} only requires~$\delta$ to vanish on the graph of~$\Gamma_u$, so the recommended discrepancy~(\ref{eq:discrepancy}) may be truncated at an arbitrarily high level without affecting the characterization of proposition~\ref{prop:char1}, hence without affecting which parameter values are compatible with the data. 

Recall the definition of~$\hat P_{UX,n}$ in~(\ref{eq:PUXn}).

\begin{assumption}\label{ass:joint convergence}
Assume\footnote{Throughout this subsection and its proof, we use~$P^{(n)}$ with a slight abuse of notation. It denotes the data law when the $(\tilde{u}^{(n)})_{n\ge 1}$ is deterministic. When $(\tilde{u}^{(n)})_{n\ge 1}$ is randomly generated, it denotes the natural product extension~$P^{(n)}\otimes Q_U^{\otimes n}$.}
\begin{equation}
\label{eq:simulator-approximation}
\forall \varepsilon > 0,\quad  
\lim_{n\to\infty}P^{(n)}\left\{ \rho_{\mathcal Q} \left(\hat P_{UX,n},Q_U\times\hat P_{X,n}\right) > \varepsilon \right\}  = 0
\end{equation}
\end{assumption}

Assumption~\ref{ass:joint convergence} is the sample counterpart of the independence restriction~$U\perp\!\!\!\perp X$ of definition~\ref{def:DGP}: the simulated pairs~$(\tilde u_i,X_i)_{i\leq n}$ must reproduce, in large samples, the product of~$Q_U$ with the empirical distribution of the covariates. If~$\tilde u^{(n)}$ consists of i.i.d.\ draws from~$Q_U$ independent of the observed row, the condition follows from assumption~\ref{ass:triangular_array}(ii) and a conditional triangular-array law of large numbers. If~$\tilde u^{(n)}$ is deterministic, as with the low discrepancy sequences we recommend, convergence of its marginal empirical distribution to~$Q_U$ is not sufficient: what matters is that the pairing of the~$i$th point of the sequence with the~$i$th observed covariate not be systematic, since an ordering of the sequence aligned with the ordering of the covariates would induce spurious dependence between~$\tilde u_i$ and~$X_i$. Lemma~\ref{lem:joint convergence} in the appendix gives one primitive sufficient condition for~(\ref{eq:simulator-approximation}) in a setting where the covariate process is stationary and strong mixing with absolutely summable mixing weights, including i.i.d covariate process as a special case. 

\begin{theorem}[Consistency]
\label{thm:new_consistency}
Fix~$\theta\in\Theta$ and~$\alpha\in(0,1)$. Suppose Assumptions~\ref{ass:discrepancy}-\ref{ass:joint convergence} hold. 
Then,
\begin{eqnarray}
\label{eq:new-consistency}
\lim_{n\to\infty}
P^{(n)}
\left\{
T_n(\theta)>c_{n,1-\alpha}^0
\right\}
& = & 1.
\end{eqnarray}
\end{theorem}

Theorem~\ref{thm:new_consistency} states that the conservative region~$CR_n^0$ excludes~$\theta$ with probability approaching one along any separated array. Under assumption~\ref{ass:discrepancy0}, (\ref{eq:compatible}) and~(\ref{eq:conserv}) hold, so that the exact critical value~$c_{n,1-\alpha}(\theta)$ is uniformly no larger than~$c_{n,1-\alpha}^0$ and~$\theta$ eventually falls outside~$CR_n$ as well.

The result requires neither stationarity, nor ergodicity, nor any mixing or independence restriction on the outcomes. 
Consistency is thereby obtained under the same agnosticism about the selection mechanism as size control in theorem~\ref{thm:size}, which is the property we argued in the introduction to be especially desirable in incomplete models.

Theorem~\ref{thm:new_consistency} is a fixed-alternative statement: the separation in~(\ref{eq:law-separation}) holds at a scale that does not shrink with~$n$. Sharpening it to local alternatives would require comparing the rate at which the criterion evaluated at~$\hat P_n$ vanishes with the rate at which~$c_{n,1-\alpha}^0$ does, which we do not pursue here. What can be said is that the separation requirement is not an artifact of our particular choice of statistic or critical value: it delimits the arrays that any procedure operating on the empirical distribution alone can hope to reject. Whether it is stronger than what other tests with the same finite-sample guarantee could achieve is the question we take up in section~\ref{sec:optimality}.

\subsection{Maximal consistency under exchangeability}
\label{sec:optimality}

The consistency results of section~\ref{sec:consistency} describe which data-generating processes our test eventually rejects. They are silent on whether a different procedure with the same finite-sample guarantees could reject more. In this section, we show that, when the data-generating process $P^{(n)}$ is the marginal distribution for first $n$ observations of an infinitely exchangeable process, the answer is negative: every infinitely exchangeable data-generating process that is consistently rejected by some test with finite-sample size control is also consistently rejected by our test. In this sense, our procedure achieves the strongest consistency attainable under finite-sample size control.

\begin{definition}[Exchangeability]
\label{def:exchangeable}
A distribution~$P^{(n)}$ on~$\mathcal Y^n\times\mathcal X^n$ is {\em finitely exchangeable} if, whenever~$(Y^{(n)},X^{(n)})\sim P^{(n)}$,
\begin{eqnarray*}
((Y_{\sigma(1)},X_{\sigma(1)}),\ldots,(Y_{\sigma(n)},X_{\sigma(n)})) & \overset{d}{=} & ((Y_1,X_1),\ldots,(Y_n,X_n))
\end{eqnarray*}
for every permutation~$\sigma$ of~$\{1,\ldots,n\}$. We denote~$\mathcal E^{(n)}$ the collection of finitely exchangeable distributions on~$\mathcal Y^n\times\mathcal X^n$. 
A distribution~$P^{(\infty)}$ on~$(\mathcal Y\times\mathcal X)^\infty$ is {\em (infinitely) exchangeable} if its first-$n$ marginal belongs to~$\mathcal E^{(n)}$ for every~$n\geq1$. We denote~$\mathcal E^{(\infty)}$ the collection of infinitely exchangeable distributions on~$(\mathcal Y\times\mathcal X)^\infty$.
\end{definition}
Exchangeability requires the joint distribution of the observed pairs to be invariant to a relabeling of the observational units. Infinite exchangeability is a natural sampling framework for cross-sectional data. The structural restrictions in definition~\ref{def:DGP} are themselves symmetric in the observational units, so that exchangeability disciplines the selection mechanism, by requiring the dependence it induces to be symmetric across units, without restricting the strength of that dependence. In particular, exchangeable data-generating processes generalize i.i.d.~sampling and accommodate selection mechanisms driven by common latent factors.

Fix~$\theta\in\Theta$ and~$\alpha\in(0,1)$. We compare our procedure to arbitrary, possibly randomized, tests of the null hypothesis of data/model compatibility~(\ref{eq:null}). A {\em test} is a Borel measurable map~$\phi_n:\mathcal Y^n\times\mathcal X^n\rightarrow[0,1]$, interpreted as the conditional probability of rejecting~$H_0(\theta)$ given the data. A {\em test sequence}~$\phi:=(\phi_n)_{n\geq1}$ controls size in finite samples over exchangeable compatible distributions if
\begin{eqnarray}
\label{eq:ex-size}
\sup_{P^{(n)}\in\mathcal P_\theta^{(n)}\cap\mathcal E^{(n)}}\mathbb E_{P^{(n)}}\left[\phi_n\right] \; \leq \; \alpha, & & \mbox{for every } n\geq1.
\end{eqnarray}
Our procedure defines such a test sequence. Let~$\phi^{OT}:=(\phi_n^{OT})_{n\geq1}$ be defined by
\begin{eqnarray}
\label{eq:OTtest}
\phi_n^{OT}(y^{(n)},x^{(n)}) & := & \mathbb P\left( \; T_n(\theta)>c_{n,1-\alpha}^0 \; \left\vert \; (Y^{(n)},X^{(n)})=(y^{(n)},x^{(n)}) \; \right.\right),
\end{eqnarray}
the conditional probability that the test statistic exceeds the parameter-free critical value of section~\ref{sec:parameter-free} given the data. The conditioning in~(\ref{eq:OTtest}) integrates out the potential simulation randomness in the construction of~$T_n(\theta)$ and~$c_{n,1-\alpha}^0$, which is independent of the data. When the approximating sequence~$\tilde u^{(n)}$ is deterministic and no other simulation randomness is involved, $\phi_n^{OT}$ reduces to the indicator of the rejection event~$\{T_n(\theta)>c_{n,1-\alpha}^0\}$. Under assumptions~\ref{ass:discrepancy} and~\ref{ass:discrepancy0}, theorem~\ref{thm:size} and~(\ref{eq:conserv}) yield~$\mathbb E_{P^{(n)}}[\phi_n^{OT}]\leq\alpha$ for {\em every}~$P^{(n)}\in\mathcal P_\theta^{(n)}$, exchangeable or not. In particular, $\phi^{OT}$ satisfies~(\ref{eq:ex-size}).

For any test sequence~$\phi$, define the collection of infinitely exchangeable data-generating processes against which~$\phi$ is consistent,
\begin{eqnarray}
\label{eq:rej-set}
\mathcal R_\theta(\phi) & := & \left\{ P^{(\infty)}\in\mathcal E^{(\infty)}: \; \mathbb E_{P^{(n)}}\left[\phi_n\right]\rightarrow1 \mbox{ as } n\rightarrow\infty \right\},
\end{eqnarray}
where~$P^{(n)}$ denotes the first-$n$ marginal of~$P^{(\infty)}$. The following theorem shows that no test sequence with finite-sample size control can be consistent against an exchangeable data-generating process unless our test is.

\begin{theorem}[Maximal consistency under exchangeability]
\label{thm:optimality}
Fix~$\theta\in\Theta$ and~$\alpha\in(0,1)$, and suppose~$\mathcal P_\theta\ne\varnothing$. Suppose assumptions~\ref{ass:discrepancy}, \ref{ass:discrepancy0} and~\ref{ass:reg} hold, and that the approximating sequence~$(\tilde u_i)_{i\geq1}$ in the construction of the test statistic is either an i.i.d. sequence of draws from~$Q_U$ independent of the data, or a deterministic sequence whose empirical distribution converges weakly to~$Q_U$. Then, for every test sequence~$\phi$ satisfying~(\ref{eq:ex-size}),
\begin{eqnarray}
\label{eq:optimality}
\mathcal R_\theta(\phi) & \subseteq & \mathcal R_\theta(\phi^{OT}).
\end{eqnarray}
\end{theorem}

Because the null hypothesis~(\ref{eq:null}) leaves the selection mechanism unrestricted, some exchangeable data-generating processes that are incompatible with the model at~$\theta$ cannot be consistently rejected by {\em any} test sequence with finite-sample size control. Theorem~\ref{thm:optimality} shows that our procedure consistently rejects every one that can be: whenever some test controlling size in finite samples 
rejects~$H_0(\theta)$ with probability approaching one along an exchangeable data-generating process, so does ours.\footnote{The theorem is stated for the conservative test based on the parameter-free critical value, and it extends to the test underlying our proposed confidence region~$CR_n$. 
}

One qualification is in order. Theorem~\ref{thm:optimality} compares tests along the first-$n$ marginals of {\em infinitely} exchangeable processes. A finitely exchangeable distribution on~$\mathcal Y^n\times\mathcal X^n$ need not arise this way, i.e., it need not be infinitely extendible (see \cite{DF:1980}). The maximal consistency statement is therefore geared to settings where the~$n$ observed units are drawn from a much larger exchangeable system, and does not cover systems intrinsically of size~$n$, such as a fully observed network of~$n$ nodes. Note, however, that the finite-sample size control of theorem~\ref{thm:size} requires no extendibility, since it holds over all of~$\mathcal P_\theta^{(n)}$. The qualification only concerns the class of alternatives over which the power comparison is made.
The consideration-set design of section~\ref{sec:MC-consideration} is a concrete
example: its selection rule is symmetric across units, so every~$P^{(n)}$ is
finitely exchangeable, but the rule depends on the whole sample of size~$n$, so
the array is not the sequence of first-$n$ marginals of an infinitely exchangeable
process.

\section{Refinements}
\label{sec:refine}

\subsection{Parametric latent variables}
\label{sec:parametric}

Most structural models of interest involve latent variables, whose distribution depends on a vector of unknown parameters. The distribution of the latent variables may also depend on the exogenous variable~$X$.
The structural model then stipulates that~$(Y,X,U)\in\mathcal Y \times \mathcal X \times \mathcal U$ satisfies support constraint $(Y,X,U) \in \Gamma(\theta)\subseteq \mathcal Y \times \mathcal X \times \mathcal U$, and~$U$ has distribution~$Q_{U\vert X;\theta}$ conditionally on~$X$. For ease of notation, we use the same~$\theta$ notation for the parameter of the latent variable distribution and the parameter of the correspondence~$\Gamma$, even though they will generally be disjoint.

The objective of this section is twofold. First we show that the procedure we proposed is without loss of generality, because the model with parametric latent variable distribution can always be transformed into a model with a latent variable distribution that is independent of parameters and covariates. The transformation is particularly straightforward when the vector of latent variables is multivariate normal. Second, when the transformation is more involved, we propose a variant of the proposed inference method that doesn't require the transformation to a model with fixed latent variable distribution.

\subsubsection{Transformation to parameter-free latent distribution}

The basic ingredient in the reformulation is a known transformation that recovers the vector of unobservable variables~$U$ from a vector of latent variables~$U^\ast$ with a distribution~$Q_U^\ast$, which is independent of~$X$ and~$\theta$. We fix the distribution~$Q^\ast_U$ on~$\mathcal U^\ast$. We then find a function~$h$ such that the random vector~$U:=h(U^\ast,X;\theta)$ has distribution~$Q_{U\vert X;\theta}$. When~$U$ is scalar, the conditional quantile transform is an example of such a function~$h$. In Example~\ref{ex:airline}, $Q_{U\vert X;\theta}$ is a multivariate normal with mean zero and covariance matrix~$\Sigma$. In that case, we can simply let~$Q^\ast_U$ be the standard multivariate normal and~$h$ be defined by~$U=\Sigma^{\frac{1}{2}}U^\ast$. More generally, such a transformation always exists, as long as~$Q_U^\ast$ is chosen to be absolutely continuous, by Theorem~2.1 of \cite{carlier2016vector}. It can also be computed as the solution of an optimal-transport problem.
Given the~$h$ function above, which is known, the structural incomplete model can be reformulated as the combination of the support constraint~$(Y,X,U^\ast)\in\Gamma^\ast(\theta)$, where
\begin{eqnarray*}
\Gamma^\ast(\theta) & := & \left\{ (y,x,u^\ast): \, (y,x,h(u^\ast,x;\theta)) \in \Gamma(\theta) \right\},
\end{eqnarray*}
and the marginal constraint~$U^\ast\sim Q_U^\ast$ and~$U^\ast\independent X$.

\subsubsection{Inference procedure without transformation}

For cases, where the transformation~$h$ is difficult to compute, we propose a variant of our inference procedure without transformation. Inference proceeds as in Section~\ref{sec:inference} with the following two modifications to account for the fact that the latent vectors~$U_i$ in sample~$U^{(n)}$ are drawn independently from~$Q_{U\vert X_i;\theta}$ conditionally on~$X^{(n)}$.
\begin{itemize}
    \item The discretization of~$Q_{U\vert X;\theta}$ for the definition of the test statistic relies on a sample~$(\tilde u_1,\ldots,\tilde u_n)$ which is obtained 
    \begin{itemize}
        \item either randomly, by drawing each~$\tilde u_i$ independently from~$Q_{U\vert X_i;\theta}$,
        \item or deterministically, by choosing a deterministic sequence~$\xi^{(n)}:=(\xi_1,\ldots,\xi_n)$ of points in~$[0,1]^{d_U}$ in such a way that its empirical distribution approximates the distribution of the uniform on~$[0,1]^{d_U}$ well. Each element of that sequence is then transformed using a map that pushes the uniform~$U[0,1]^{d_U}$ to~$Q_{U\vert X_i;\theta}$ instead of~$Q_U$ in the main procedure.
    \end{itemize}
    \item A Monte Carlo latent sample~$\tilde U^{\prime(n)}$ is now defined as a collection~$(\tilde U_1^\prime,\ldots,\tilde U_n^\prime)$ of independent vectors conditional on~$X^{(n)}:=(X_1,\ldots,X_n)$ such that for each~$i\leq n$, $\tilde U_i^\prime$ is drawn from the conditional distribution~$Q_{U\vert X_i;\theta}$. 
\end{itemize}

\subsection{Discrete outcomes} 
\label{sec:discrete}

We now propose a refinement of our procedure, which bypasses the need for an approximation of the distribution~$Q_U$ of latent variables.\footnote{We are grateful to Francesca Molinari for suggesting this refinement. The usual disclaimer applies.} This refinement applies to models, where the space~$\mathcal Y$ of endogenous variables is finite. This refinement is easier to implement with parametric unobservable distribution, so we will use the framework and notation of Section~\ref{sec:parametric}, where the distribution of latent variables~$U$ conditional on~$X$ is denoted~$Q_{U\vert X;\theta}$. We start with the traditional illustration of incomplete models with finite outcomes.

\begin{example}[Entry game]\label{ex:entry}
    Let~$Y=(Y^0,Y^1)\in\{0,1\}^2$ be a pure strategy Nash equilibrium profile in a 2-by-2 perfect information game with payoffs~$\pi^j:=Y^j(U^j-\theta Y^{1-j})$, $j\in\{0,1\}$. Assume that the distribution~$Q_U$ of~$U:=(U^0,U^1)$ is standard bivariate normal. For any~$u\in\mathbb R^2$, and~$y\in\{0,1\}^2$, $\Gamma(\theta)$ is the collection of pairs~$(y,u)$ such that~$y$ is a pure-strategy Nash equilibrium profile when~$U$ takes the value $u$.
\end{example}

With finite outcomes, the sample analogue~$\mathcal D(Q_{U\vert X;\theta}\times \hat P_{X,n},\hat P_n;\theta)$ of the characterization in proposition~\ref{prop:char1} can be computed without relying on an approximation of distribution~$Q_U$. Call~$\mathcal Y(u,x;\theta)$ the set of~$y$'s predicted by~$u$. Precisely, $\mathcal Y(u,x;\theta)$ is defined by the fact that~$y^\prime\in\mathcal Y(u,x;\theta)$ if and only if~$(y^\prime,x,u)\in\Gamma(\theta)$. Since~$\mathcal Y$ is finite, for any given~$x$ and~$\theta$, $\mathcal Y(u,x;\theta)$, as a subset of~$\mathcal Y$, can only take a finite number of distinct values. Call them~$\mathcal Y^1(x;\theta),\ldots,\mathcal Y^K(x;\theta)$.\footnote{For the rest of the section, we assume~$K$ is independent of~$x$ for ease of notation.} Now call~$(\mathcal U^1(x;\theta),\ldots,\mathcal U^K(x;\theta))$ the partition of~$\mathcal U$ defined for each~$k\leq K$ by~$u\in\mathcal U^k(x;\theta)$ if and only if~$\mathcal Y(u,x;\theta)=\mathcal Y^k(x;\theta)$. Distribution~$Q_{U\vert X;\theta}$ on~$\mathcal U$ induces mass~$q^k(x;\theta):=Q_{U\vert X;\theta}(\mathcal U^k(x;\theta)\vert X=x;\theta)$ on each element~$\mathcal U^k(x;\theta)$ of this partition. Unlike \cite{GRS:2022}, we allow this partition to depend on~$X$, so that~$X$ can have many support points, or be continuous and multivariate.

We can therefore use the sample analogue~$\mathcal D(Q_{U\vert X;\theta}\times \hat P_{X,n},\hat P_n;\theta)$ of the characterization in proposition~\ref{prop:char1} as a test statistic without need of an approximation. By construction, all values of~$u$ within a element~$\mathcal U^k(x;\theta)$ predict the same set~$\mathcal Y^k(x;\theta)$. Therefore, we can replace the distribution~$Q_{U\vert X;\theta}\times \hat P_{X,n}$ with the probability mass function with~$nK$ support points. 
Let index~$i$ no longer be the sample index. Instead, let it range from~$1$ to~$nK$. Define~$l_i := \lceil i / K \rceil$, and~$k_i := 1+((i-1) \mbox{ mod } K)$. The probability mass function that characterizes~$Q_{U\vert X;\theta}\times \hat P_{X,n}$ has~$nK$ support points~$\mathcal U^{k_i}(X_{l_i};\theta)$ and probability~$q^{k_i}(X_{l_i};\theta)/n$, where~$1\leq l_i\leq n$ and~$1\leq k_i \leq K$ for each~$i\leq nK$, and~$(l_i,k_i)\neq (l_{i'},k_{i'})$, all~$i\ne i'$ (so as to range over all~$nK$ possible pairs~$(k,l)$). 

One possible way to construct the test statistic would be to choose a point~$u_{k_i}(X_{l_i};\theta)$ in each~$\mathcal U^{k_i}(X_{l_i};\theta)$ and use discrepancy~(\ref{eq:discrepancy}). However, such a definition would make the test statistic, and hence the resulting inference, depend on the arbitrary choice of representative~$u_{k_i}(X_{l_i};\theta)$ in each~$\mathcal U^k(X_{l_i};\theta)$. Instead, we use the fact that all~$u\in\mathcal U^k(X_{l_i};\theta)$ predict the same outcome~$\mathcal Y^k(X_{l_i};\theta)$, and use discrepancy~(\ref{eq:alternative discrepancy}) to define the test statistic. Here, discrepancy~(\ref{eq:alternative discrepancy}) can be written
\begin{eqnarray*}
    &&\delta((u_{k_i}(X_{l_i};\theta),X_{l_i}),(Y_j,X_j);\theta) \\
    && \hskip50pt =\;\inf_{\tilde y_i\in\mathcal Y^{k_i}(X_{l_i};\theta)} ((\tilde y_i,X_{l_i})-(Y_j,X_j))\tilde\Sigma^{-1}((\tilde y_i,X_{l_i})-(Y_j,X_j))^{\top},
\end{eqnarray*}
where~$\tilde\Sigma$ is defined as in~(\ref{eq:alternative discrepancy}).

The test statistic is: 
\begin{eqnarray*}
    T_n(\theta) & = & \min_{\pi\geq0}\sum_{ij}\pi_{ij}
    \;\delta((u_{k_i}(X_{l_i};\theta),X_{l_i}),(Y_j,X_j);\theta) \\
    \mbox{subject to:} && \sum_{i=1}^{nK}\pi_{ij} = \frac{1}{n},\;\; \sum_{j=1}^n\pi_{ij} = \frac{1}{n}q^{k_i}(X_{l_i};\theta),\;\; i\leq nK,j\leq n.
\end{eqnarray*}

Similarly, the critical values are obtained as the quantiles of the following statistic.
\begin{eqnarray*}
    \tilde T_n(\theta) & = & \max_{\tilde y^{(n)}\in\mathcal Y^{n
    }}\min_{\pi\geq0}\sum_{ij}\pi_{ij}
    \;\delta((u_{k_i}(X_{l_i};\theta),X_{l_i}),(\tilde y_j,X_j);\theta)\\
    \mbox{subject to:} && \sum_{i=1}^{nK}\pi_{ij} = \frac{1}{n},\;\; \sum_{j=1}^n\pi_{ij} = \frac{1}{n}q^{k_i}(X_{l_i};\theta),\;\;\tilde y_j\in\mathcal Y^{\tilde k_j}(X_{j};\theta),\;\; i\leq nK,j\leq n,
\end{eqnarray*}
where, for each~$j\leq n$, $\tilde k_j$ is a random draw from the distribution with probability mass function~$(k,q^k(X_j;\theta))_{k=1}^{K}$, and the draws are independent conditionally on~$X^{(n)}$. As before, the quantiles of statistic~$\tilde T_n$ are approximated with~$S$ Monte Carlo samples~$(\tilde k_1^s,\ldots,\tilde k_n^s)$.

\begin{continued}[Example~\ref{ex:entry} continued:]
Consider the 2-by-2 entry game with~$\theta>0$. The five possible values for~$\Gamma_y(u;\theta)$ are the~$K=5$ predicted combinations of Nash equilibrium in pure strategies, i.e., $\mathcal Y^1:=\{(0,0)\}$, $\mathcal Y^2:=\{(0,1)\}$, $\mathcal Y^3:=\{(0,1),(1,0)\}$, $\mathcal Y^4:=\{(1,0)\}$ and~$\mathcal Y^5:=\{(1,1)\}$. For each~$k$, $\mathcal U^k(\theta)$ is the region of~$u=(u^1,u^2)\in\mathbb R^2$ such that the Nash equilibrium set in pure strategies is~$\mathcal Y^k$ when profits are~$\pi^j:=Y^j(u^j-\theta Y^{1-j})$, $j\in\{0,1\}$. Hence, $\mathcal U^1(\theta)=[-\infty,0]^2$, $\mathcal U^2(\theta)=[-\infty,\theta]\times[0,+\infty)\backslash\mathcal U^3(\theta)$, $\mathcal U^3(\theta)=[0,\theta]^2$, $\mathcal U^4(\theta)=[0,+\infty)\times[-\infty,\theta]\backslash\mathcal U^3(\theta)$ and~$\mathcal U^5(\theta)=[\theta,+\infty)^2$. 

Since~$U$ is bivariate standard normal, $q^k(\theta)$ is the integral of a bivariate normal on region~$\mathcal U^k(\theta)$. For instance, $q^3(\theta)=(\Phi(\theta)-\Phi(0))^2$. Denote~$u_k(\theta)$ an arbitrary element of~$\mathcal U^k(\theta)$. Since the model has no covariates, the test statistic is the optimal transport of discrete distribution~$(q^k(\theta))_{k=1}^K$ to the distribution~$\hat P_n$ with cost~$C_{kj}(\theta)=\delta(u_k(\theta),Y_j;\theta)$, $k\leq K,j\leq n$, i.e., \begin{eqnarray*}
    T_n(\theta) & = & \min_{\pi\geq0}\sum_{kj}\pi_{kj}\;C_{kj}(\theta). \\
    \mbox{subject to:} && \sum_{k=1}^{K}\pi_{kj} = \frac{1}{n},\;\; \sum_{j=1}^n\pi_{kj} = q^k(\theta),\;\; k\leq K,j\leq n.
\end{eqnarray*}
The critical values are obtained as the quantiles of the following statistic.
\begin{eqnarray*}
    \tilde T_n(\theta) & = & \max_{\tilde y^{(n)}\in\mathcal Y^{n
    }}\min_{\pi\geq0}\sum_{kj}\pi_{kj}\;\delta( u_k(\theta),\tilde y_j;\theta). \\
    \mbox{subject to:} && \sum_{k=1}^{K}\pi_{kj} = \frac{1}{n},\;\; \sum_{j=1}^n\pi_{kj} = q^k(\theta),\;\;\tilde y_j\in\mathcal Y^{\tilde k_j}(\theta),\;\; k\leq K,j\leq n,
\end{eqnarray*}
where, for each~$j\leq n$, $\tilde k_j$ is a random draw from the distribution with probability mass function~$(k,q^k(\theta))_{k=1}^{K}$, and the draws are mutually independent. As before, the quantiles of statistic~$\tilde T_n$ are approximated with~$S$ Monte Carlo samples~$(\tilde k_1^{(s)},\ldots,\tilde k_n^{s})$.
\end{continued}

\section{Simulation evidence}
\label{sec:MC}
This section contains two simulation exercises. The first simulation exercise considers a setting in which existing moment-inequality methods are asymptotically valid. We compare their size control and power and also report computing time and sensitivity to the choice of discrepancy~$\delta$ in the test statistic. The second simulation design uses the latent consideration-set model of Example~\ref{ex:consideration} to illustrate how dependence in the selection rule can invalidate conventional moment-inequality inference, while our procedure continues to control size and retains nontrivial power.

\subsection{Regression with interval-censored covariates}
\label{sec:MC-interval}
\subsubsection*{Simulation design}

We analyze the performance of our inference procedure with a simulation design based on Example~\ref{ex:interval}. 
We test~$H_0:\theta=1$. In the size analysis, the true value of~$\theta$ is set to~$1$. In the power analysis, it ranges between~$-1$ and~$2.5$. In each simulation replication, we draw an i.i.d. sample~$((\underline W_1,\overline W_1),\ldots,(\underline W_n,\overline W_n))$ of replications of~$(\min\{W_1,W_2\}-c/2,\max\{W_1,W_2\}+c/2)$, where~$c\geq0$,
\[
\left(\begin{array}{c}
     W_1  \\
     W_2
\end{array}\right)
\sim
N\left(\left(\begin{array}{c}
     0  \\
     0
\end{array}\right), 
\left(\begin{array}{cc}
   1  & \rho \\
   \rho  & 1
\end{array}\right)
\right),
\]
and~$\rho$ takes values~$0$, $0.5$ and~$1$.
In one simulation design (hereafter {\em random} or R), each~$W^\ast_j$ is selected randomly (i.e., drawn independently from a uniform distribution) from~$[\underline W_j,\overline W_j]$, for all~$j\leq n$. In a second simulation design (hereafter {\em worst case} or WC), each~$W^\ast_j$ is selected from~$[\underline W_j,\overline W_j]$, for all~$j\leq n$, so as to maximize the value of the resulting test statistic. In the latter design, exact size control is expected. In this design, the vector~$X_j$ of exogenous variables contains~$\underline{W}$ and~$\overline{W}$.
A sample of latent variables~$(U_1,\ldots,U_n)$ is drawn independently from~$N(0,1)$ and~$Y_j$ is set equal to~$\theta W^\ast_j+U_j$ for each~$j\leq n$.

\subsubsection*{Competing inference strategies}

The inference strategy proposed in Section~\ref{sec:inference} requires a choice of discrepancy~$\delta$ for the test statistic. This is akin to the choice of statistic (Cramer-von Mises vs. Kolmogorov-Smirnov, for instance) in other proposals. We compare the performance of our procedure with three competing discrepancies: 
\begin{enumerate}
    \item Strategy OTE, using discrepancy~(\ref{eq:discrepancy}) without weighting, i.e., with~$\hat\Sigma$ replaced by the identity matrix;
    \item Strategy OTW, using discrepancy~(\ref{eq:discrepancy});
    \item Strategy OTY, using discrepancy~(\ref{eq:alternative discrepancy}).
\end{enumerate}
Given the choice of discrepancy, the numerical implementation of the procedure is discussed in appendix~\ref{sec:num}.

We compare results using our inference strategy and results using existing strategies explicitly designed for cases with many moment inequalities, with an emphasis on incomplete models, i.e., \cite{CCK:2019} (hereafter CCK) and \cite{andrews2017inference} (hereafter AS).\footnote{We are grateful to the authors for sharing their simulation code.} Both strategies involve three steps:
\begin{enumerate}
    \item Transform the model into a conditional moment inequality model using the strategy proposed in \cite{BMM:2011} and \cite{GH:2011}, and a discretization of the outcome space~$\mathcal Y$. Based on Theorems~1 and~4 in \cite{GH:2011}\footnote{Theorem~4 on core determining classes in \cite{GH:2011} is stated for a discrete outcome space, but it extends to~$\mathcal Y=\mathbb R$ with a straightforward adjustment of the proof.}, the following collection of conditional moment inequalities characterizes the set of parameter values compatible with the distribution of the observables:
    \begin{eqnarray}\label{eq:Artstein}
        \Phi(y-\theta\overline W)\leq\mathbb E[1\!\{Y\leq y\}\vert \underline W,\overline W]\leq \Phi(y-\theta\underline W), \mbox{ for all }y\in\mathbb R,
    \end{eqnarray}
where~$\Phi$ is the cdf of the standard normal distribution. We discretize the outcome space using~$L+1$ equally spaced cutoffs on~$[-2,2]$ to obtain a finite collection of conditional moment inequalities
     \begin{eqnarray*}
        &&\hskip20pt\Phi(-2+4l/L-\theta\overline W)\leq\mathbb E[1\!\{Y\leq -2+4l/L\}\vert \underline W,\overline W]\leq \Phi(-2+4l/L-\theta\underline W),
    \end{eqnarray*} 
    for~$l=0,\ldots,L$. We report results with $L = 25$ and $L = 100$ in simulations.
\item Transform the conditional moment inequalities into unconditional ones, using the strategy proposed in Section~3 of AS. Hence each conditional moment inequality is replaced with a collection of unconditional moment inequalities according to~(3.1) and~(3.2) page~278 of AS. As in (3.6) of AS, we use $\mathcal G_{\text{c-cubes}}$ of $r=1,\dots,r_{\text{max}}$ with $r_{\text{max}} = 3$ or $10$ in simulations.
    \item In the third step, we perform inference according to each alternative in the following way. 
    \begin{enumerate}
        \item We perform inference according to AS using the CvM test statistic~(3.7) on page~279, with function~$S_1$ defined in~(3.9) and GMS critical values. Tuning parameters are selected using the rule of thumb proposed in AS. Precisely, we set~$\kappa_n=(0.3\ln n)^{1/2}$, $B_n=(0.4\ln n/\ln\ln n)^{1/2}$ and~$\varepsilon=5/100$.
	\item We perform inference according to CCK using the recommended test statistic~(13) page~1877 and the recommended bootstrapped two-step critical values~(39) page~1886 with~$1,000$ bootstrap samples (with Efron bootstrap). The tuning parameter is selected using the rule of thumb proposed in Section~6.2 page~1896 of CCK. Precisely, we set~$\beta=0.001$. In addition, we also apply the same regularized standard deviation~$\sqrt{s^2+0.05}$ to CCK, where~$s^2$ is the sample variance of the corresponding moment, as we did for AS.\footnote{The original CCK procedure uses ordinary sample standard deviations. We modify its implementation by applying the same additive variance regularization as in AS. Without this regularization, CCK exhibts significant overrejection in our simulations.}
    \end{enumerate}
\end{enumerate}

\subsubsection*{Size control}

Sample sizes range over~$n\in\{10,50,100,500\}$. Nominal levels range over~$1-\alpha\in\{0.90,0.95,0.99\}$. Coverage probabilities are based on quantiles of~$\tilde T_n(\theta)$ as critical values. For each sample size, nominal level and choice of critical value, we conduct $5,000$ replications of the test. For each test, the critical values are based on~$S=1000$ Monte Carlo samples. 

Table~\ref{table:OTcoverage} shows coverage probabilities for our testing procedure under three different values of the correlation parameter~$\rho$, two values of~$c$ and three different choices of discrepancy for the test statistic. Coverage probabilities are close to the nominal levels in all cases and all sample sizes. As expected from the theory, these results are consistent with the finite-sample size-control result across the different specifications of the correlation~$\rho$, the degree of incompleteness~$c$ and the choice of distance.

\begin{table}[h]
\caption{Coverage probabilities: ``$1-\alpha$'' columns report coverage probabilities based on quantiles of~$\tilde T_n(\theta)$ as critical values, for nominal levels~$0.90$, $0.95$ and~$0.99$ and true value~$\theta=1$. OTE, OTW and OTY indicate the procedure with unweighted discrepancy in the latent variable space, the weighted discrepancy in the latent variable space, and the weighted discrepancy in the outcome space respectively. The data is simulated with~$\rho=0$, $0.5$ and~$1$, and worst case~$W^\ast$ from~$[\underline W,\overline W]$.}
\begin{center}
\begin{tabular}{c c r r r r r r r r r} 
\toprule 
&& \multicolumn{9}{c}{$1-\alpha$}  \\ 
\cmidrule(l){3-11} 
&& \multicolumn{3}{c}{0.90} & \multicolumn{3}{c}{0.95} & \multicolumn{3}{c}{0.99} \\ \\ 
 & & OTE & OTW & OTY & OTE & OTW & OTY & OTE & OTW & OTY \\ \\
$n$ & $(\rho,c)$ &  &  &  &  &  &  &  &  & \\ 
  \midrule 
10  & (0,1) &  0.903  & 0.903 &0.905& 0.951 & 0.948 & 0.951 & 0.989 & 0.990 & 0.989 \\ \\ 
  & (0.5,1) & 0.904 & 0.898 &0.902& 0.949 & 0.946 & 0.947& 0.988 & 0.990 & 0.991 \\ \\ 
  & (1,0) & 0.897 & 0.899 &0.900& 0.947 &  0.947 &0.948& 0.991 & 0.990 & 0.989\\ \\ 
50  & (0,1) & 0.902 & 0.903 &0.906& 0.951 & 0.953 & 0.955& 0.991 & 0.992 & 0.992\\ \\ 
  & (0.5,1) & 0.902 & 0.904 &0.902& 0.950 & 0.951 & 0.953& 0.991 & 0.991 & 0.990\\ \\ 
  & (1,0) & 0.899 & 0.899 &0.899& 0.954 & 0.953 & 0.953& 0.992 & 0.990 & 0.990 \\ \\ 
100  & (0,1) & 0.902 &  0.904 &0.903& 0.953 & 0.951 & 0.950& 0.989 & 0.988 & 0.987\\ \\ 
  & (0.5,1) & 0.904 & 0.902 &0.903& 0.952 & 0.950 & 0.947& 0.991 & 0.990 & 0.988 \\ \\ 
  & (1,0) & 0.899 & 0.903 &0.904& 0.950 & 0.952 & 0.952& 0.992 & 0.991 & 0.991\\ \\ 
500  & (0,1) & 0.899 & 0.898 &0.898& 0.947 & 0.951 & 0.952& 0.991 & 0.990 & 0.990\\ \\ 
  & (0.5,1) & 0.901 & 0.897 &0.899& 0.951 & 0.946 & 0.948& 0.990 & 0.990 &0.988\\ \\ 
  & (1,0) & 0.898 & 0.898 &0.898& 0.948 & 0.947 & 0.947& 0.988 & 0.988 & 0.988\\ \\ 
\bottomrule 
\end{tabular}
\end{center}
\label{table:OTcoverage}
\end{table}

Table~\ref{table:compare coverage} compares the coverage probabilities and computation times of OTW, AS, and CCK. We use the random (R) simulation design with $\rho = 1$ and~$c=0$, which is the most challenging case for size control. For AS and CCK, we vary $L\in\{25,100\}$ and $r_{\max}\in\{3,10\}$. As expected, our method based on optimal transport is computationally attractive for smaller sample sizes, but less so for higher sample sizes. Coverage probabilities are close to nominal level for our procedure. AS tends to overreject across the reported configurations. CCK is generally closer to nominal size, but can either slightly overreject or underreject in small samples. For example, at~$n=100$, its coverage is~$0.942$ for~$L=100$ and~$r_{\max}=3$, and~$0.957$ for~$L=25$ and~$r_{\max}=10$. Computational performance depends on both sample size and grid specification. 

\begin{table}[htbp]
\begingroup
\centering
\setlength{\extrarowheight}{0pt}
\setlength{\tabcolsep}{11pt}
\renewcommand{\arraystretch}{1.15}
\caption{Coverage probabilities and computation times: The ``cp'' columns report coverage probabilities at nominal level~$0.95$. The ``ct'' columns report the running times in seconds for all $5{,}000$ tests, run in parallel on a Mac Studio with M4 Max (16 CPU cores), excluding the time on starting multi-thread workers and data generation. The design has $(\rho, c) =(1,0)$, with true and tested coefficient~$\theta=1$. Each test uses $1{,}000$ bootstrap resamples or Monte Carlo calibration draws. AS and CCK use an additive variance regularizer of~$0.05$. }
\label{table:compare coverage}
\begin{tabular}{@{}lrrrrrr@{}}
\toprule
& \multicolumn{2}{c}{$n=100$} & \multicolumn{2}{c}{$n=500$} & \multicolumn{2}{c}{$n=1{,}000$} \\
\cmidrule(lr){2-3}\cmidrule(lr){4-5}\cmidrule(lr){6-7}
Method & cp & ct & cp & ct & cp & ct \\
\midrule
OTW & 0.952 & 20.5 & 0.947 & 471.7 & 0.950 & 2524.9 \\
\addlinespace[0.6em]
\multicolumn{7}{@{}c@{}}{$L=25$, $r_{\max}=3$} \\
AS & 0.935 & 5.7 & 0.939 & 23.4 & 0.938 & 49.4 \\
CCK & 0.954 & 1.5 & 0.951 & 5.7 & 0.951 & 11.5 \\
\addlinespace[0.6em]
\multicolumn{7}{@{}c@{}}{$L=25$, $r_{\max}=10$} \\
AS & 0.936 & 25.4 & 0.939 & 76.8 & 0.938 & 142.2 \\
CCK & 0.957 & 4.8 & 0.951 & 18.8 & 0.953 & 42.8 \\
\addlinespace[0.6em]
\multicolumn{7}{@{}c@{}}{$L = 100$, $r_{\max}=3$} \\
AS & 0.929 & 19.7 & 0.944 & 86.1 & 0.939 & 170.3 \\
CCK & 0.942 & 3.5 & 0.951 & 13.1 & 0.953 & 26.0 \\
\addlinespace[0.6em]
\multicolumn{7}{@{}c@{}}{$L = 100$, $r_{\max}=10$} \\
AS & 0.932 & 79.2 & 0.941 & 274.2 & 0.937 & 511.2 \\
CCK & 0.946 & 16.3 & 0.952 & 48.2 & 0.954 & 91.3 \\
\bottomrule
\end{tabular}
\par\vspace{0.5em}
\endgroup
\end{table}

\subsubsection*{Power analysis}
In our power analysis, we fix nominal size to~$\alpha:=0.05$, and the null hypothesis to~$H_0:\theta=1$. We vary the true value of~$\theta$ from~$-1$ to~$2.5$, and for each true value of~$\theta$, we obtain rejection frequencies from~$5,000$ replications of the dgp simulation and testing procedure, and~$1,000$ Monte Carlo samples in each instance of the testing procedure. With these rejection frequencies, we trace power curves. Unless specified otherwise, the sample size is~$n=500$, $c=1$, the correlation parameter~$\rho$ is set to~$0$, the DGP involves the random selection of~$W^\ast$ from~$[\underline W,\overline W]$. The benchmark procedure is OTW, i.e., our testing procedure with the recommended weighted discrepancy in the latent variable space in the definition of the test statistic.

    \begin{figure}[h]
	\centering
	\includegraphics[width=0.7\linewidth]{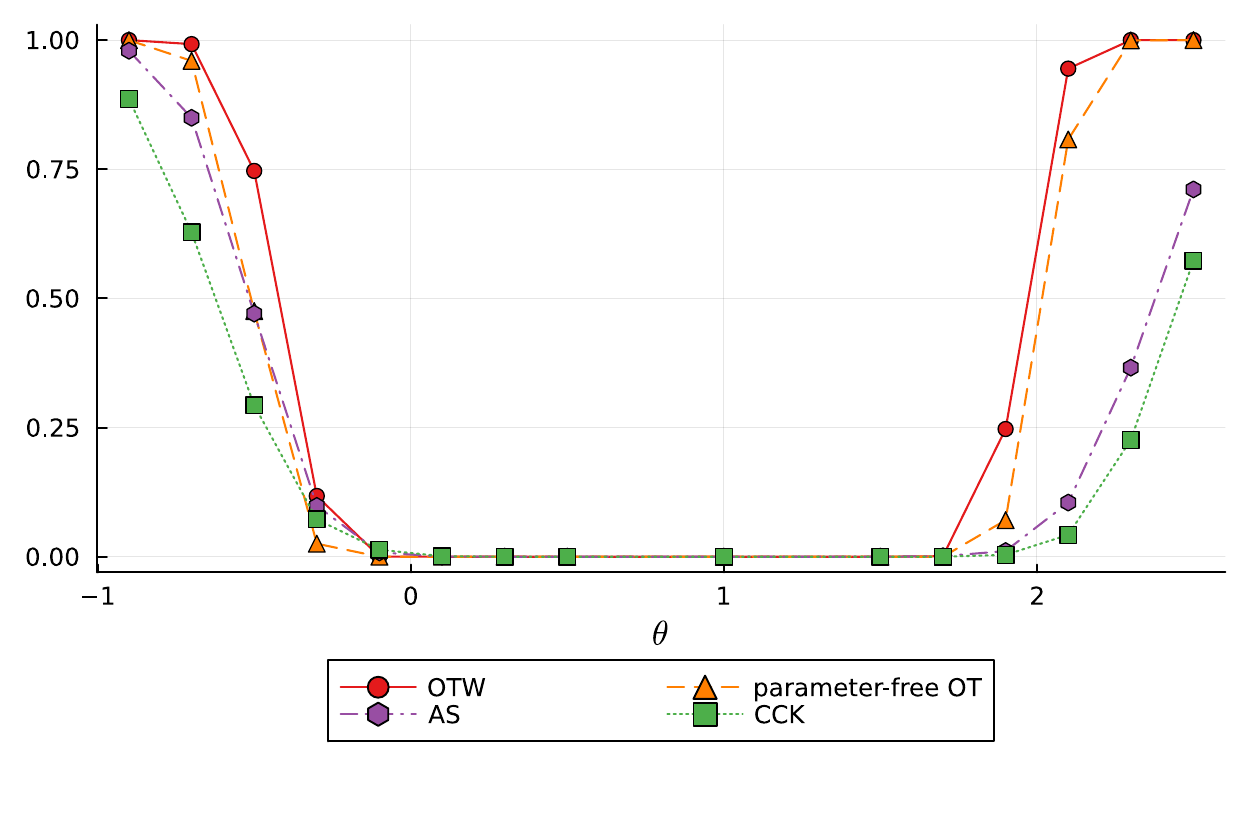}
        \caption{Power comparisons with competing  procedures. The sample size is~$n=500$, the DGP is R with~$\rho=0$ and~$c=1$. The solid red curve with circles, the dashed orange curve with triangles, the semi-dashed purple curve with hexagons and the dotted green curve with squares are the power curves for~OTW, parameter-free OT, AS and CCK respectively. AS and CCK use~$L=100$ and~$r_{\max}=3$.}
	\label{fig:power_curve_compare}
    \end{figure}

Figure~\ref{fig:power_curve_compare} compares the power curves for our procedure, AS and CCK. Our procedure clearly dominates both AS and CCK. More surprisingly, the parameter-free procedure, which we expected to be quite conservative, is quite competitive relative to AS and clearly dominates CCK.

    \begin{figure}[h]
	\centering
	\includegraphics[width=0.7\linewidth]{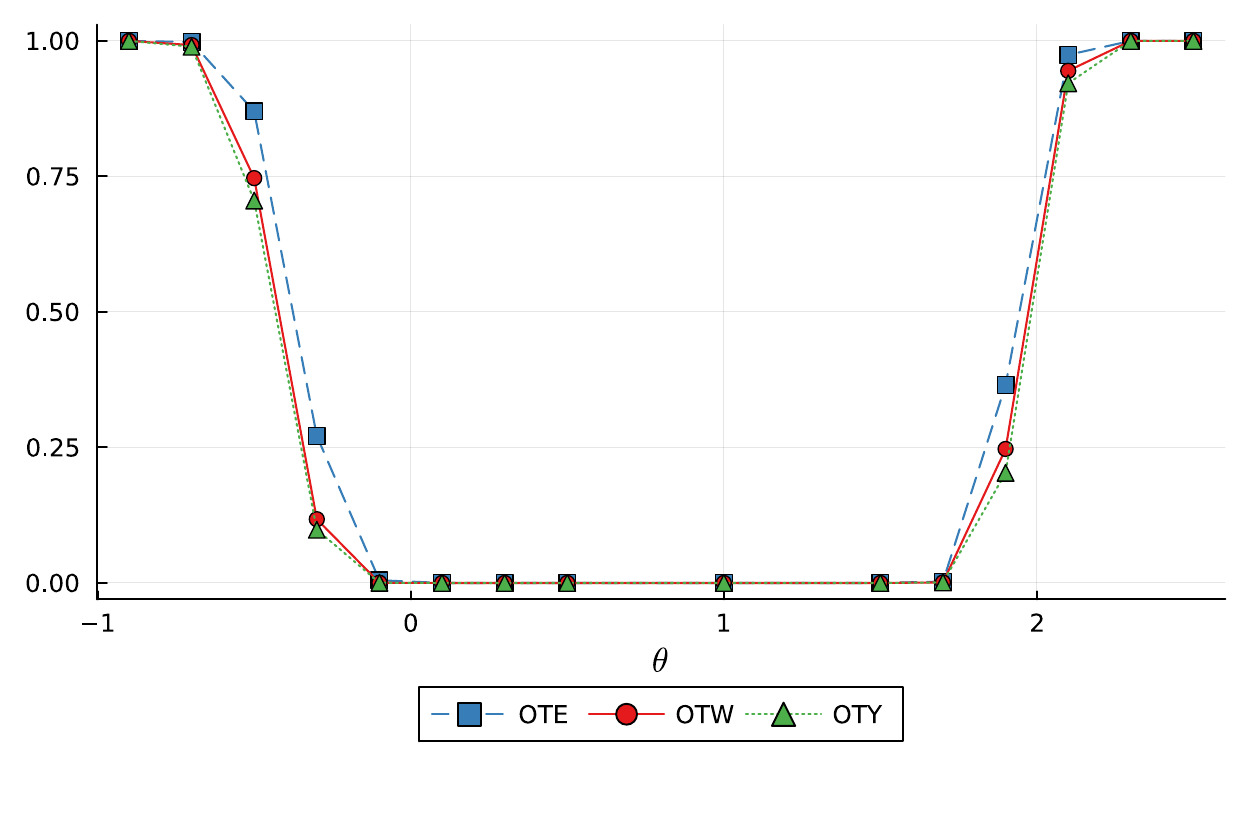}
        \caption{Power curve and discrepancy. The sample size is~$n=500$, the DGP is R with~$\rho=0$ and $c=1$. The dashed blue curve with squares, the solid red curve with circles and the dotted green curve with triangles are the power curves with the procedures OTE, OTW and OTY respectively.}
	\label{fig:power_curve_d}
    \end{figure}

Figure~\ref{fig:power_curve_d} shows power curves for all three choices of discrepancy in the definition of the test statistic, i.e., unweighted OTE, the recommended weighted OTW and weighted discrepancy on the outcome space OTY. We don't see a significant difference in power performance between OTW and OTY, although the power of OTE is slightly better.

    \begin{figure}[htbt]
	\centering
	\includegraphics[width=0.7\linewidth]{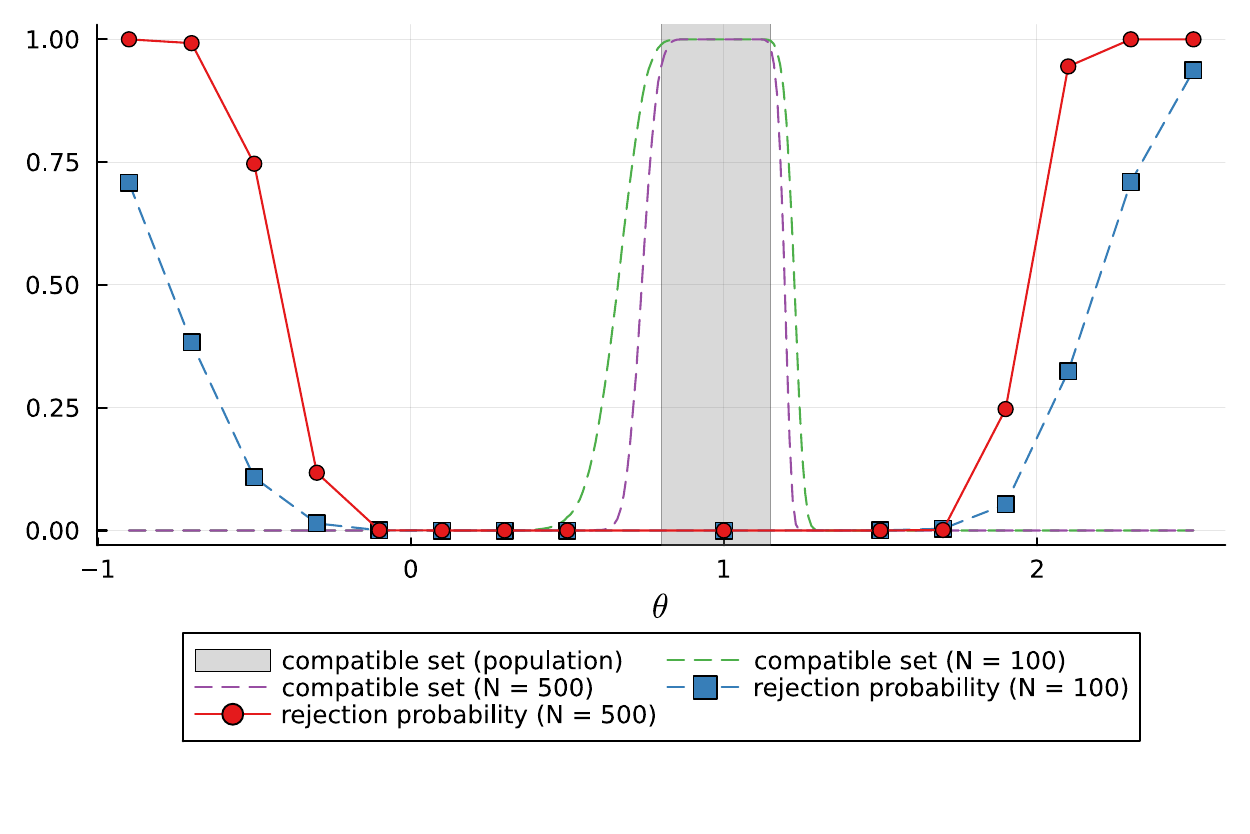}
        \caption{Power curve and model compatibility. The sample sizes are~$n=100$ and~$500$, the DGP is R with~$\rho=0$, $c=1$, and the procedure is OTW. The solid red curve with circles (resp. dashed blue curve with squares) is the power curve for sample size~$500$ (resp.~$100$). The  dashed purple (resp. green) curve gives at each true value~$\theta$ the probability that~$1$ is deemed compatible with the model based on~$500$ (resp.~$100$) random values of~$X$. The shaded grey region is the set of true values~$\theta_0$ for which the null value~$1$ is compatible with the true data-generating process.}
	\label{fig:power_curve}
    \end{figure}

Finally, figure~\ref{fig:power_curve} achieves two objectives. First, we trace power curves for~$n=100$ and~$n=500$ to see the effect of sample size on power. The effect of sample size is significant, but the procedure still has good power for sample sizes as low as~$n=100$. Second, we try to disentangle the effects of partial identification and sampling uncertainty. To that end, we approximate the set of parameter values compatible with the distribution of the observables, based on the characterization from \cite{BMM:2011} and \cite{GH:2011}. Applied to the present simulation design, this characterization is specified in~(\ref{eq:Artstein}). The grey area in figure~\ref{fig:power_curve} covers the set of values of~$\theta$ such that the null value~$1$ is compatible with the distribution of the observables when~$\theta$ is the true value. We also disentangle the specific effect of sampling of covariates in the following way. We approximate this compatible set with random draws of size~$n=100$ and~$500$ of the conditioning variables. We trace two curves, the dashed green curve for~$n=100$ and the dashed purple curve for~$n=500$. At each value of~$\theta$, the dashed curve gives the probability that~$1$ satisfies~(\ref{eq:Artstein}) based on a random draw of the conditioning covariates (drawn according to their known DGP). We find that the true compatible set accounts for a small portion of the indeterminacy. However, the compatible set approximated with the available sample of covariates accounts for a large proportion of the indeterminacy.

\subsection{Dependent selection in a latent consideration-set model}
\label{sec:MC-consideration}

The objective of this simulation exercise is to illustrate the role of the selection mechanism in inference. Even when covariates and latent variables are i.i.d., a common selection rule can make the selected outcomes dependent across observations. We show that treating those outcomes as i.i.d. can produce substantial size distortions, whereas our procedure continues to control size because it does not restrict the joint selection mechanism.

\subsubsection*{Simulation design}

We specialize Example~\ref{ex:consideration} to three products, $\mathcal D=\{1,2,3\}$, no covariates, and a minimum consideration-set size~$\kappa=2$. We adopt the generic parameter notation of the paper, 
and specify utilities as
\[
W(d,u_i;\theta)=\theta_d+u_i(d),
\qquad d\in\mathcal D,
\]
where the coordinates~$u_i(d)$ are i.i.d.~$N(0,1)$ across consumers and products. We test the true value~$\theta_0=(0,0,0)$. Since utility shocks are continuous, preference rankings of products are strict almost surely. Let~$d_i^w$ denote consumer~$i$'s least preferred product. The set of products that can be chosen from some consideration set of size at least two is
\begin{eqnarray}
\label{eq:cs-prediction-set}
D_{2,i}^\ast=\mathcal D\setminus\{d_i^w\}.
\end{eqnarray}
In the notation of the general framework, $D_{2,i}^\ast$ is the outcome consistent with the structural model at~$u_i$: $\Gamma_y(u_i;\theta_0)=D_{2,i}^\ast$.

The dependent selection mechanism that we construct here is a centralized ``Trending Now'' rule. To give the rule an economic interpretation, suppose a platform sells all three products, observes the latent rankings in the full sample, and can influence individual consideration sets through advertising or a recommendation system. The platform computes the share of consumers for whom each product is inducible and selects the product with the largest share:
\begin{eqnarray}
\label{eq:cs-trending-score}
\widehat S_{n,d}=\frac{1}{n}\sum_{i=1}^n1\!\{d\in D_{2,i}^\ast\},
\qquad
d_n^{\mathrm{tr}}\in\argmax_{d\in\mathcal D}\widehat S_{n,d}.
\end{eqnarray}
Ties in the definition of~$d_n^{\mathrm{tr}}$ are broken using randomness independent of the utilities. Thus, $d_n^{\mathrm{tr}}$ is the product that can be induced for the largest share of consumers and is the platform's natural candidate to promote as ``trending.''
The platform then gives consumer~$i$ the consideration set
\begin{eqnarray}
\label{eq:cs-trending-menu}
G_{n,i}=\begin{cases}
\{d_n^{\mathrm{tr}},d_i^w\},&d_n^{\mathrm{tr}}\in D_{2,i}^\ast,\\
D_{2,i}^\ast,&d_n^{\mathrm{tr}}=d_i^w,
\end{cases}
\qquad
Y_{n,i}=\argmax_{d\in G_{n,i}}W(d,u_i;\theta_0).
\end{eqnarray}
Thus, if the trending product can be induced, it is paired with the consumer's worst product and is chosen. Otherwise, the consumer is shown her two best products and chooses her most preferred product. Every menu has size two and~$Y_{n,i}\in D_{2,i}^\ast$, so this joint selection mechanism is admissible under Definition~\ref{def:DGP}. It is nevertheless dependent across individuals: the common promoted product~$d_n^{\mathrm{tr}}$ is computed from the latent rankings of the entire sample. For each sample size~$n$, the platform recomputes~$d_n^{\mathrm{tr}}$ from the full latent sample.

\subsubsection*{Sharp moment inequalities and selection bias}

Theorem~3.1 in \cite{BCMT:2021} gives a sharp determining class indexed by subsets of~$\mathcal D$ with cardinality smaller than~$\kappa$. Since~$\kappa=2$, only the three singleton inequalities are needed. At~$\theta_0$, each product belongs to the top two with probability~$2/3$, so, for each~$i\leq n$, the sharp marginal inequalities are
\begin{eqnarray}
\label{eq:cs-moment-inequalities}
\mathbb P(Y_{n,i}=d)\leq\mathbb P(d\in D_{2,i}^\ast)=\frac{2}{3},
\qquad d\in\mathcal D,
\end{eqnarray}
or, equivalently, $\mathbb E[1\!\{Y_{n,i}=d\}-2/3]\leq0$. As restrictions on a generic one-observation law, these inequalities sharply characterize~$\mathcal P_{\theta_0}$. These moment inequalities remain valid even when there is dependence across observations. Indeed, their validity follows pointwise from~$1\!\{Y_{n,i}=d\}\leq1\!\{d\in D_{2,i}^\ast\}$ and does not require independent selection.

The ``Trending Now'' rule nevertheless creates a winner-selection effect in the finite-sample moments. Define~$\widehat p_{n,d}=n^{-1}\sum_{i=1}^n1\!\{Y_{n,i}=d\}$. Since each~$D_{2,i}^\ast$ contains two products, $\sum_{d\in\mathcal D}\widehat S_{n,d}=2$. Equations~(\ref{eq:cs-trending-score})-(\ref{eq:cs-trending-menu}) therefore imply
\begin{eqnarray}
\label{eq:cs-selection-bias}
\widehat p_{n,d_n^{\mathrm{tr}}}
=\widehat S_{n,d_n^{\mathrm{tr}}}
=\max_{d\in\mathcal D}\widehat S_{n,d}
\geq\frac{2}{3},
\qquad
\widehat p_{n,d}\leq\frac{1}{3}\quad\mbox{for }d\neq d_n^{\mathrm{tr}}.
\end{eqnarray}
By label symmetry, $\mathbb P(Y_{n,i}=d)=1/3$ for every fixed product~$d$, so the population restrictions in \eqref{eq:cs-moment-inequalities} are satisfied. In every realized sample, however, the coordinate corresponding to the promoted product is selected to be on or above its moment-inequality boundary, while the other two empirical moments are bounded above by~$-1/3$. An inference procedure calibrated under i.i.d. sampling does not take this sample-wide selection into consideration and can consequently over-reject. Appendix~\ref{sec:cs-size-distortion} formalizes the resulting size distortion for the procedure of \cite{AS:2010}, which we call AS2010 to distinguish it from the Andrews--Shi procedure considered above.

\subsubsection*{Implementation of our procedure}

We implement the parameter-dependent procedure with two discrepancies.\footnote{Although~$\mathcal Y$ is finite here, so that the refinement of section~\ref{sec:discrete} applies, we implement the general simulation-based statistic of section~\ref{sec:test}, since it is the version to which the consistency results of sections~\ref{sec:consistency}
and~\ref{sec:optimality} apply.} The latent-space discrepancy is as follows,
\[
\delta_U(u,d;\theta)=\inf_{u^\prime:d\in D_2^\ast(u^\prime;\theta)}\|u-u^\prime\|^2.
\]
In this additive finite-product specification, the region $\{u^\prime:d\in D_2^\ast(u^\prime;\theta)\}$ is a finite union of polyhedral regions, and $\delta_U(u,d;\theta)$ has the closed form
\[
\delta_U(u,d;\theta) =\frac{1}{2}\max\left(0\ ,\  \min_{c\neq d}\{\theta_c+u(c)\}-\{\theta_d+u(d)\} \right)^2.
\]
Thus, the discrepancy is zero if~$d$ is not strictly least preferred under~$(u,\theta)$. Otherwise, the discrepancy is one half of the squared gap between the two lowest total utilities. The outcome-space discrepancy is~$\delta_Y(u,d;\theta)=1\!\{d\notin D_2^\ast(u;\theta)\}$. We refer to the resulting procedures as~OT$_U$ and~OT$_Y$, respectively. Appendix~\ref{sec:cs-computation} derives both discrepancies for the general finite-product case.

Following the notation of Section~\ref{sec:test}, for each~$n$ we draw one fixed reference latent sample~$\tilde u^{(n)}=(\tilde u_1,\ldots,\tilde u_n)$, shared by both procedures, and form the~$n\times3$ table with entries~$c_{id}=\delta_U(\tilde u_i,d;\theta_0)$ or~$c_{id}=\delta_Y(\tilde u_i,d;\theta_0)$. Because there are no covariates, the full~$n\times n$ cost matrix satisfies~$C_{ij}=c_{i,Y_{n,j}}$. The observed test statistic therefore depends on the data only through the three product counts and is computed as one capacitated optimal-transport problem.

For each Monte Carlo draw used to construct the critical value, an independent latent sample~$\tilde U^{\prime(n)}$ produces admissible sets~$\mathcal S_j=D_2^\ast(\tilde U_j^\prime;\theta_0)$, $j\leq n$. The critical statistic solves
\begin{eqnarray}
\label{eq:cs-max-min-main}
\tilde T_n(\theta_0)
=\max_{\tilde y_j\in\mathcal S_j,\,j\leq n}
\min_{\pi\in\Pi_n}\sum_{i,j=1}^n\pi_{ij}c_{i,\tilde y_j}.
\end{eqnarray}
A direct calculation would enumerate~$2^n$ admissible outcome profiles. Instead, we aggregate profiles into product-count vectors. The inner transport value is convex in the count vector, so its maximum over the feasible-count polytope is attained at a vertex. Each such vertex is generated by a strict global priority ordering of the products. Consequently, solving $\tilde{T}_n(\theta_0)$ requires at most~$3!=6$ aggregated transport problems. Appendix~\ref{sec:cs-computation} gives the exact reductions and computational details.\footnote{This reduction algorithm extends to finite-outcome models without covariates. For~$J$ outcomes it replaces a problem exponential in~$n$ with at most~$J!$ transport problems, although the factorial dependence on~$J$ eventually becomes limiting. With observation-specific covariates, outcome counts are no longer sufficient. Finite-outcome max--min problems can instead be formulated as mixed-integer programs based on the assignment dual.}

\subsubsection*{Results}

Table~\ref{tab:cs-size} reports rejection frequencies. AS2010 is the one-sided max test with generalized moment selection and an i.i.d. nonparametric bootstrap. The capacity probabilities in~(\ref{eq:cs-moment-inequalities}) are computed exactly by enumerating the six strict product rankings.

\begin{table}[htbp]
\centering
\caption{Rejection probabilities under dependent ``Trending Now'' selection}
\label{tab:cs-size}
\small
\setlength{\tabcolsep}{4pt}
\begin{tabular}{@{}c r r r r r r r r r@{}}
\toprule
&\multicolumn{3}{c}{$\alpha=0.10$}
&\multicolumn{3}{c}{$\alpha=0.05$}
&\multicolumn{3}{c}{$\alpha=0.01$}\\
\cmidrule(lr){2-4}\cmidrule(lr){5-7}\cmidrule(l){8-10}
$n$ & OT$_U$ & OT$_Y$ & AS2010 & OT$_U$ & OT$_Y$ & AS2010 & OT$_U$ & OT$_Y$ & AS2010\\
\midrule
10    & 0.032 & 0.017 & 0.164 & 0.032 & 0.017 & 0.046 & 0.000 & 0.000 & 0.046\\
50    & 0.091 & 0.072 & 0.312 & 0.042 & 0.032 & 0.117 & 0.005 & 0.005 & 0.018\\
100   & 0.073 & 0.072 & 0.242 & 0.044 & 0.043 & 0.136 & 0.008 & 0.009 & 0.023\\
200   & 0.107 & 0.090 & 0.326 & 0.056 & 0.048 & 0.141 & 0.010 & 0.009 & 0.034\\
500   & 0.096 & 0.096 & 0.307 & 0.045 & 0.045 & 0.161 & 0.011 & 0.011 & 0.035\\
1,000 & 0.089 & 0.089 & 0.297 & 0.043 & 0.042 & 0.145 & 0.009 & 0.010 & 0.033\\
\bottomrule
\end{tabular}
\vspace{0.5em}
\begin{minipage}{\textwidth}
\footnotesize
\emph{Notes}: The design has three products, consideration sets of size two, no covariates, zero mean utilities, and i.i.d.~$N(0,1)$ utility shocks. Entries are based on~$5{,}000$ DGP replications. OT$_U$ and OT$_Y$ use parameter-dependent critical values, one common fixed reference sample for each~$n$, and~$100{,}000$ common Monte Carlo latent draws. Their finite-Monte-Carlo p-values include ties in the upper tail and use the plus-one correction; the same draws are reused across DGP replications. AS2010 uses hard GMS with~$\kappa_n=\sqrt{\log n}$ and~$1{,}000$ i.i.d. nonparametric bootstrap draws for each dataset; its p-value includes bootstrap ties and has no plus-one correction.
\end{minipage}
\end{table}

The rejection frequencies of AS2010 display the predicted size distortion. At the~$5\%$ nominal level, they rise from~$0.117$ at~$n=50$ to values between~$0.136$ and~$0.161$ for~$n\geq100$, approximately three times the nominal level. At the~$10\%$ and~$1\%$ levels, the corresponding limits are~$0.2978$ and~$0.0300$, again approximately three times their nominal levels. In contrast, both OT procedures track the nominal levels without systematic over-rejection.

The OT procedures are conservative when the sample is very small. At~$n=10$, for example, OT$_U$ rejects with probability~$0.032$ at both the~$10\%$ and~$5\%$ nominal levels, and OT$_Y$ rejects with probability~$0.017$ at both levels. This occurs because, with discrete outcomes and a fixed reference sample, the statistic~$\tilde T_n(\theta_0)$ has finite support: it depends on finitely many admissible-set patterns and product-count vectors. Its cdf therefore has atoms and does not satisfy the continuity condition for equality in Theorem~\ref{thm:size}. A nonrandomized test that includes ties in the upper tail cannot split an atom at the critical value and consequently under-rejects. This discreteness is most pronounced at small~$n$; as~$n$ grows, the support becomes finer and rejection frequencies move closer to their nominal levels. 

    \clearpage
    \begin{figure}[h]
    \centering
    \includegraphics[width=0.49\linewidth,trim=0 455 603 43,clip]{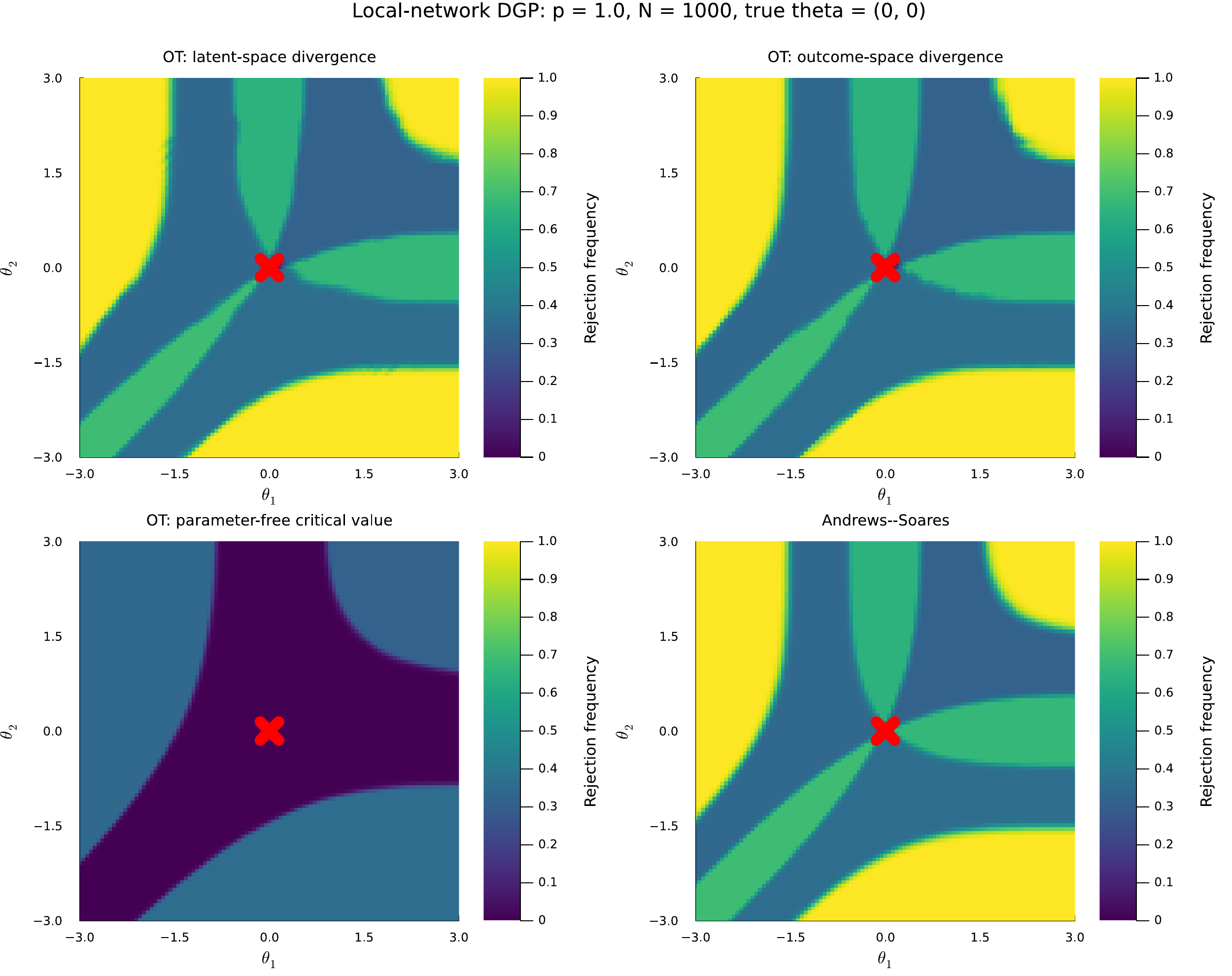}\hfill
    \includegraphics[width=0.49\linewidth,trim=597 455 0 43,clip]{CS_illustration_p1.pdf}

    \medskip

    \includegraphics[width=0.49\linewidth,trim=597 7 0 497,clip]{CS_illustration_p1.pdf}
    \caption{Test inversion under dependent ``Trending Now'' selection. The panels display OT$_U$, OT$_Y$, and AS2010, from left to right and then below. The sample size is~$n=1{,}000$, the true parameter is~$\theta_0=(0,0,0)$, and candidate values~$\theta=(\theta_1,\theta_2,0)$ are evaluated on a~$101\times101$ grid over~$[-3,3]^2$. Each cell reports the rejection frequency at the~$5\%$ nominal level over~$1{,}000$ DGP replications; equivalently, one minus the reported value is its pointwise inclusion frequency in the inverted confidence region. The red cross marks~$\theta_0$. OT$_U$ and OT$_Y$ use~$1{,}000$ Monte Carlo critical draws, while AS2010 uses~$1{,}000$ bootstrap draws and~$100{,}000$ latent draws to approximate its capacity probabilities. The same DGP samples are used across candidate values and procedures. 
    }
    \label{fig:cs-inversion}
    \end{figure}

Figure~\ref{fig:cs-inversion} complements the size results by comparing the three procedures under test inversion. We normalize~$\theta_3=0$ and evaluate null values~$\theta=(\theta_1,\theta_2,0)$ over~$[-3,3]^2$, while holding the DGP fixed at~$\theta_0=(0,0,0)$. Each panel reports the rejection frequency across repeated samples, so one minus the reported value is the frequency with which a given parameter value belongs to the corresponding inverted confidence region. The rejection profiles of OT$_U$ and OT$_Y$ are broadly similar to that of AS2010 over the displayed parameter range, even though AS2010 excludes the true value too often under the dependent selection rule. Thus, in this design, the finite-sample size control of the OT procedures does not come from near-universal nonrejection: their implied confidence regions are not significantly wider than that of the size-distorted AS2010. 
Taken together, this exercise shows that sample-wide dependent selection can invalidate moment-inequality inference calibrated under i.i.d. sampling, while the sample-level finite-sample guarantee of our procedure remains intact.

\section{Empirical illustration}
\label{sec:empirical}

We illustrate our testing procedure with an application to the data and results in \cite{CMT:2021}, described in example~\ref{ex:airline}. We replicated their model exactly and used our methodology on the same data.\footnote{We replicated table~5 page~3023 in \cite{CMT:2021} to check that our modeling and coding replicates their paper.} We first recall the details of the structural model in \cite{CMT:2021}. We then explain our empirical exercise and discuss results.

\subsection{Structural model description}
\label{sec:simulation-DGP}
The sample size~$n$ is the number of regional markets in which~$J$ firms, indexed by~$j$, potentially operate. In each regional market, indexed by~$i$, the firms simultaneously decide whether or not to enter, and the firms who enter compete in prices. 

\subsubsection{Exogenous variables}
Exogenous variables are the following. The size of market~$i$ is~$M_i$. The identity of potential entrants\footnote{An airline is considered a potential entrant if it is serving at least one market out of both of the endpoint airports.} in market~$i$ is collected in~$\mathcal G_i$.
Each firm~$j\in \mathcal{G}_i$ is associated with three types of exogenous covariates in each market~$i$: First, the vector of demand relevant firm characteristics is~$A_{ij}$ (in \cite{CMT:2021}, the notation is~$X$). Second, the vector of production cost relevant firm characteristics is~$W_{ij}$. Finally, $Z_{ij}$ is the vector of characteristics relevant to the fixed cost of entry. 

We call~$X_{i}:=(M_i,\mathcal G_i,A_{i},W_{i},Z_{i})$ the vector of all exogenous variables in market~$i$, with~$A_i:=(A_{ij})_{j}$, $W_i:=(W_{ij})_{j}$ and~$Z_i:=(Z_{ij})_{j}$.

\subsubsection{Latent variables}
We follow the model structure and parametric specifications in \cite{CMT:2021}. Firm~$j$ has fixed cost of entry~$\exp(\gamma Z_j+\nu_j+\nu_c)$, where~$\gamma$ is an unknown vector of parameters of interest, $\nu_c$ is a common entry-cost shock, and~$\nu_j$ is an idiosyncratic entry-cost shock. Both common and idiosyncratic shocks are normally distributed latent variables and independent of each other. Firm~$j$ has marginal unit cost of production~$\exp(\delta W_j+\eta_j)$, where~$\delta$ is an unknown vector of parameters of interest and~$\eta_j$ is a normally distributed latent variable. The market share is generated by a nested logit demand model.\footnote{The two nests in the nested logit structure are the outside option on the one hand, and the collection of all modeled options on the other hand.} Specifically, consumer~$l$'s indirect utility from choosing the outside option is~$\epsilon_{l0}$, whereas their utility from choosing firm $j$'s product is
$u_{lj} =  \beta A_j - \rho P_j + \xi_{j} + (1 - \lambda)\epsilon_{lj},$
where~$P_j$ is the product price, $\rho$ and~$\lambda$ are unknown parameters of interest, $\xi_j$ is a normally distributed latent variable, and~$\epsilon_{l0}$ and~$(\epsilon_{lj})_j$ are independent type~I extreme value preference shocks. 

The vector of latent variables~$U:=(\xi, \eta,\nu,\nu_c)$ is assumed to follow the normal distribution~$N(0,\Sigma)$ where~$\xi = (\xi_j)_j$, $\eta = (\eta_j)_j$, $\nu = (\nu_j)_j$ and
	\begin{equation*}
	\Sigma = 
	\begin{bmatrix}
		\sigma^2_{\xi}\cdot I & \sigma_{\xi \eta} \cdot I & \sigma_{\xi\nu}\cdot I & 0\\
		\sigma_{\xi\eta}\cdot I & \sigma^2_{\eta} \cdot I & \sigma_{\eta\nu}\cdot I & 0\\
		\sigma_{\xi\nu}\cdot I & \sigma_{\eta\nu} \cdot I & \sigma^2_{\nu}\cdot I & 0\\
        0&0&0& \sigma^2_{\nu_c}
	\end{bmatrix}.
	\end{equation*}

\subsubsection{Endogenous variables}
We now describe the support restriction on the joint distribution of~$(Y,X,U)$ derived from the structural entry and competition model in \cite{CMT:2021}. The endogenous variable~$D_j$ is equal to~$1$ if firm $j$ is a potential entrant and it chooses to enter the market, and~$0$ otherwise. Firms with $D_j = 1$ then choose price~$P_j$ and realize their market share~$S_j$. Market shares~$S:=(S_j)_{j}$ are determined by the entry profile~$D:=(D_j)_{j}$ and the price chosen by firms who entered. Given the nested logit specification, market shares are given by
\begin{equation}\label{eq:demand_parametric}
	S_j = 
    \begin{cases}
      \frac{\Lambda^{1-\lambda}}{1 + \Lambda^{1-\lambda}} \; S^\dagger_j, &\text{if }D_j = 1 \vspace{1em}\\
      0 &\text{if }D_j = 0
    \end{cases}
\end{equation}
where
\begin{equation*}
\Lambda := \sum_{j: D_j = 1} \Lambda_j, \hskip5pt \Lambda_j := \exp\{(\beta A_j - \rho P_j + \xi_{j}) / (1-\lambda)\} \mbox{ and }S^\dagger_j := \Lambda_j/\Lambda.
\end{equation*}
For firm $j$ with $D_j = 1$, its profit maximizing price for firm~$j$ satisfies the following first order condition:
\begin{eqnarray}
\label{eq:pricing_parametric}
P_j - \exp(\delta W_j + \eta_j) & = & \frac{1 - \lambda}{\rho \left[ 1 - \lambda S^\dagger_j  - (1-\lambda)S_j \right]}.
\end{eqnarray}
Equations \eqref{eq:demand_parametric} and \eqref{eq:pricing_parametric} are joint equations for prices and market shares of firms that enter the market. The solution of these equations, together with the normalization that $P_j = 0$ and $S_j = 0$ if $D_j = 0$, determines~$P$ and~$S$ as functions of~$(D,A,\xi)$. Finally, firm~$j$ enters the market if and only if it makes nonnegative profit, i.e.,
\begin{eqnarray}
(P_j - \exp(\delta W_j + \eta_j))MS_j - \exp(\gamma Z_j + \nu_j + \nu_c) & \geq & 0.
\label{eq:entry_parametric}
\end{eqnarray}
The entry profile~$D$ is determined as a pure strategy Nash equilibrium of the full information simultaneous entry game, with payoffs given by~(\ref{eq:entry_parametric}) for a firm who enters, and normalized to zero otherwise.

Finally, the parameter vector~$\theta$ consists of all $\beta$, $\gamma$, $\delta$, $\lambda$, $\rho$, and~$\Sigma$ introduced above.

\subsection{Data}

We use data from \cite{CMT:2021} and repeat the description here for the reader's convenience. The data are drawn from the second quarter of 2012’s Airline Origin and Destination Survey, the T-100 Domestic Segment Data Set’s
Aviation Support Tables, available from the Department of Transportation’s
National Transportation Library, and the US Census for demographic
data. The basic unit of observation is an airline in a market (a market carrier).
A market is a unidirectional trip between two airports, irrespective
of intermediate transfer points. The data set includes the markets
between the top 100 USMSAs ranked by their population. There are 8,163 unidirectional
markets. There are five carriers in the
data set: AA, DL, UA, US, WN, and an aggregate of low-cost carriers denoted LCC. The
LCCs include Alaska, JetBlue, Frontier, Allegiant, Spirit, Sun Country, and
Virgin. There are 22,445 market-carrier
observations for which prices and market shares are observed. An airline is considered a potential entrant if it is serving at least
one market out of both of the endpoint airports. 

Demand-relevant firm characteristics~$A_{ij}$ include origin presence, which is defined as the number of markets served by an airline out of the origin airport, the distance between the origin and destination airports, and a dummy for WN and LCC. Fixed-entry-cost relevant firm characteristics~$Z_{ij}$ include nonstop origin (the number of nonstop routes that
an airline serves out of the origin airport) and nonstop destination (the
number of nonstop routes that an airline serves out of the destination airport).
The marginal-cost relevant firm characteristics~$W_{ij}$ are distance between origin and destination, and a dummy for LCCs and Southwest.

\subsection{Implementation of the inference}
For each value of~$\theta$, we construct a low-discrepancy sequence\footnote{We use the generalized golden sequence from the \texttt{GoldenSequences.jl} Julia package.}~$(a_i)_{i=1}^n$ on~$[0,1]^{19}$, where the dimension~$19$ corresponds to the dimension of the latent vector~$U=(\xi,\eta,\nu,\nu_c)$. We transform each~$a_i$ componentwise by the standard normal quantile function and then apply the linear transformation induced by~$\Sigma$:
\[
\tilde u_i := \Sigma^{1/2}\Phi^{-1}(a_i),
\]
where~$\Phi^{-1}$ is applied componentwise and~$\Sigma^{1/2}$ is obtained by the Cholesky decomposition. The resulting sequence~$(\tilde u_i)_{i=1}^n$ approximates the distribution~$Q_{U\mid\theta}$ of the latent vector~$U$ under parameter value~$\theta$.

We use a discrepancy defined in the outcome space, as in~\eqref{eq:alternative discrepancy}. Because the dimension of~$(Y_i,X_i)$ depends on the set of potential entrants~$\mathcal G_i$, and because~$\mathcal G_i$ varies across markets, we restrict matching to markets with the same set of potential entrants. For~$(i, j)$ with $\mathcal G_i=\mathcal G_j$, the cost matrix is defined \footnote{To guard against near singularity of~$\tilde\Sigma_{\mathcal G_i}$, we use a ridge-regularized inverse instead of $\tilde\Sigma_{\mathcal G_i}^{-1}$. Specifically, if~$V\Lambda V^{\top}$ is the eigenvalue decomposition of~$\tilde\Sigma_{\mathcal G_i}$, we set~$\tilde\Sigma_{\mathcal G_i,\varepsilon}^{-1}:=V(\Lambda+\varepsilon I)^{-1}V^{\top}$. In the empirical implementation, we set $\varepsilon=10^{-5}$.}
\[
 C_{ij}(\theta) \coloneqq \inf_{(y,x)\in\Gamma_y(\tilde u_i,X_i;\theta)}
        \left(
        [(y-Y_j),(x-X_j)]
        \tilde\Sigma_{\mathcal G_i}^{-1}
        [(y-Y_j),(x-X_j)]^{\top}
        \right)
\]
For~$(i, j)$ with $\mathcal G_i\neq\mathcal G_j$, set~$C_{ij}(\theta)=+\infty$. Here,~$\tilde\Sigma_{\mathcal G_i}$ is the simulated covariance matrix of~$(Y_j,X_j)$ for markets with potential-entrant set~$\mathcal G_i$ computed using a uniform selection rule, as described in Section~\ref{sec:test}. Setting~$C_{ij}(\theta)=+\infty$ when~$\mathcal G_i\neq\mathcal G_j$ prevents the optimal-transport problem from matching observations that belong to different market configurations and therefore have different dimensions.

We compute the parameter-dependent critical value using~$1{,}000$ Monte Carlo latent samples. The numerical procedures used to compute the test statistic and the critical values are described in Appendix~\ref{sec:num}.

\subsection{Confidence set and results}
We use the following stochastic search algorithm to approximate the confidence set. The algorithm assumes that the confidence set is connected. Let~$\hat{\theta}^0$ be a parameter value that is not rejected by our procedure at the~$5\%$ significance level, and initialize $\mathcal C^0=\{\hat{\theta}^0\}$. For~$l=1,\ldots,L$, we perform the following steps:

\begin{enumerate}
    \item Draw~$1000$ random directions~$r_k$, $k=1,\ldots,1000$, with unit norm. Independently draw~$1000$ parameter values~$\theta_k$ from~$\mathcal C^{l-1}$ with replacement. This yields the pairs~$((r_k,\theta_k):k=1,\ldots,1000)$.

    \item For each~$k=1,\ldots,1000$, conduct a one-dimensional search starting from~$\theta_k$ in the direction~$r_k$. We first test~$\theta_k+r_k$ at the~$5\%$ significance level. If this value is not rejected, we test~$\theta_k+2r_k$ and continue doubling the step size until a rejected value is found. Once a rejected value is found, we use bisection between the last nonrejected value and the first rejected value to approximate the boundary of the confidence region in that direction. If~$\theta_k+r_k$ is rejected immediately, the bisection starts between~$\theta_k$ and~$\theta_k+r_k$.\footnote{The model imposes a negative price effect in the demand equation and requires~$\lambda\in[0,1]$. Whenever a proposed point~$\theta_k+\tau r_k$ violates these restrictions, we replace it by~$\theta_k+\bar\tau_k r_k$, where~$\bar\tau_k$ is the largest scale along the ray in direction~$r_k$ satisfying these restrictions.} For each pair~$(r_k,\theta_k)$, we conduct up to $10$ tests. The bisection may stop early if the distance between the upper and lower bounds on the scale parameter falls below~$10^{-15}$.

    \item Define~$\mathcal C^l$ as the union of~$\mathcal C^{l-1}$ and all nonrejected parameter values found in round~$l$.
\end{enumerate}

We run the algorithm for~$L=10$ rounds. Since each round uses~$1000$ directions and up to 10 tests per direction, the procedure could evaluate up to~$100{,}000$ parameter values. Because some searches terminate early, the actual number of evaluated parameter values is~$93{,}252$. On a cluster with~$496$ vCPUs, the total running time is~$105.33$ hours. We then project the final approximation~$\mathcal C^L$ coordinatewise and report the resulting intervals in Column~(1) of Table~\ref{tab:empirical_CS}. For comparison, Column~(2) reproduces the intervals reported in Column~(3) of Table~4 in \cite{CMT:2021}.

\begin{table}[htbp]
\centering
\caption{Coordinate projections of 95\% confidence regions}
\label{tab:empirical_CS}
\begin{tabularx}{\textwidth}{@{}X C{4cm} C{4cm}@{}}
\toprule
Parameter 
& \makecell[c]{Finite-sample OT\\(1)} 
& \makecell[c]{CMT\\(2)} \\
\midrule
\multicolumn{3}{@{}l}{\textit{A. Demand}} \\
Price (\$100)        & $[-2.581, -1.437]$ &  $[-1.557, -1.488]$\\
$\lambda$            & $[0.000, 0.505]$   & $[0.186, 0.206]$\\
Distance             & $[-0.365, 1.119]$  & $[0.724, 0.793]$\\
Origin presence      & $[0.365, 1.965]$   & $[1.688, 1.752]$\\
LCC                   & $[1.660, 3.192]$   & $[0.080, 0.273]$\\
WN                   & $[-2.209, -0.221]$ & $[-0.029, 0.128]$\\
Constant             & $[-4.486, -2.581]$ & $[-4.683, -4.587]$\\
\addlinespace
\multicolumn{3}{@{}l}{\textit{B. Marginal Cost}} \\
Distance             & $[-0.286, 0.386]$  & $[0.083, 0.094]$\\
LCC                  & $[-0.554, 0.777]$  & $[-0.027, 0.054]$\\
WN                   & $[-1.057, 0.198]$  & $[-0.079, -0.017]$\\
Constant             & $[4.948, 5.654]$   & $[5.132, 5.179]$\\
\addlinespace
\multicolumn{3}{@{}l}{\textit{C. Fixed Cost}} \\
Nonstop origin       & $[-1.420, 0.143]$  & $[-0.387, -0.327]$ \\
Nonstop destination  & $[-2.474, -0.881]$ & $[-1.538, -1.473]$\\
Constant             & $[-0.329, 1.178]$  & $[1.227, 1.315]$\\
\addlinespace
\multicolumn{3}{@{}l}{\textit{D. Variance-Covariance}} \\
Variance demand                         & $[3.867, 41.209]$   & $[1.736, 1.876]$ \\
Variance marginal cost                  & $[0.224, 2.725]$    & $[0.330, 0.353]$\\
Variance fixed cost                     & $[12.683, 172.861]$ & $[14.640, 15.636]$\\
Demand-marginal cost covariance         & $[0.584, 10.103]$   & $[0.470, 0.512]$\\
Demand-fixed cost covariance            & $[-2.288, 4.277]$   & $[0.674, 0.829]$\\
Marginal cost-fixed cost covariance     & $[-0.526, 1.720]$   & $[-0.709, -0.659]$\\
\bottomrule
\end{tabularx}
\vspace{0.5em}
\begin{minipage}{\textwidth}
\footnotesize
\emph{Notes}: Column~(1) reports the coordinate projections of the nonempty $95\%$ confidence region obtained by our finite-sample optimal-transport procedure. Column~(2) reproduces the coordinate projections reported in column~3 of table~4 of \cite{CMT:2021}.
\end{minipage}
\end{table}

Table~\ref{tab:empirical_CS} shows that our finite-sample-valid confidence region leads to projections that differ from those reported in \cite{CMT:2021}. In some dimensions, such as the price coefficient in the demand equation, the projected intervals overlap and deliver similar results. In other dimensions, including the fixed effects of low-cost carriers and Southwest in demand, the projected intervals do not overlap. \cite{CMT:2021} obtain their intervals by applying the methodology of \cite{CHT:2007}. However, they recenter the criterion function around the minimum value they find and interpret the resulting region as a confidence region for a pseudo-true value. See their Online Appendix~B for further details. 
The discrepancies in Table~\ref{tab:empirical_CS} may reflect small-sample distortions in the asymptotic procedure of~\cite{CMT:2021}, its lack of robustness to nonstationarity and strong dependence arising from dependent equilibrium selection, or the possibility that the reported parameter vector does not globally minimize their criterion function.
We do not attempt to distinguish between these explanations. The exercise nevertheless shows that finite-sample-valid inference can be implemented in a highly complex empirical structural model and can lead to materially different conclusions in this application.

\section*{Discussion}

We have proposed a procedure to compute confidence regions in incomplete models with exact coverage in finite samples. Compared to existing approaches, our procedure has many advantages, some straightforward and others more subtle. First, finite-sample validity avoids reliance on asymptotic approximations, which are often suspect. It also removes the need for user-chosen tuning parameters, that inference results are often very sensitive to. Second, our procedure removes the need for transforming conditional into unconditional moment inequalities, and for reducing the very large number of moment inequalities with complex and model-specific core determining classes. Third, finite-sample validity allows us to conduct inference in models, where the specification depends on the sample size. This includes possible future applications to games on networks and network formation games, when a single network is observed. In such cases, the support constraint in the model specification depends on the sample size, and so does the dimension of the latent variable, which involves an individual's neighbors in the network.
Finally, although we haven't developed it here, our method extends to specifications where the structural support constraint is individual-specific, thereby allowing us to conduct inference with the structural vector autoregressions proposed in \cite{GK:2021} and \cite{GKR:2021}.
This paper has contributed to a growing literature that shows how optimal-transport theory provides a rich set of tools in econometrics in general, and incomplete models in particular. We expect these tools to underlay an extension to semiparametric  incomplete models with independence constraints.

\begin{appendix}

\section{Proofs of results in the main text}
\label{sec:proofs}

\begin{proof}[Proof of Proposition~\ref{prop:char1}]  

Under assumption~\ref{ass:discrepancy}, the minimum in~(\ref{eq:char1}) is attained by theorem~1.3 of \cite{Villani:2003}.
First show that~$P_0\in\mathcal P_\theta$ implies~$\mathcal D(Q_U\times P_{X,0},P_0;\theta)=0$. 
If~$P_0\in\mathcal P_\theta$, then there exists a joint probability~$\tilde \pi$ over $\mathcal Y\times\mathcal X\times\mathcal U$ with marginals~$P_{0}$ and~$Q_U$, such that $\mathbb E_{\tilde \pi}1\{(Y,X,U) \notin \Gamma(\theta)\}=0$ and~$U\independent X$. The latter implies the existence of a probability~$\pi$ on $\mathcal U\times \mathcal X \times \mathcal Y\times \mathcal X$, which is in $\mathcal M(Q_U\times P_{X,0},P_0)$, with~$\pi(X=X^\prime)=1$, and such that $\pi((U,X) \notin \Gamma_u(Y,X^\prime;\theta))=0$. This implies~$\mathbb E_\pi \delta\left( (U,X),(Y,X^\prime);\theta\right)=0$. Hence,~$\mathcal D(Q_U\times P_{X,0},P_0;\theta)=0$.
Now show the converse. If~$\mathcal D(Q_U\times P_{X,0},P_0;\theta)=0$, then, since the minimum is attained,
there exists a random vector~$(U,X,Y,X^\prime)$ such that~$(U,X)$ has distribution~$Q_U\times P_{X,0}$, $(Y,X^\prime)$ has distribution~$P_0$, such that
$\delta((U,X),(Y,X^\prime);\theta)=0$ almost surely. This implies
$X^\prime=X$ and~$(Y,X,U) \in \Gamma(\theta)$ almost surely, and hence~$P_0\in\mathcal P_\theta$.
\end{proof}

\begin{proof}[Proof of Theorem~\ref{thm:size}]

We fix an arbitrary~$\theta$ such that~$\mathcal P_\theta^{(n)}\ne\varnothing$ and some~$\alpha\in(0,1)$.
  
Take an arbitrary distribution $P^{(n)}\in\mathcal P_\theta^{(n)}$. Let~$(Y^{(n)}, X^{(n)})$ be a random vector distributed according to~$P^{(n)}$. 
Let~$T_n(\theta)$ be the resulting test statistic. 
By the definition of $\mathcal{P}_\theta^{(n)}$, there exists a random vector $U^{(n)}$ such that $(Y_i, X_i, U_i)\in \Gamma(\theta)$ and $U_i \sim Q_U$ almost surely for each $i$. Because $(Y_i, X_i) \in \Gamma_y(U_i, X_i;\theta)$, we know that~$Y^{(n)}$ belongs to the set~$\Gamma_y^{(n)}(U^{(n)},X^{(n)};\theta)$ defined in~(\ref{eq:Gamma(n)}). Therefore,
  \begin{eqnarray}
  \label{eq:size-proof}
    T_n(\theta) = \mathcal D_n(C(\theta)) \leq \sup_{\tilde y^{(n)}\in\Gamma_y^{(n)}(U^{(n)},X^{(n)};\theta)}\mathcal D_n(C(\tilde y^{(n)};\theta)).
  \end{eqnarray}
By Definition~\ref{def:MC}, $(U^{(n)},X^{(n)})$ and~$(\tilde U^{\prime(n)},X^{(n)})$ are identically distributed. Hence, the right-hand side of~(\ref{eq:size-proof}) and~$\tilde T_n(\theta)$ are identically distributed. Therefore,~(\ref{eq:size}) follows from~(\ref{eq:size-proof}).
 
\subsubsection*{Proof that~(\ref{eq:size}) holds as an equality}
Fix~$\epsilon>0$. We show below that for any~$\beta\in(0,1)$, there exists some $P^{(n)}$ in~$\mathcal P_\theta^{(n)}$ such that
  \begin{eqnarray}
  \label{eq:almost_exact}
  P^{(n)}\left(T_n(\theta)\leq c_{n,1-\beta}(\theta) - \epsilon \; \right) \leq 1-\beta.
  \end{eqnarray}
Suppose the cdf of $\tilde T_n(\theta)$ is continuous and increasing in a neighborhood of $c_{n,1-\alpha}(\theta)$.  For any small enough $\zeta > 0$, $c_{n,1-\alpha + \zeta}(\theta) - c_{n, 1-\alpha} (\theta)> 0$. Let $\epsilon = c_{n,1-\alpha + \zeta}(\theta) - c_{n, 1-\alpha}(\theta)$. Then, \eqref{eq:almost_exact} applied to~$\beta=\alpha - \zeta$ implies that there exists some $P^{(n)}$ in~$\mathcal P_\theta^{(n)}$ such that
\begin{eqnarray*}
P^{(n)} \left(T_n(\theta)\leq c_{n,1-\alpha}(\theta) \; \right)
& = & P^{(n)} \left(T_n(\theta)\leq c_{n,1-\alpha + \zeta}(\theta) - \epsilon  \; \right) \\ & \le & 1-\alpha + \zeta.
\end{eqnarray*}
The above inequality holds for arbitrary small $\zeta > 0$, and the result follows.

\subsubsection*{Proof of~(\ref{eq:almost_exact})}
By assumption, $\mathcal P_\theta^{(n)}$ is nonempty under the null hypothesis. Hence, there exists a marginal distribution~$P_{X,n}$ such that~$\Gamma_y(U,X;\theta)$ is almost surely non-empty if~$X\sim P_{X,n}$ and~$U\sim Q_U$. Let~$(U^{(n)},X^{(n)})$ be a vector of~$n$ i.i.d. draws from~$P_{X,n} \times Q_U$. Write~$U^{(n)}:=(U_1,\ldots,U_n)$ and~$X^{(n)}=(X_1,\ldots,X_n)$.
We will construct a map~$\varphi: \mathcal U^n\times\mathcal X^n \rightarrow \mathcal Y^n$ such that the distribution~$P^{(n)}$ of~$(\varphi(U^{(n)},X^{(n)}),X^{(n)},U^{(n)})$ is in~$\mathcal P_\theta^{(n)}$, and satisfies~(\ref{eq:almost_exact}). Note that restriction~(3) in the definition of~$\mathcal P_\theta^{(n)}$ (Definition~\ref{def:DGP}) is satisfied by the construction of~$(U^{(n)},X^{(n)})$.

In addition, the map~$\varphi$ we construct must satisfy the following.

\vskip10pt
\noindent (i)~It must be measurable. To show this, we will rely on a classical theorem on the existence of measurable selections of correspondences, namely Proposition~7.50 
page~184 of \cite{BS:96}. 

\vskip10pt
\noindent (ii)~It must be a selection from the correspondence~$\Gamma^{(n)}(U^{(n)},X^{(n)};\theta)$ defined in~(\ref{eq:Gamma(n)}),
so support restriction~(2) in the definition of~$\mathcal P_\theta$ (Definition~\ref{def:DGP}) is satisfied. This will be imposed in the construction.

\vskip10pt
\noindent (iii)~The distribution~$P^{(n)}$ of~$(\varphi(U^{(n)},X^{(n)}),X^{(n)},U^{(n)})$ must satisfy~(\ref{eq:almost_exact}).  
By definition of~$T_n(\theta)$ and~$\tilde T_n(\theta)$, (\ref{eq:almost_exact}) is satisfied if~$Y^{(n)}:=\varphi(U^{(n)},X^{(n)})$ satisfies
\begin{eqnarray}\label{eq:Phi}
    \mathcal D_n(C(Y^{(n)};\theta)) \geq \sup_{\tilde y^{(n)}\in\Gamma_y^{(n)}(U^{(n)},X^{(n)};\theta)}\mathcal D_n(C(\tilde y^{(n)};\theta)) \; - \; \epsilon .
\end{eqnarray}
In the display above, as defined in~(\ref{eq:cost}), $C(\tilde y^{(n)};\theta)$ is the cost matrix with~$(i,j)$th component~$\delta((\tilde u_i, X_i), (\tilde y_j, X_j);\theta)$.

\vskip20pt

Define the correspondence~$\Phi:\mathcal X^n\times\mathcal U^n\rightrightarrows\mathcal Y^n$ by
  \begin{eqnarray*}
\Phi\left(U^{(n)},X^{(n)}\right) & := & \left\{ y^{(n)} \in \Gamma_y^{(n)}(U^{(n)},X^{(n)};\theta) : \mbox{ (\ref{eq:Phi}) holds} \right\}.
  \end{eqnarray*}
We fulfill requirements~(i), (ii), and~(iii) by showing that~$\Phi$ admits a measurable selection~$\varphi$. 
This follows directly from Proposition~7.50 page~184 of \cite{BS:96}: The correspondence~$\Phi$ admits a universally measurable selection~$\varphi$ on~$\mathcal X^n\times\mathcal U^n$. 
We have therefore proved that the distribution~$P^{(n)}$ of~$(Y^{(n)},X^{(n)})$ is in~$\mathcal P_\theta^{(n)}$, and satisfies~(\ref{eq:almost_exact}) as desired.
\end{proof}

\begin{proof}[Proof of Theorem~\ref{thm:new_consistency}]
For~$p\in\mathcal P$, write~$p_X$ for its~$\mathcal X$-marginal and define
\begin{equation}
D_\theta(p) \coloneqq \mathcal D(Q_U\times p_X,p;\theta).
\end{equation}
Propositions~\ref{prop:char1} and~\ref{prop:iid} imply that
\begin{eqnarray}
\label{eq:criterion-zero-set-proof}
D_\theta(p)=0 & \Leftrightarrow & p\in\mathcal P_\theta.
\end{eqnarray}

We first establish three consequences of the assumptions. First, the map~$D_\theta$ is continuous for weak convergence. Indeed, if~$p_m$ converges weakly to~$p$, then~$(p_m)_X$ converges weakly to~$p_X$, and hence~$Q_U\times(p_m)_X$ converges weakly to~$Q_U\times p_X$. Assumption~\ref{ass:reg} and Theorem~5.20 in~\cite{Villani:2009} then give~$D_\theta(p_m)\to D_\theta(p)$. Together with~(\ref{eq:criterion-zero-set-proof}), this implies that for every weakly compact~$\mathcal K\subseteq\mathcal P$ and every~$\varepsilon>0$ such that the constraint set is nonempty,
\begin{equation}
\label{eq:compact-modulus}
\kappa_{\theta,\mathcal K}(\varepsilon)\coloneqq \inf_{\substack{p\in\mathcal K:\\
\operatorname{dist}_{\mathcal P}(p,\mathcal P_\theta)\geq\varepsilon}}
D_\theta(p)
\;>\;0.
\end{equation}

Second, we want to show 
\begin{equation}
\label{eq:criterion-approximation}
T_n(\theta)-D_\theta(\hat P_n)  \to  0 \qquad\text{in~$P^{(n)}$-probability}.
\end{equation}

To see why \eqref{eq:criterion-approximation} is true, fix an arbitrary weakly compact set~$\mathcal K\subseteq\mathcal P$. For every~$a>0$, there exists~$b>0$ such that the following statement is true:
\begin{eqnarray}
\label{eq:uniform-criterion-approximation}
p\in\mathcal K,\quad
\rho_{\mathcal Q}(q,Q_U\times p_X)<b
\quad \text{ would imply }\quad
\left|\mathcal D(q,p;\theta)-D_\theta(p)\right|<a.
\end{eqnarray}
If this claim failed, there would be some~$a>0$ and sequences~$p_m\in\mathcal K$ and~$q_m\in\mathcal Q$ for which the distance on the left converges to zero while the absolute difference on the right is at least~$a$. Compactness gives a subsequence~$p_m\Rightarrow p\in\mathcal K$. Along that subsequence, both~$q_m$ and~$Q_U\times(p_m)_X$ converge weakly to~$Q_U\times p_X$, contradicting Theorem~5.20 in~\cite{Villani:2009}. 

Fix an arbitrary $\eta > 0$. Let $\mathcal{K}_\eta$ be the compact set supplied by \eqref{eq:law-tightness}. For any $a > 0$, let $b$ be value in \eqref{eq:uniform-criterion-approximation} with $\mathcal{K} = \mathcal{K}_\eta$. Assumption \ref{ass:joint convergence} and the construction of $\mathcal{K}_\eta$ implies 
\begin{equation*}
  \liminf_{n\to \infty} P^{(n)}\left\{ \hat{P}_n \in \mathcal{K}_\eta\text{ and } \rho_{\mathcal{Q}}(\hat{P}_{UX,n}, Q_U\times \hat{P}_{X,n}) < b \right\} \ge 1 - \eta.
\end{equation*}
Statement \eqref{eq:uniform-criterion-approximation}  then implies that 
\begin{equation}\label{eq:uniform_approx_D}
\liminf_{n\to \infty} P^{(n)}\left\{ |\mathcal D(\hat P_{UX,n},\hat P_n;\theta) - D_\theta(\hat{P}_n)| < a \right\} \ge 1-\eta.
\end{equation}
Because~$T_n(\theta)=\mathcal D(\hat P_{UX,n},\hat P_n;\theta)$, and because \eqref{eq:uniform_approx_D} holds for arbitrarily small $a$ and $\eta$, we have established \eqref{eq:criterion-approximation}.

Third, Lemma~\ref{lem:critical-value-collapse} stated immediately after this proof gives~(\ref{eq:critical-collapse}).

We now prove the result. Fix~$\eta>0$. By~(\ref{eq:law-separation}), choose~$\varepsilon>0$ such that
\begin{eqnarray}
\label{eq:near-core-bound}
\limsup_{n\to\infty}
P^{(n)}
\left\{
\operatorname{dist}_{\mathcal P}(\hat P_n,\mathcal P_\theta)
\leq\varepsilon
\right\}
& < & \eta.
\end{eqnarray}
Choose~$\mathcal K_\eta$ according to~(\ref{eq:law-tightness}). 

If
$\left\{ p\in\mathcal K_\eta: \operatorname{dist}_{\mathcal P}(p,\mathcal P_\theta) \geq\varepsilon \right\}$
is empty, then~(\ref{eq:near-core-bound}) and~(\ref{eq:law-tightness}) immediately imply that the upper limit of the nonrejection probability is at most~$2\eta$, because 
\begin{eqnarray*}
  &&\limsup_{n\to\infty} P^{(n)} \left\{ T_n(\theta)\leq c_{n,1-\alpha}^0 \right\} \\
  &\le&\limsup_{n\to\infty} P^{(n)} \left\{ T_n(\theta)\leq c_{n,1-\alpha}^0 \text{ and }\operatorname{dist}_{\mathcal P}(\hat{P}_n,\mathcal P_\theta) < \varepsilon \right\} +\limsup_{n\to\infty} P^{(n)}\left\{ \operatorname{dist}_{\mathcal P}(\hat{P}_n,\mathcal P_\theta) \ge \varepsilon \right\} \\
  &\le&\limsup_{n\to\infty} P^{(n)} \left\{ \operatorname{dist}_{\mathcal P}(\hat{P}_n,\mathcal P_\theta) < \varepsilon \right\} +\limsup_{n\to\infty} P^{(n)}\left\{ \hat{P}_n\notin \mathcal{K}_\eta \right\} \\
  &\le& 2\eta.
\end{eqnarray*}

Now, suppose $\left\{ p\in\mathcal K_\eta: \operatorname{dist}_{\mathcal P}(p,\mathcal P_\theta) \geq\varepsilon \right\}$ is nonempty. Let~$\kappa:=\kappa_{\theta,\mathcal K_\eta}(\varepsilon)>0$ be defined by~(\ref{eq:compact-modulus}). On the event
\begin{eqnarray*}
\left\{
\hat P_n\in\mathcal K_\eta,\quad
\operatorname{dist}_{\mathcal P}(\hat P_n,\mathcal P_\theta)\ge\varepsilon,\quad
|T_n(\theta)-D_\theta(\hat P_n)|<\kappa/3,\quad
c_{n,1-\alpha}^0<\kappa/3
\right\},
\end{eqnarray*}
we have $D_\theta(\hat{P}_n) \ge \kappa > 0$, $T_n(\theta) > D_\theta(\hat{P}_n) - \kappa/3$ and $c^0_{n,1-\alpha} < \kappa/3$, so that $T_n(\theta)>c_{n,1-\alpha}^0$. Therefore,
\begin{eqnarray*}
&&P^{(n)}
\left\{
T_n(\theta)\leq c_{n,1-\alpha}^0
\right\}\\
& \leq &
P^{(n)}
\left\{
\operatorname{dist}_{\mathcal P}(\hat P_n,\mathcal P_\theta)
\leq\varepsilon
\right\}
+
P^{(n)}\{\hat P_n\notin\mathcal K_\eta\}\\
&&+
P^{(n)}
\left\{
|T_n(\theta)-D_\theta(\hat P_n)|
\geq\kappa/3
\right\}
+
P^{(n)}
\left\{
c_{n,1-\alpha}^0\geq\kappa/3
\right\}.
\end{eqnarray*}
Equations~(\ref{eq:criterion-approximation}) and~(\ref{eq:critical-collapse}) make the last two terms converge to zero. Equations~(\ref{eq:near-core-bound}) and~(\ref{eq:law-tightness}) then imply
\begin{eqnarray*}
\limsup_{n\to\infty}
P^{(n)}
\left\{
T_n(\theta)\leq c_{n,1-\alpha}^0
\right\}
& \leq & 2\eta.
\end{eqnarray*}
Since~$\eta>0$ is arbitrary,~(\ref{eq:new-consistency}) follows. 
\end{proof}

\begin{lemma}[Vanishing parameter-free critical value]
\label{lem:critical-value-collapse}
Fix~$\alpha\in(0,1)$. Let~$\mathcal F_n^0$ be the sigma-field generated by~$X^{(n)}$ and by a possibly random $\tilde{u}^{(n)}$. Thus,~$\mathcal F_n^0=\sigma(X^{(n)})$ when~$\tilde u^{(n)}$ is deterministic, whereas~$\mathcal F_n^0=\sigma(X^{(n)},\tilde u^{(n)})$ when $\tilde u^{(n)}$ is drawn randomly independently. Suppose Assumption~\ref{ass:discrepancy0}, Assumption~\ref{ass:triangular_array}(ii), and Assumption~\ref{ass:joint convergence} hold, with a fixed discrepancy~$\delta^0$. Recall that $c_{n,1-\alpha}^0$ is the conditional~$1-\alpha$ quantile of~$\tilde T_n^0$ given~$\mathcal F_n^0$. Then,
\begin{equation}
\label{eq:critical-collapse}
c_{n,1-\alpha}^0
\longrightarrow
0
\qquad\mbox{in~$P^{(n)}$-probability}.
\end{equation}
\end{lemma}

\begin{proof}
We use~$P^{(n)}$ according to the convention in Assumption~\ref{ass:joint convergence}: if the primary sequence~$\tilde u^{(n)}$ is random, its randomness is already included in~$P^{(n)}$. Only for this proof, enlarge the corresponding probability space by an independent reference sample
\[
\tilde U^{\prime(n)}
=
(\tilde U'_1,\ldots,\tilde U'_n)
\sim
Q_U^{\otimes n},
\]
and denote the resulting product extension by
\[
\overline P^{(n)}
\coloneqq
P^{(n)}\otimes Q_U^{\otimes n}.
\]
Thus, under~$\overline P^{(n)}$, the conditional distribution of~$\tilde U^{\prime(n)}$ given all the variables on the original probability space is~$Q_U^{\otimes n}$. The measures~$P^{(n)}$ and~$\overline P^{(n)}$ induce the same law for~$X^{(n)}$ and for every other random element on the original probability space.

To make explicit the meaning of the theoretical critical value, note that
\begin{equation}
\label{eq:conditional-critical-value}
c_{n,1-\alpha}^0
=
\inf\left\{
t\in\mathbb Q:
\overline P^{(n)}
\left(
\tilde T_n^0\leq t
\mid\mathcal F_n^0
\right)
\geq
1-\alpha
\right\}.
\end{equation}
A regular conditional distribution exists because the underlying state spaces are Polish. Taking the infimum over~$\mathbb Q$ makes~$c_{n,1-\alpha}^0$ explicitly~$\mathcal F_n^0$-measurable. In particular, it is a function of~$X^{(n)}$ when~$\tilde u^{(n)}$ is deterministic, and a function of~$(X^{(n)},\tilde u^{(n)})$ when the realized random primary simulator is held fixed. In either case, it is measurable on the original probability space, so its convergence can be stated under~$P^{(n)}$ alone. The additional law~$\overline P^{(n)}$ is needed only to carry the independent reference sample in the proof.

Write
\[
q_n \coloneqq \hat P_{UX,n}, \qquad
q'_n(\cdot) \coloneqq \frac{1}{n}\sum_{i=1}^n1\{(\tilde U'_i,X_i)\in\cdot\}, \qquad
r_n \coloneqq Q_U\times\hat P_{X,n}.
\]
Assumption~\ref{ass:joint convergence} gives
\begin{equation}
\label{eq:primary-simulator-approximation-proof}
\rho_{\mathcal Q}(q_n,r_n)
\longrightarrow
0
\qquad\mbox{in~$P^{(n)}$-probability}.
\end{equation}
The same convergence holds under~$\overline P^{(n)}$, since the event in~(\ref{eq:primary-simulator-approximation-proof}) does not depend on the additional reference sample.

We next show that
\begin{equation}
\label{eq:reference-simulator-approximation}
\rho_{\mathcal Q}(q'_n,r_n)
\longrightarrow
0
\qquad\mbox{in~$\overline P^{(n)}$-probability}.
\end{equation}
For every bounded continuous function~$f$ on~$\mathcal U\times\mathcal X$,
\[
\int f\,dq'_n-\int f\,dr_n
=
\frac{1}{n}\sum_{i=1}^n
\left\{
f(\tilde U'_i,X_i)
-
\int f(u,X_i)\,Q_U(du)
\right\}.
\]
Conditionally on the original probability space, the summands on the right-hand side are independent, have conditional mean zero, and are bounded in absolute value by~$2\|f\|_\infty$. Their conditional variance is therefore bounded by~$4\|f\|_\infty^2/n$. Conditional Chebyshev's inequality gives convergence to zero in~$\overline P^{(n)}$-probability for every fixed~$f$ in a countable convergence-determining class.

For completeness, we verify the tightness needed to pass from this convergence-determining class to the metric~$\rho_{\mathcal Q}$. Fix~$\eta>0$, let~$\mathcal K_\eta$ be supplied by Assumption~\ref{ass:triangular_array}(ii), and define
\[
\mathcal H_\eta
\coloneqq
\left\{
Q_U\times p_X:p\in\mathcal K_\eta
\right\}.
\]
The map~$p\mapsto Q_U\times p_X$ is weakly continuous, so~$\mathcal H_\eta$ is weakly compact, and
\[
\liminf_{n\to\infty}
\overline P^{(n)}
\{r_n\in\mathcal H_\eta\}
\geq
1-\eta.
\]
By Prokhorov's theorem, for every~$m\geq1$ there is a compact set~$L_m\subseteq\mathcal U\times\mathcal X$ such that
\[
\sup_{r\in\mathcal H_\eta}r(L_m^c)
\leq
\eta\,4^{-m}.
\]
Moreover, for every Borel set~$B\subseteq\mathcal U\times\mathcal X$,
\[
\mathbb E_{\overline P^{(n)}}
\left[
q'_n(B)\mid X^{(n)}
\right]
=
r_n(B).
\]
Consequently, on~$\{r_n\in\mathcal H_\eta\}$, Markov's inequality conditional on~$X^{(n)}$ gives
\[
\overline P^{(n)}
\left\{
q'_n(L_m^c)>2^{-m}
\mid X^{(n)}
\right\}
\leq
\eta\,2^{-m}.
\]
A union bound shows that, apart from an event with asymptotic probability at most~$2\eta$,~$r_n$ belongs to~$\mathcal H_\eta$ and~$q'_n$ belongs to
\[
\mathcal H'_\eta
\coloneqq
\left\{
q\in\mathcal Q:
q(L_m^c)\leq2^{-m}
\text{ for every }m\geq1
\right\}.
\]
The set~$\mathcal H'_\eta$ is uniformly tight and weakly closed, hence weakly compact. Thus, the laws of~$(q'_n,r_n)$ are asymptotically tight. If~(\ref{eq:reference-simulator-approximation}) failed, there would be a subsequence along which~$\rho_{\mathcal Q}(q'_n,r_n)$ remained bounded away from zero with positive probability. Asymptotic tightness would yield a further subsequence whose laws converge weakly on~$\mathcal Q\times\mathcal Q$. The convergence already proved for the countable determining class forces the limiting law to be concentrated on pairs~$(q',r)$ satisfying~$\int f\,dq'=\int f\,dr$ for every function~$f$ in that class, and hence on the diagonal~$\{(q,r):q=r\}$. Since~$\rho_{\mathcal Q}$ is continuous and vanishes on the diagonal, this is a contradiction. This proves~(\ref{eq:reference-simulator-approximation}).

Let~$\mathcal D^0$ denote the optimal-transport value generated by~$\delta^0$. We claim that, for every weakly compact~$\mathcal H\subseteq\mathcal Q$ and every~$a>0$, there exists~$b>0$ such that
\begin{equation}
\label{eq:uniform-reference-collapse}
r\in\mathcal H,\quad
\rho_{\mathcal Q}(q,r)<b,\quad
\rho_{\mathcal Q}(q',r)<b
\quad\Longrightarrow\quad
\mathcal D^0(q,q')<a.
\end{equation}
If the claim failed, there would exist sequences~$r_m\in\mathcal H$,~$q_m$, and~$q'_m$ such that both distances converge to zero while~$\mathcal D^0(q_m,q'_m)\geq a$. Along a subsequence,~$r_m$ converges weakly to some~$r\in\mathcal H$, and then both~$q_m$ and~$q'_m$ converge weakly to~$r$. Because~$\delta^0$ is bounded and continuous, Theorem~5.20 in~\cite{Villani:2009} implies
\[
\mathcal D^0(q_m,q'_m)
\longrightarrow
\mathcal D^0(r,r)
=
0,
\]
where the last equality follows from the identity coupling and~$\delta^0(v,v)=0$. This contradiction proves~(\ref{eq:uniform-reference-collapse}).

By construction,
\[
\tilde T_n^0
=
\mathcal D^0(q_n,q'_n).
\]
Apply~(\ref{eq:uniform-reference-collapse}) with~$\mathcal H=\mathcal H_\eta$, and combine~(\ref{eq:primary-simulator-approximation-proof}),~(\ref{eq:reference-simulator-approximation}), and the compact containment of~$r_n$. Since~$\eta>0$ is arbitrary,
\begin{equation}
\label{eq:reference-collapse}
\tilde T_n^0
\longrightarrow
0
\qquad\mbox{in~$\overline P^{(n)}$-probability}.
\end{equation}

It remains to pass from the reference statistic $\tilde{T}^0_n$ to its conditional quantile $c^{0}_{n,1-\alpha}$ conditional on $\mathcal{F}^0_n$. Since~$\delta^0$ is nonnegative, both~$\tilde T_n^0$ and~$c_{n,1-\alpha}^0$ are nonnegative. Fix~$t>0$. On the event~$\{c_{n,1-\alpha}^0>t\}$, choose a rational number~$s$ such that~$t<s<c_{n,1-\alpha}^0$. Equation (\ref{eq:conditional-critical-value}) then gives~$\overline P^{(n)}(\tilde T_n^0\leq s\mid\mathcal F_n^0)<1-\alpha$, and hence
\[
\{c_{n,1-\alpha}^0>t\}
\subseteq
\left\{
\overline P^{(n)}
\left(
\tilde T_n^0>t
\mid\mathcal F_n^0
\right)
>
\alpha
\right\}.
\]
Markov's inequality and the tower property therefore give
\begin{eqnarray*}
P^{(n)}
\left\{
c_{n,1-\alpha}^0>t
\right\}
& = &
\overline P^{(n)}
\left\{
c_{n,1-\alpha}^0>t
\right\}\\
& \leq &
\frac{1}{\alpha}
\overline P^{(n)}
\left\{
\tilde T_n^0>t
\right\}
\longrightarrow
0,
\end{eqnarray*}
where the convergence follows from~(\ref{eq:reference-collapse}). This proves~(\ref{eq:critical-collapse}).
\end{proof}

\begin{proof}[Proof of Theorem~\ref{thm:optimality}]

The proof relies on the representation theorem of de Finetti in the form given by \cite{HS:1955}: for every~$P^{(\infty)}\in\mathcal E^{(\infty)}$, there exists a unique Borel probability measure~$\Lambda$ on~$\mathcal P$, 
such that, for every~$n\geq1$ and every Borel set~$A\subseteq\mathcal Y^n\times\mathcal X^n$,
\begin{eqnarray}
\label{eq:deFinetti}
P^{(n)}(A) & = & \int_{\mathcal P} P^{\otimes n}(A)\;\Lambda(dP),
\end{eqnarray}
where~$P^{(n)}$ denotes the first-$n$ marginal of~$P^{(\infty)}$.

\subsubsection*{Step 1: properties of~$\phi^{OT}$}
We first record that~$\phi^{OT}$ satisfies~(\ref{eq:ex-size}) and
\begin{eqnarray}
\label{eq:iid-consistency}
\forall P\in\mathcal P\setminus\mathcal P_\theta,\quad\mathbb E_{P^{\otimes n}}\left[\phi_n^{OT}\right] & \rightarrow & 1 \quad\mbox{ as } n\rightarrow\infty.
\end{eqnarray}
For the size property, fix~$n\geq1$ and~$P^{(n)}\in\mathcal P_\theta^{(n)}$. By definition~(\ref{eq:OTtest}), $\mathbb E_{P^{(n)}}[\phi_n^{OT}]$ is the probability of the event~$\{T_n(\theta)>c_{n,1-\alpha}^0\}$, where the probability is taken over both the data and the simulation randomness. By theorem~\ref{thm:size}, the probability of~$\{T_n(\theta)>c_{n,1-\alpha}(\theta)\}$ is at most~$\alpha$, and, under assumption~\ref{ass:discrepancy0}, (\ref{eq:conserv}) yields~$c_{n,1-\alpha}(\theta)\leq c_{n,1-\alpha}^0$. Hence~$\mathbb E_{P^{(n)}}[\phi_n^{OT}]\leq\alpha$, and~(\ref{eq:ex-size}) holds a fortiori.

For~(\ref{eq:iid-consistency}), fix~$P\in\mathcal P\setminus\mathcal P_\theta$ and let the data be i.i.d.\ from~$P$. The empirical distribution~$\hat P_n$ converges weakly to~$P$ almost surely. Since~$\mathcal P_\theta$ is closed by lemma~\ref{lem:mixture}(i), $\operatorname{dist}_{\mathcal P}(P,\mathcal P_\theta)>0$, and therefore condition~(\ref{eq:law-separation}) holds. Almost-sure convergence of~$\hat P_n$ to~$P$ also implies condition~(\ref{eq:law-tightness}).

If~$(\tilde u_i)_{i\geq1}$ is an i.i.d.\ sequence of draws from~$Q_U$ independent of the data, Assumption~\ref{ass:joint convergence} follows by the same conditional law of large numbers. If~$(\tilde u_i)_{i\geq1}$ is deterministic and its empirical distribution converges weakly to~$Q_U$, Lemma~\ref{lem:joint convergence} gives~$\hat P_{UX,n}\Rightarrow Q_U\times P_X$, whereas~$\hat P_{X,n}\Rightarrow P_X$; the triangle inequality for~$\rho_{\mathcal Q}$ then gives Assumption~\ref{ass:joint convergence}. All assumptions of Theorem~\ref{thm:new_consistency} are therefore satisfied, and that theorem yields~(\ref{eq:iid-consistency}).

\subsubsection*{Step 2: mixture representation of rejection probabilities}
Fix~$P^{(\infty)}\in\mathcal E^{(\infty)}$, with first-$n$ marginals~$P^{(n)}$ and de Finetti representation~(\ref{eq:deFinetti}) for a unique mixing measure~$\Lambda$. Let~$\phi$ be an arbitrary test sequence. By lemma~\ref{lem:mixture}(ii), the map~$P\mapsto\mathbb E_{P^{\otimes n}}[\phi_n]$ is Borel measurable on~$\mathcal P$. From representation~(\ref{eq:deFinetti}), 
we have:
\begin{eqnarray}
\label{eq:mixture}
\mathbb E_{P^{(n)}}\left[\phi_n\right] & = & \int_{\mathcal P}\mathbb E_{P^{\otimes n}}\left[\phi_n\right]\;\Lambda(dP), \quad\mbox{ for every } n\geq1.
\end{eqnarray}

\subsubsection*{Step 3: consistency of a size-controlling test forces~$\Lambda(\mathcal P_\theta)=0$}
By lemma~\ref{lem:mixture}(i), $\mathcal P_\theta$ is a Borel subset of~$\mathcal P$, so that~$\Lambda(\mathcal P_\theta)$ is well defined. Let~$\phi$ be a test sequence satisfying~(\ref{eq:ex-size}). For every~$P\in\mathcal P_\theta$, proposition~\ref{prop:iid} gives~$P^{\otimes n}\in\mathcal P_\theta^{(n)}$, and~$P^{\otimes n}$ is finitely exchangeable; hence~$P^{\otimes n}\in\mathcal P_\theta^{(n)}\cap\mathcal E^{(n)}$, and~(\ref{eq:ex-size}) yields~$\mathbb E_{P^{\otimes n}}[\phi_n]\leq\alpha$. Since~$\phi_n$ takes values in~$[0,1]$, splitting the integral in~(\ref{eq:mixture}) over~$\mathcal P_\theta$ and its complement gives, for every~$n\geq1$,
\begin{eqnarray}
\label{eq:mixture-bound}
\mathbb E_{P^{(n)}}\left[\phi_n\right] & \leq & \alpha\,\Lambda(\mathcal P_\theta)+\left(1-\Lambda(\mathcal P_\theta)\right) \; = \; 1-(1-\alpha)\,\Lambda(\mathcal P_\theta).
\end{eqnarray}
Suppose now that~$P^{(\infty)}\in\mathcal R_\theta(\phi)$. The left-hand side of~(\ref{eq:mixture-bound}) converges to one, whereas the right-hand side does not depend on~$n$. Since~$1-\alpha>0$, it follows that~$\Lambda(\mathcal P_\theta)=0$.

\subsubsection*{Step 4: consistency of~$\phi^{OT}$}
Continue with~$P^{(\infty)}\in\mathcal R_\theta(\phi)$ as in step~3, so that~$\Lambda(\mathcal P_\theta)=0$, i.e., $P\in\mathcal P\setminus\mathcal P_\theta$ for~$\Lambda$-almost every~$P$. By step~1, $\mathbb E_{P^{\otimes n}}[\phi_n^{OT}]\rightarrow1$ as~$n\rightarrow\infty$ for~$\Lambda$-almost every~$P$. These expectations take values in~$[0,1]$, so dominated convergence applied to~(\ref{eq:mixture}) with~$\phi=\phi^{OT}$ yields~$\mathbb E_{P^{(n)}}[\phi_n^{OT}]\rightarrow1$, i.e., $P^{(\infty)}\in\mathcal R_\theta(\phi^{OT})$. Since~$P^{(\infty)}$ was an arbitrary element of~$\mathcal R_\theta(\phi)$, (\ref{eq:optimality}) follows.
\end{proof}

\section{Additional results and lemmas required for the proofs}
\label{sec:lemmas}

\begin{proposition}\label{prop:iid}
    If~$P_0^{(n)}$ is i.i.d. with marginals~$P_0$, then~$P_0^{(n)}\in\mathcal P_\theta^{(n)}$ if and only if~$P_0\in\mathcal P_\theta$.
\end{proposition}

\begin{proof}[Proof of Proposition~\ref{prop:iid}] 
Let~$P_0^{(n)}$ be i.i.d. with marginals~$P_0$. By definition~\ref{def:DGP}, $P_0^{(n)}\in\mathcal P_\theta^{(n)}$ if and only if there exists a random vector~$(Y^{(n)},X^{(n)},U^{(n)})$ such that~$(Y^{(n)},X^{(n)})\sim P_0^{\otimes n}$, $U^{(n)}\sim Q_U^{\otimes n}$, $U^{(n)}\independent X^{(n)}$ and~$(Y_i,X_i,U_i)\in\Gamma(\theta)$ almost surely for each~$i$. In turn, this is equivalent to the existence of a random vector~$(Y,X,U)$ such that~$(Y,X)\sim P_0$, $U\sim Q_U$, $U\independent X$ and~$(Y,X,U)\in\Gamma(\theta)$ almost surely, which is the statement~$P_0\in\mathcal P_\theta$, and the result follows.
\end{proof}

\begin{proposition}\label{prop:closed}
    Let~$\Gamma$ be a subset of~$\mathcal Y\times\mathcal X\times\mathcal U$. There exists a function~$\delta: ((u,x),(y,x'))\in(\mathcal U\times\mathcal X)\times(\mathcal Y\times\mathcal X)\mapsto\delta((u,x),(y,x'))$ satisfying
    \begin{eqnarray}\label{eq:closed set}
        \delta\geq0, \mbox{ l.s.c., and }[\delta((u,x),(y,x'))=0 \iff (x=x' \mbox{ and }(y,x,u)\in\Gamma)], 
    \end{eqnarray}
    if and only if~$\Gamma$ is closed.
\end{proposition}

\begin{proof}[Proof of proposition~\ref{prop:closed}]
    Suppose~$\delta$ is a function satisfying~(\ref{eq:closed set}). Since \(\delta\) is lower semi-continuous and nonnegative, $A:=\{((u,x),(y,x')):\delta((u,x),(y,x'))\le 0\}$ is closed. By assumption, $A=\{((u,x),(y,x')):x=x',\ (y,x,u)\in\Gamma\}.$ Define the continuous map~$S(y,x,u)=((u,x),(y,x)).$ Then~$(y,x,u)\in\Gamma\iff S(y,x,u)\in A,$ hence~$\Gamma=S^{-1}(A).$ Since \(A\) is closed and \(S\) is continuous, \(\Gamma\) is closed.

    Conversely, suppose~$\Gamma$ is closed. Define~$A:=\{((u,x),(y,x')):x=x',\ (y,x,u)\in\Gamma\}$. We show that the indicator function~$1_{A^c}$ satisfies~(\ref{eq:closed set}). By construction, $1_{A^c}\geq0$ and $1_{A^c}((u,x),(y,x'))=0$ if and only if~$x=x'$ and~$(y,x,u)\in\Gamma$. There remains to show that~$1_{A^c}$ is lower semi-continuous, or, equivalently, that~$A$ is closed. This follows from the fact that~$A:=\{((u,x),(y,x')): x=x'\}\cap T^{-1}(\Gamma)$, where~$T((u,x),(y,x'))=(y,x,u)$ is continuous, and the diagonal~$\{((u,x),(y,x')): x=x'\}$ is closed.
\end{proof}

\begin{lemma}\label{lem:joint convergence}
Suppose~$(X_i)_{i\geq1}$ is stationary with marginal distribution~$P_{X,0}$ and is strong mixing with absolutely summable mixing weights. Suppose~$(\tilde{u}_i)_{i\geq1}$ is a deterministic sequence such that its empirical distribution~$\hat{Q}_n$ converges weakly to~$Q_U$. Then, the joint empirical distribution~$\mu_n$ of~$(\tilde{u}_i,X_i)_{i\leq n}$ converges weakly to~$Q_U\times P_{X,0}$ almost surely.
\end{lemma}

\begin{proof}[Proof of lemma~\ref{lem:joint convergence}]
Take an arbitrary~$f\in C_b(\mathcal{U}\times \mathcal{X})$ and define
\[
    m_f(u) \coloneqq \mathbb E[f(u,X_1)] = \int_{\mathcal{X}} f(u,x)\,P_{X,0}(\ud x), \qquad u\in \mathcal{U}.
\]
Since $f$ is bounded and continuous, $m_f$ is bounded and continuous. Indeed, if $u_k\to u$, then $f(u_k,X_1)\to f(u,X_1)$ almost surely, and dominated convergence gives $m_f(u_k)\to m_f(u)$.

Now decompose
\begin{align*}
    &\frac{1}{n}\sum_{i=1}^n f(\tilde{u}_i,X_i) - \int\int f(u,x)\,P_{X,0}(\ud x)\,Q_U(\ud u) \\
    &\qquad =
    \underbrace{\frac{1}{n}\sum_{i=1}^n \left\{f(\tilde{u}_i,X_i)-m_f(\tilde{u}_i)\right\}}_{A_n(f)}
    +
    \underbrace{\left[ \frac{1}{n}\sum_{i=1}^n m_f(\tilde{u}_i) - \int m_f(u)\,Q_U(\ud u) \right]}_{B_n(f)}.
\end{align*}
Since $m_f\in C_b(\mathcal{U})$ and $\hat{Q}_n$ converges weakly to $Q_U$, we know $B_n(f)\to 0$.
It remains to control $A_n(f)$. Let~$Z_i := f(\tilde{u}_i,X_i)-m_f(\tilde{u}_i),\;i\geq 1.$ Let~$M\coloneqq \norm{f}_\infty$. We have~$|Z_i|\leq 2M$ for all~$i\geq1$ and~$(Z_i)_{i\geq1}$ is a mean-zero strong mixing sequence with mixing weights dominated by those of~$(X_i)_{i\geq1}$. 

Let $\mathcal F_0\subset C_b(\mathcal U\times\mathcal X)$ be a countable
convergence-determining class for probability measures on the separable metric
space $\mathcal U\times\mathcal X$. For each $f\in\mathcal F_0$, the preceding
argument and Corollary~3.2(ii) of \cite{rio2017asymptotic} imply that~$A_n(f)\longrightarrow 0$
almost surely,
whereas~$\widehat Q_n\Rightarrow Q_U$ and $m_f\in C_b(\mathcal U)$ imply that~$B_n(f)\longrightarrow 0.$
Since $\mathcal F_0$ is countable, there exists an event of probability one on
which
\[
\int f\,d\mu_n
\longrightarrow
\int f\,d\bigl(Q_U\otimes P_{X,0}\bigr)
\]
for every $f\in\mathcal F_0$. Since $\mathcal F_0$ is convergence determining,
it follows that~$\mu_n\Rightarrow Q_U\otimes P_{X,0}$ almost surely as desired.
\end{proof}

\begin{lemma}\label{lem:mixture}
Suppose assumption~\ref{ass:discrepancy} holds. Then:
\begin{enumerate}[label=\textnormal{(\roman*)}]
    \item $\mathcal P_\theta$ is a closed, hence Borel, subset of~$\mathcal P$;
    \item for every~$n\geq1$ and every Borel function~$\phi_n:\mathcal Y^n\times\mathcal X^n\rightarrow[0,1]$, the map~$P\mapsto\mathbb E_{P^{\otimes n}}[\phi_n]$ is Borel measurable on~$\mathcal P$.
\end{enumerate}
\end{lemma}

\begin{proof}[Proof of lemma~\ref{lem:mixture}]
We first prove~(i). By propositions~\ref{prop:iid} and~\ref{prop:char1}, $P\in\mathcal P_\theta$ if and only if~$\mathcal D(Q_U\times P_X,P;\theta)=0$, where~$P_X$ denotes the~$X$-marginal of~$P$. The map~$P\mapsto Q_U\times P_X$ is weakly continuous: if~$P_k\rightarrow P$ weakly, then, for any bounded continuous~$f$ on~$\mathcal U\times\mathcal X$, the function~$x\mapsto\int f(u,x)\,Q_U(du)$ is bounded and continuous by dominated convergence, so that~$\int f\,d(Q_U\times P_{X,k})\rightarrow\int f\,d(Q_U\times P_X)$. Under assumption~\ref{ass:discrepancy}, the cost~$\delta(\cdot,\cdot;\theta)$ is nonnegative and lower semi-continuous, and Kantorovich duality (theorem~1.3 of \cite{Villani:2003}) gives
\begin{eqnarray*}
\mathcal D(Q,P;\theta) & = & \sup\left\{ \int f\,dQ+\int g\,dP: \; (f,g)\in C_b(\mathcal U\times\mathcal X)\times C_b(\mathcal Y\times\mathcal X), \; f\oplus g\leq\delta(\cdot,\cdot;\theta) \right\},
\end{eqnarray*}
where~$(f\oplus g)(v,w):=f(v)+g(w)$. Each admissible pair~$(f,g)$ defines a weakly continuous function of~$(Q,P)$, so that, as a supremum of continuous functions, $(Q,P)\mapsto\mathcal D(Q,P;\theta)$ is lower semi-continuous for the weak topology. Composing with the continuous map~$P\mapsto(Q_U\times P_X,P)$, the map~$P\mapsto\mathcal D(Q_U\times P_X,P;\theta)$ is lower semi-continuous on~$\mathcal P$. Since the discrepancy is nonnegative, $\mathcal P_\theta=\{P\in\mathcal P:\mathcal D(Q_U\times P_X,P;\theta)\leq0\}$, which is closed as the lower level set of a lower semi-continuous function, and~(i) follows.

We now prove~(ii). As in section~\ref{sec:optimality}, we identify~$\mathcal Y^n\times\mathcal X^n$ with~$(\mathcal Y\times\mathcal X)^n$ through the canonical reordering of coordinates. Let~$\mathcal H$ be the collection of bounded Borel functions~$h$ on~$\mathcal Y^n\times\mathcal X^n$ such that~$P\mapsto\int h\,dP^{\otimes n}$ is Borel measurable on~$\mathcal P$. The collection~$\mathcal H$ is a vector space, contains the constant functions, and is closed under monotone limits of uniformly bounded sequences, since pointwise limits preserve Borel measurability. Let~$\mathcal M$ be the collection of functions of the form~$h(w_1,\ldots,w_n)=\prod_{i\leq n}h_i(w_i)$ with~$h_i\in C_b(\mathcal Y\times\mathcal X)$ for each~$i\leq n$. The collection~$\mathcal M$ is closed under multiplication, and, for~$h\in\mathcal M$, $\int h\,dP^{\otimes n}=\prod_{i\leq n}\int h_i\,dP$ is a finite product of weakly continuous functions of~$P$, hence continuous, so that~$\mathcal M\subseteq\mathcal H$. By the functional monotone class theorem, $\mathcal H$ contains every bounded~$\sigma(\mathcal M)$-measurable function. Since~$C_b(\mathcal Y\times\mathcal X)$ generates the Borel sigma-field of~$\mathcal Y\times\mathcal X$, and the~$n$-fold product of the Borel sigma-field of the separable metric space~$\mathcal Y\times\mathcal X$ is the Borel sigma-field of the product space, $\sigma(\mathcal M)$ is the Borel sigma-field of~$\mathcal Y^n\times\mathcal X^n$. Claim~(ii) follows by taking~$h=\phi_n$.
\end{proof}

\section{Conditional size control}
\label{sec:add}

In this section, we provide size control results that are analogous to Theorem~\ref{thm:size}, except for the conditioning on the sample~$X^{(n)}$ of realized covariates. The latter enables us to use the data-driven discrepancy~(\ref{eq:discrepancy}). Denote~$\mathcal R^{(n)}$ the set of Borel probability measures on~$\mathcal X^n$. Define the following set of conditional distributions:
\begin{eqnarray*}
\mathcal P^{(n)}(X^{(n)}) & := & \left\{ P^{(n)}_{Y^{(n)}\vert X^{(n)}}:\; \exists P^{(n)}_X\in\mathcal R^{(n)}, \mbox{ s.t. }
P^{(n)}_{Y^{(n)}\vert X^{(n)}}\times P^{(n)}_X \in \mathcal P^{(n)}_\theta \right\}.
\end{eqnarray*}

\begin{theorem}
\label{thm:cond}
For all~$\theta\in\Theta$, all~$\alpha\in(0,1)$ and all~$n\in\mathbb N$ such that~$\mathcal P_\theta^{(n)}$ is nonempty, the confidence region~$CR_n$ defined in~(\ref{eq:CR}) has correct coverage probability,
\begin{eqnarray}
\inf_{P^{(n)}_{Y^{(n)}\vert X^{(n)}}\in\mathcal P^{(n)}(X^{(n)})} P^{(n)}_{Y^{(n)}\vert X^{(n)}} \left( \; T_n(\theta)\leq c_{n,1-\alpha}(\theta) \; \vert X^{(n)} \; \right) & \geq & 1-\alpha,
\end{eqnarray}
with equality if the cumulative distribution function of~$\tilde T_n(\theta)$ conditional on~$X^{(n)}$ is continuous and increasing in a neighborhood of~$c_{n,1-\alpha}(\theta)$.
\end{theorem}

\begin{proof}[Proof of Theorem~\ref{thm:cond}]
Let~$X^{(n)}$ be the sample of observed covariates. We fix an arbitrary~$\theta$ such that~$\mathcal P_\theta^{(n)}$ is nonempty and an arbitrary~$\alpha\in(0,1)$.
Take an arbitrary distribution $P^{(n)}_{Y^{(n)}\vert X^{(n)}}$ in $\mathcal P^{(n)}(X^{(n)})$, and let~$Y^{(n)}$ be a random vector distributed according to~$P^{(n)}_{Y^{(n)}\vert X^{(n)}}$. 
Let~$T_n(\theta)$ be the test statistic constructed from~$(Y^{(n)},X^{(n)})$. 
The proof then proceeds as in Theorem~\ref{thm:size}.
\end{proof}

\section{Numerical implementation}
\label{sec:num}

\subsection*{Test statistic}

Computation of the test statistic requires computing the simulated i.i.d. sequence or low discrepancy sequence~$\tilde u^{(n)}$, computing cost matrix~(\ref{eq:CM}), and solving optimization problem~$(\ref{eq:OT})$. 
The sequence~$\tilde u^{(n)}=(\tilde u_1,\ldots,\tilde u_n)$ is constructed for each~$i=1,\ldots,n,$ as follows. A deterministic sequence~$\xi^{(n)}:=(\xi_1,\ldots,\xi_n)$ of points in~$[0,1]^{d_U}$ is derived in such a way that its empirical distribution approximates the distribution of the uniform on~$[0,1]^{d_U}$ well. Such a sequence is called a quasi-random or low discrepancy sequence.\footnote{See for instance \cite{niederreiter1992random}. Sobol and Halton  sequence generators are available in most packages.} 
Each element of that sequence is then transformed using a map that pushes the uniform ~$U[0,1]^{d_U}$ to~$Q_U$. In many cases, this map can be very simple. For instance, if~$U$ has independent marginals, the componentwise quantile function is suitable. If~$Q_U$ is a multivariate normal, we can use the composition of quantiles of the standard normal distribution with the linear transformation from the multivariate standard normal to~$Q_U$, as we do in the simulations. More generally, we can set~$\tilde u_i := \nabla\psi_U(\xi_i),$ where~$\nabla\psi_U$ is the unique gradient of a convex function that pushes the uniform~$U[0,1]^{d_U}$ to~$Q_U$ (see \cite{McCann:95} and \cite{CGHH:2017}).

The cost matrix~$C(\theta)$ is computed with respect to the discrepancy~(\ref{eq:discrepancy}) or (\ref{eq:alternative discrepancy}) depending on the application. Discrepancy~(\ref{eq:alternative discrepancy}) is preferred to~(\ref{eq:discrepancy}) only in cases where~$\Gamma_u(y,x;\theta)$ is much more costly to compute than~$\Gamma_y(u,x;\theta)$.
Optimization problem~$(\ref{eq:OT})$ is a discrete optimal-transport problem, which is a special kind of linear programming problem. There is a large literature on its implementation, reviewed in part in \cite{PC:2019}. 
Discrete optimal-transport problems are equivalent to \emph{assignment} problems, for which many
efficient algorithms exist in the literature. Efficient ready-to-use implementations abound. In the simulations and empirical application, we use the Julia implementation assignment.jl of a variant of the shortest path algorithm in \cite{jonker1987shortest}.

\subsection*{Critical values}

The generic simulation procedure to compute critical value~$c_{n,1-\alpha}$ is the following.
\begin{enumerate}
\item Generate~$S=1,000$ independent Monte Carlo latent samples~$\tilde U^{(s)}:=(\tilde U_j^s)_{j\leq n}$.
\item For each~$s\in\{1,\ldots,S\}$, compute 
\begin{eqnarray}\label{eq:T_tilde_computation_123}
    \tilde T_n^s(\theta) & = & \sup_{\tilde y\in\Gamma_y^{(n)}(\tilde U^{(s)},X^{(n)};\theta)}\mathcal D_n(C(\tilde y;\theta)), 
\end{eqnarray}
and let~$\tilde T^{(s)}_n(\theta)$, $s=1,\ldots,S,$ be the order statistics.
\item The critical value~$c_{n,1-\alpha}(\theta)$ is approximated with
$\hat c_{n,1-\alpha}(\theta):=\tilde T_n^{(\lceil S(1-\alpha)\rceil)}(\theta).$
\end{enumerate}

In step~(2) above, we use the following algorithm for statistic~$\tilde T_n(\theta)$.
First, by the Kantorovich duality of optimal transport, we have:
\begin{eqnarray*}
    \mathcal D_n(C(\tilde y;\theta)) & = & \sup_{\alpha,\beta} \left( \frac{1}{n}\sum_i\alpha_i+\frac{1}{n}\sum_j\beta_j\right) \mbox{ s.t. }\alpha_i+\beta_j\leq C_{ij}(\tilde y;\theta).
\end{eqnarray*}
We replace the latter with the regularized version
\begin{eqnarray*}
    V_\varepsilon(\tilde y^{(n)};\theta) & := & \max_{\alpha,\beta} \left( \frac{1}{n}\sum_i\alpha_i+\frac{1}{n}\sum_j\beta_j-\frac{1}{2\varepsilon}\sum_{i,j}\max\left(0,  \alpha_i+\beta_j - C_{ij}(\tilde y^{(n)};\theta) \right)^2\right),
\end{eqnarray*}
which is a continuously differentiable concave maximization problem. The approximation error is controlled by the following:\footnote{In the simulations, we set~$\varepsilon$ such that~$\varepsilon/(2n) = 0.0005$.} 
\begin{eqnarray*}
    \mathcal D_n(C(\tilde y^{(n)};\theta)) & \in & \left[V_\varepsilon(\tilde y^{(n)};\theta)-\frac{\varepsilon}{2n},V_\varepsilon(\tilde y^{(n)};\theta)\right],
\end{eqnarray*}

Since~$\mathcal D_n(C(\tilde y^{(n)};\theta))\leq V_\varepsilon(\tilde y^{(n)};\theta)$, using~$V_\varepsilon$ in the computation of the simulated critical values is conservative and preserves finite-sample validity. As such, $\tilde T_n^{s}(\theta)$ can be approximated by the maximum of~$V_\varepsilon(\tilde y^{(n)};\theta)$ over~$\tilde y^{(n)}\in\Gamma^{(n)}_y(\tilde U^{(s)},X^{(n)};\theta)$. This computation involves a single joint maximization over~$(\alpha,\beta,\tilde y^{(n)})$.

The implementation depends on the structure of~$\Gamma_y^{(n)}(\tilde U^{(s)},X^{(n)};\theta)$.
\begin{itemize}
    \item If the components of~$\tilde y_j$ are continuous as in our simulation exercises, we solve the $\max_{\tilde{y}^{(n)}}V_\epsilon(\tilde{y}^{(n)}; \theta)$ using L-BFGS methods from multiple starting points and use it approximate $\tilde T_n^{s}(\theta)$. In the interval-regression simulations, we solve L-BFGS optimization from two starting points and keep the larger solution.

    \item If~$\tilde y_j$ is discrete and the total cardinality of~$\Gamma_y^{(n)}(\tilde U^{(s)},X^{(n)};\theta)$ is small, as is almost always the case in our empirical application, we compute~\eqref{eq:T_tilde_computation_123} directly by enumeration over~$\Gamma_y^{(n)}(\tilde U^{(s)},X^{(n)};\theta)$.

    \item If~$\tilde y_j$ is discrete but the cardinality of~$\Gamma_y^{(n)}(\tilde U^{(s)},X^{(n)};\theta)$ is too large for enumeration\footnote{We set a cutoff cardinality of~$10,000$ in our implementation.}, we use a convexified relaxation. Let~$\tilde y_{j,k}$ denote the~$k$th equilibrium outcome associated with observation~$j$ in Monte Carlo sample~$s$, for~$k=1,\ldots,K_j$. Let

    \[
    \Delta_j\coloneqq \left\{ \omega_j\in\mathbb R_+^{K_j}:\sum_{k=1}^{K_j}\omega_{j,k}=1 \right\}
    \]
    be the probability simplex over the equilibrium outcomes associated with observation~$j$. We then solve
    \begin{eqnarray*}
              \overline{T}^s_n(\theta)
              &=&
              \sup_{\alpha,\beta,\omega}
              \left[
              \frac{1}{n}\sum_i\alpha_i+\frac{1}{n}\sum_j\beta_j
               -
               \frac{1}{2\varepsilon} \sum_{i,j} \max\left\{ 0,\,\alpha_i+\beta_j-
               \sum_{k=1}^{K_j}\omega_{j,k} C_{ij}(\tilde y_{j,k};\theta)
               \right\}^2
               \right] \\
       && \mbox{s.t. } \omega_j\in \Delta_j
       \quad\text{for all }j\leq n.
    \end{eqnarray*}
    This relaxation replaces the discrete selection of an equilibrium outcome for each observation by a convex combination over that observation's equilibrium outcomes. The objective is jointly concave in~$(\alpha,\beta,\omega)$, so the problem is computationally tractable. Because the vertices of~$\Delta_j$ correspond to the original discrete equilibrium selections, the relaxation satisfies $\overline{T}^s_n(\theta)\geq \tilde T_n^s(\theta)$.
    Consequently, using~$\overline{T}^s_n(\theta)$ to compute the simulated critical value may be conservative, but it preserves size control.
\end{itemize}

\subsection*{Weighting matrices in the interval-regression simulations}
In the interval-regression simulations, OTE uses identity weighting. OTW uses the block-diagonal precision matrix~$\operatorname{diag}(1,A_X)$, where~$\hat\Sigma_X$ is the sample covariance matrix of~$(\underline W,\overline W)$, computed with denominator~$n-1$. If the condition number of~$\hat\Sigma_X$ is below~$10^{10}$, we set~$A_X=\hat\Sigma_X^{-1}$. Otherwise, we set $A_X=\operatorname{diag}(1/\widehat{\operatorname{Var}}(\underline W),0)$. For OTY, at each candidate value~$\theta$, we calculate the covariance matrix~$\tilde\Sigma$ of~$(Y,\underline W,\overline W)$ analytically under a working model that assigns probability~$1/n$ to each observed endpoint pair, draws~$W^\ast$ uniformly within the corresponding interval, and sets~$Y=\theta W^\ast+U$, with~$U\sim N(0,1)$ independent of the endpoints and~$W^\ast$. As discussed in the paper, the theoretical size-control result remains valid when this working model differs from the actual data generating process. To guard against rank deficiency of $\tilde\Sigma$, we use $(\tilde\Sigma+10^{-12}I_3)^{-1}$ instead of $\tilde\Sigma^{-1}$ directly.

\section{Additional details for the consideration-set simulation}
\label{sec:cs-appendix}

\subsection{Size distortion under the ``Trending Now'' selection mechanism}
\label{sec:cs-size-distortion}

We maintain the notation of Section~\ref{sec:MC-consideration}. Let
\[
\widehat p_{n,d}=\frac{1}{n}\sum_{i=1}^n1\!\{Y_{n,i}=d\},
\qquad
\widehat g_{n,d}=\frac{2}{3}-\widehat p_{n,d},
\qquad
\widehat\sigma_{n,d}=\{\widehat p_{n,d}(1-\widehat p_{n,d})\}^{1/2}.
\]
The hard-GMS implementation of \cite{AS:2010} retains the moments
\begin{eqnarray}
\label{eq:cs-gms-set}
\widehat{\mathcal B}_n
=\left\{d\in\mathcal D:
\frac{\sqrt n\,\widehat g_{n,d}}{\widehat\sigma_{n,d}}
\leq\kappa_n\right\},
\end{eqnarray}
where~$\kappa_n=\sqrt{\log n}$ in the simulations. Its max statistic is
\begin{eqnarray}
\label{eq:cs-as-statistic}
T_n^{AS}
=\left[
\max\left\{0,\max_{d\in\mathcal D}
\frac{\sqrt n\,(\widehat p_{n,d}-2/3)}{\widehat\sigma_{n,d}}
\right\}
\right]^2.
\end{eqnarray}
Each bootstrap draw is studentized using its own bootstrap standard deviation. The code uses the signed-infinity extension of the studentized statistic when a sample or bootstrap variance is zero. This convention matters in very small samples but is asymptotically irrelevant here.

\begin{proposition}
\label{prop:cs-as-distortion}
Suppose~$\kappa_n\rightarrow\infty$ and~$\kappa_n/\sqrt n\rightarrow0$. Under the ``Trending Now'' DGP,
\[
\mathbb P\{\widehat{\mathcal B}_n=\{d_n^{\mathrm{tr}}\}\}\longrightarrow1
\]
and
\begin{eqnarray}
\label{eq:cs-as-limit}
T_n^{AS}\quad\Longrightarrow\quad
\left(\max_{d\in\mathcal D}Z_d\right)^2,
\qquad
(Z_1,Z_2,Z_3)\sim N(0,\Omega),
\end{eqnarray}
where
\[
\Omega=\begin{pmatrix}
1&-1/2&-1/2\\
-1/2&1&-1/2\\
-1/2&-1/2&1
\end{pmatrix}.
\]
If the number of i.i.d. bootstrap draws tends to infinity so that the simulated critical value converges to its ideal GMS-bootstrap counterpart, the limiting rejection probability at nominal size~$\alpha\in(0,1/2)$ is
\begin{eqnarray}
\label{eq:cs-as-rejection-limit}
3\alpha-3\mathbb P\{Z_1>z_{1-\alpha},Z_2>z_{1-\alpha}\},
\end{eqnarray}
where~$z_{1-\alpha}$ is the~$1-\alpha$ standard-normal quantile and~$\operatorname{corr}(Z_1,Z_2)=-1/2$. The expression in~(\ref{eq:cs-as-rejection-limit}) is~$0.2978$, $0.1498$, and~$0.0300$ when~$\alpha$ is~$0.10$, $0.05$, and~$0.01$, respectively.
\end{proposition}

\begin{proof}
Equation~(\ref{eq:cs-selection-bias}) implies~$\widehat p_{n,d_n^{\mathrm{tr}}}\geq2/3$, so the winning moment is always retained by~(\ref{eq:cs-gms-set}). For every~$d\neq d_n^{\mathrm{tr}}$, the same equation gives~$\widehat p_{n,d}\leq1/3$, hence~$\widehat g_{n,d}\geq1/3$. Since~$\widehat\sigma_{n,d}\leq1/2$, the standardized slackness of every nonwinning moment is at least~$2\sqrt n/3$. The condition~$\kappa_n/\sqrt n\rightarrow0$ therefore implies that both nonwinning moments are eventually discarded.

Define the latent inducibility indicators by
\[
A_{i,d}=1\!\{d\in D_{2,i}^\ast\},\qquad d\in\mathcal D.
\]
Their mean vector is~$(2/3,2/3,2/3)$, their variances are~$2/9$, and their pairwise covariances are~$-1/9$. Since~$\widehat p_{n,d}-2/3\leq-1/3$ for each nonwinning product, whereas~$\widehat p_{n,d_n^{\mathrm{tr}}}-2/3\geq0$, equations~(\ref{eq:cs-trending-score}) and~(\ref{eq:cs-selection-bias}) give
\[
\sqrt{T_n^{AS}}
=\frac{\sqrt n\{\max_d\widehat S_{n,d}-2/3\}}
{\widehat\sigma_{n,d_n^{\mathrm{tr}}}}.
\]
The law of large numbers gives~$\widehat\sigma_{n,d_n^{\mathrm{tr}}}\rightarrow_p\sqrt2/3$. The multivariate central limit theorem applied to~$(A_{i,1},A_{i,2},A_{i,3})$ and the continuous mapping theorem then yield~(\ref{eq:cs-as-limit}). Notice that~$Z_1+Z_2+Z_3=0$ almost surely, so~$\max_dZ_d\geq0$.

Once GMS retains only~$d_n^{\mathrm{tr}}$, the i.i.d. bootstrap treats the selected coordinate as a single binding moment. Its ideal critical value for~$T_n^{AS}$ therefore converges to~$z_{1-\alpha}^2$. Rejection occurs asymptotically when~$\max_dZ_d>z_{1-\alpha}$. Because~$z_{1-\alpha}>0$ and~$Z_1+Z_2+Z_3=0$, all three coordinates cannot exceed the threshold. Inclusion--exclusion and symmetry yield~(\ref{eq:cs-as-rejection-limit}).
\end{proof}

\subsection{Discrepancies and exact computation of the max--min statistic}
\label{sec:cs-computation}

We first describe the two discrepancies for a general finite product set~$\mathcal D=\{1,\ldots,J\}$ and minimum consideration-set size~$\kappa$. Write~$Q=J-\kappa+1$. For a latent vector~$u$ and parameter~$\theta$, write~$v_d(u;\theta)=\theta_d+u(d)$ for the total utility of product~$d$. Under strict utilities, a product is inducible if and only if it is among the consumer's top~$Q$ products. Let
\[
\mathcal A(u;\theta)=\{d\in\mathcal D:v_d(u;\theta)\mbox{ is among the top }Q\mbox{ total utilities}\}.
\]
For the latent-space discrepancy, fix product~$d$, let~$a=v_d(u;\theta)$, and order the other total utilities as~$b_{(1)}\leq\cdots\leq b_{(J-1)}$. Product~$d$ is inducible if it weakly dominates at least~$\kappa-1$ competitors. Its squared Euclidean distance to the rationalizing region is therefore
\begin{eqnarray}
\label{eq:cs-latent-divergence}
\delta_U(u,d;\theta)
=\min_{t\in\mathbb R}
\left\{(t-a)^2+
\sum_{\ell=1}^{\kappa-1}(b_{(\ell)}-t)_+^2\right\},
\qquad (x)_+=\max\{x,0\}.
\end{eqnarray}
Only the~$\kappa-1$ lowest-utility competitors can enter the least-cost adjustment. The minimizer follows from a one-dimensional active-set calculation. If~$A(t)=\{\ell\in\{1,\ldots,\kappa-1\}:b_{(\ell)}>t\}$ is the active set, then
\[
t=\frac{a+\sum_{\ell\in A(t)}b_{(\ell)}}{1+|A(t)|},
\]
and the inequalities defining~$A(t)$ identify the solution. In the simulation, $J=3$ and~$\kappa=2$. For any~$\theta$, the discrepancy has the closed form
\[
\delta_U(u,d;\theta)
=\frac12\left[\min_{c\neq d}v_c(u;\theta)-v_d(u;\theta)\right]_+^2,
\]
It is zero unless~$d$ is strictly least preferred under~$(u,\theta)$; if it is, the discrepancy is one half of the squared gap between the two lowest total utilities. The outcome-space discrepancy is simply
\begin{eqnarray}
\label{eq:cs-outcome-divergence}
\delta_Y(u,d;\theta)=1\!\{d\notin\mathcal A(u;\theta)\}.
\end{eqnarray}

We now give the exact max--min calculation. Fix the reference latent sample and let~$C=(c_{id})$ be either discrepancy's~$n\times J$ cost table. For a possibly fractional product-count vector~$r=(r_1,\ldots,r_J)$ with nonnegative entries summing to~$n$, define the aggregated transport value
\begin{eqnarray}
\label{eq:cs-aggregated-transport}
\phi_C(r)=\frac{1}{n}\min_{x\geq0}
\sum_{i=1}^n\sum_{d=1}^Jc_{id}x_{id}
\quad\mbox{s.t.}\quad
\sum_{d=1}^Jx_{id}=1,\quad
\sum_{i=1}^nx_{id}=r_d.
\end{eqnarray}
If~$z_d=\sum_{j=1}^n1\!\{Y_j=d\}$, the observed test statistic is~$T_n=\phi_C(z)$. For one critical-value draw, let~$\mathcal S_j$ be the set of outcomes predicted by the~$j$th Monte Carlo latent vector. The outer maximization in~(\ref{eq:cs-max-min-main}) permits any~$y_j\in\mathcal S_j$. Write
\[
r(y)=\sum_{j=1}^ne_{y_j},
\qquad
\mathcal R=\{r(y):y_j\in\mathcal S_j,\ j\leq n\},
\]
where~$e_d$ is the~$d$th coordinate vector.

\begin{proposition}[Exact priority reduction]
\label{prop:cs-priority-reduction}
Let~$\mathfrak S_J$ be the collection of strict priority orderings of~$\mathcal D$. For~$\rho\in\mathfrak S_J$, let~$\operatorname{top}_\rho(\mathcal S)$ be the highest-priority element of~$\mathcal S$, and define
\[
r^\rho=\sum_{j=1}^ne_{\operatorname{top}_\rho(\mathcal S_j)}.
\]
If the target-column cost depends on~$j$ only through its selected label, as in~$C_{ij}(y_j)=c_{i,y_j}$, then
\begin{eqnarray}
\label{eq:cs-priority-reduction}
\max_{y_j\in\mathcal S_j,\,j\leq n}
\min_{\pi\in\Pi_n}\sum_{i,j}\pi_{ij}c_{i,y_j}
=\max_{r\in\mathcal R}\phi_C(r)
=\max_{\rho\in\mathfrak S_J}\phi_C(r^\rho).
\end{eqnarray}
Thus the max--min problem is solved exactly by checking at most~$J!$ count vectors.
\end{proposition}

\begin{proof}
Columns carrying the same product label have identical costs, so aggregating their transport flows gives the first equality in~(\ref{eq:cs-priority-reduction}). The transport dual is
\[
\phi_C(r)=\frac{1}{n}\max_{\alpha,\beta}
\left\{\sum_i\alpha_i+\sum_dr_d\beta_d:
\alpha_i+\beta_d\leq c_{id}\right\}.
\]
It follows that~$\phi_C$ is convex in~$r$. The convex hull of the feasible counts is the Minkowski sum
\[
\operatorname{co}(\mathcal R)
=\sum_{j=1}^n\operatorname{co}\{e_d:d\in\mathcal S_j\}.
\]
Every vertex of this polytope is generated by a strict global priority: maximizing a generic linear functional over the Minkowski sum selects its highest-priority admissible product separately from each~$\mathcal S_j$. Conversely, a small perturbation of a linear functional exposing any vertex makes its coordinates strictly ordered without changing the selected vertex. Since a convex function attains its maximum over a finite polytope at a vertex, the last equality in~(\ref{eq:cs-priority-reduction}) follows. Every priority-induced vertex is an original integer count vector, so this argument neither interchanges the maximum and minimum nor relaxes the outer selection problem.
\end{proof}

The implementation of Proposition~\ref{prop:cs-priority-reduction} enumerates the~$J!$ priorities, forms~$r^\rho$ for each, discards duplicate count vectors, solves~(\ref{eq:cs-aggregated-transport}), and retains the largest value. Because~$J=3$ and every~$\mathcal S_j$ is a two-product set in the simulation, every priority-induced count vector has at most two positive entries. With one active product the transport value is a direct sum. If products~$a$ and~$b$ are active and~$r_a=m$, order~$\Delta_i=c_{ia}-c_{ib}$ as~$\Delta_{(1)}\leq\cdots\leq\Delta_{(n)}$. Then the inner problem has the closed form
\begin{eqnarray}
\label{eq:cs-two-label-transport}
\phi_C(r)=\frac{1}{n}\left\{
\sum_{i=1}^nc_{ib}+\sum_{\ell=1}^m\Delta_{(\ell)}
\right\}.
\end{eqnarray}
Hence OT$_U$ checks at most six count vectors and solves each by a sort.

For OT$_Y$, an additional simplification is available. Let~$\mathcal A_i$ denote the reference admissible set and define, for nonempty~$B\subseteq\mathcal D$,
\[
H_{\mathcal A}(B)=\sum_{i=1}^n1\!\{\mathcal A_i\cap B\neq\varnothing\},
\qquad
H_{\mathcal S}(B)=\sum_{j=1}^n1\!\{\mathcal S_j\cap B\neq\varnothing\}.
\]
The minimum number of unit-cost matches under~(\ref{eq:cs-outcome-divergence}) is the largest Hall deficiency. Maximizing that deficiency over admissible selections gives
\begin{eqnarray}
\label{eq:cs-outcome-max-min}
\tilde T_{n,Y}
=\frac{1}{n}\max_{\varnothing\neq B\subseteq\mathcal D}
\left[H_{\mathcal S}(B)-H_{\mathcal A}(B)\right]_+.
\end{eqnarray}
For~$J\leq20$, the code evaluates the~$2^J-1$ nonempty subsets directly; this requires only seven subsets in the simulation. For larger~$J$, the implementation falls back on the priority reduction when enumeration of the priority orderings remains feasible.

Proposition~\ref{prop:cs-priority-reduction} applies to the case without any covariates. Observation-specific covariates generally produce costs~$c_{ijd}$, so product counts cease to be sufficient. With finite outcomes, the exact max--min problem can then instead be written as the mixed-integer program
\begin{equation}
\label{eq:cs-covariate-milp}
\begin{aligned}
&\max_{\alpha,\beta,z}\quad
\frac{1}{n}\left(\sum_i\alpha_i+\sum_j\beta_j\right)\\
&\quad\mbox{s.t.}\quad
\alpha_i+\beta_j\leq\sum_{d\in\mathcal S_j}c_{ijd}z_{jd},
\qquad
\sum_{d\in\mathcal S_j}z_{jd}=1,
\qquad z_{jd}\in\{0,1\}.
\end{aligned}
\end{equation}
This formulation follows from the assignment dual and remains exact, but it has at most~$nJ$ binary selectors and~$n^2$ dual constraints.

\end{appendix}

\bibliographystyle{abbrvnat}
\bibliography{Wasserstein}

\end{document}